\documentclass{egpubl}
\usepackage{egsr2022}
\SpecialIssuePaper

\usepackage[T1]{fontenc}
\usepackage{dfadobe}  
\usepackage{cite}

\usepackage[dvipsnames]{xcolor}

\BibtexOrBiblatex
\electronicVersion
\PrintedOrElectronic
\ifpdf \usepackage[pdftex]{graphicx} \pdfcompresslevel=9
\else \usepackage[dvips]{graphicx} \fi

\usepackage{booktabs} %
\usepackage{diagbox}
\usepackage{subfig}
\usepackage{multirow}
\usepackage{mathrsfs}
\usepackage{graphicx}
\usepackage{enumitem}
\usepackage{soul}
\usepackage[normalem]{ulem}
\usepackage{egweblnk} 

\begin{document}
\captionsetup{labelfont=bf,textfont=it}

\author[Y. Hu, M. Ha\v{s}an, P. Guerrero, H. Rushmeier \& V. Deschaintre]
{\parbox{\textwidth}{\centering Yiwei Hu$^{1,2}$, Milo\v{s} Ha\v{s}an$^{1}$, Paul Guerrero$^{1}$, Holly Rushmeier$^{2}$ and Valentin Deschaintre$^{1}$ }\\
{\parbox{\textwidth}{\centering $^1$ Adobe Research \\
         $^2$ Yale University \\
       }
}
}
       
\title{Controlling Material Appearance by Examples}
\teaser{
\includegraphics[width=\linewidth]{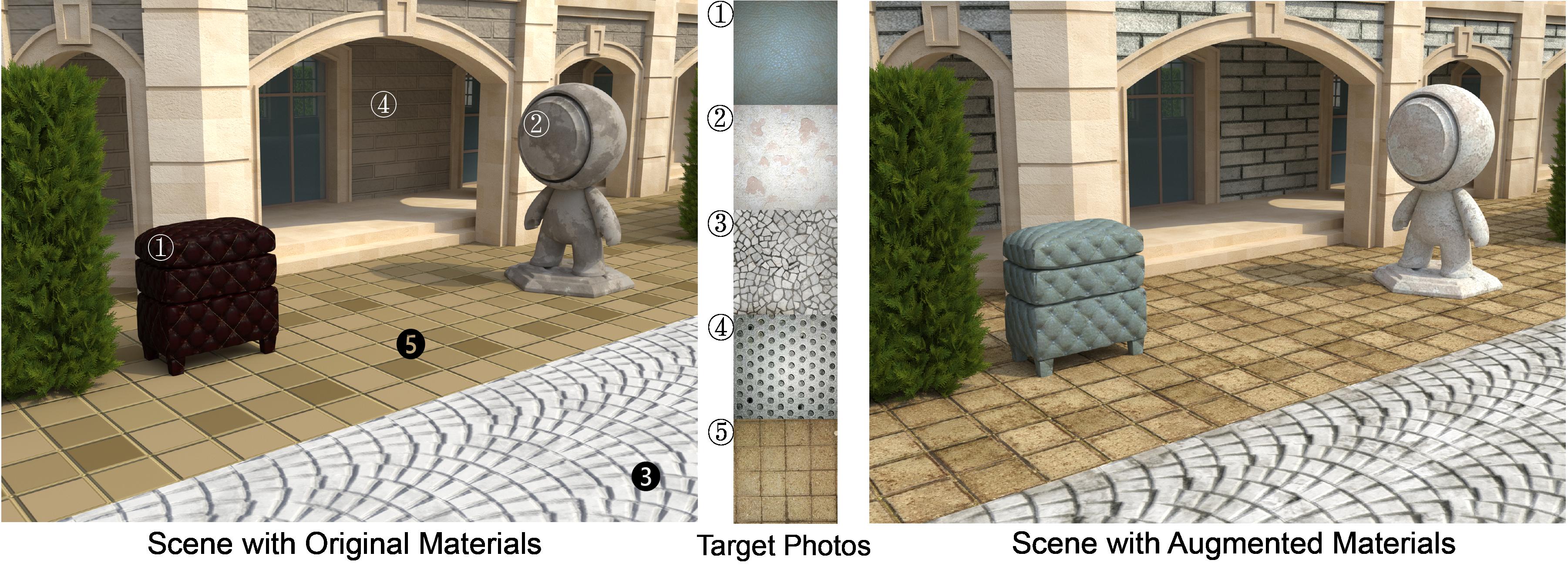}
\centering
\caption{We propose a method to control and improve the appearance of an existing material (left) by transferring the appearance of materials in one or multiple target photo(s) (center) to the existing material. The augmented material (right) combines the coarse structure from the original material with the fine-scale appearance of the target(s) and preserve the input tileability. Our method can also transfer appearance from materials from different types by spatial control. This enables a simple workflow to make existing materials more realistic using readily-available images or photos. 
}
\label{fig:teaser}
}

\maketitle
\begin{abstract}
Despite the ubiquitous use of materials maps in modern rendering pipelines, their editing and control remains a challenge. In this paper, we present an example-based material control method to augment input material maps based on user-provided material photos. We train a tileable version of MaterialGAN and leverage its material prior to guide the appearance transfer, optimizing its latent space using differentiable rendering. Our method transfers the micro and meso-structure textures of user provided target(s) photographs, while preserving the structure and quality of the input material. We show our methods can control existing material maps, increasing realism or generating new, visually appealing materials. 


\end{abstract}

\section{Introduction}
Realistic materials are one of the key components of vivid virtual environments. Materials are typically represented by a parametric surface reflectance model and a set of 2D texture maps (material maps) where each pixel represents a spatially varying parameter of the model (e.g. albedo, normal, and roughness values). This representation is ubiquitous because of its compactness and ease of visualization. It is also used by most recent light-weight acquisition methods~\cite{Deschaintre18, Deschaintre19, Gao19, Guo20}. Material maps are however difficult to edit. In addition to the significant artistic expertise required to create realistic material detail, tools such as Photoshop~\shortcite{photoshop} are designed for natural images rather than parameter maps and make it hard to account for the correlations that exist between individual material maps.
Previous work proposed propagating local edits globally through segmentation and similarity~\cite{AppProp} or providing an SVBRDF~\footnote{Spatially-Varying Bidirectional Reflectance Distribution Function} exemplar to transfer material properties~\cite{Deschaintre20}, but these edits remain limited to spatially constant material parameter modifications. Procedural materials~\cite{SubstanceDes, hu2019, Shi20, hu2022} also enable material map editing, but are challenging to design and can lack realism if the procedural model is not expressive and realistic enough ~\cite{hu2019}.

In this paper, we propose a method to control
the appearance of material maps using photographs or online textures.
Given a set of input material maps and target photographs (or textures),
we transfer micro- and meso-scale texture details from the target photographs to the input material maps. We preserve the large-scale structures of the input material, but augment the fine-scale material appearance based on the target photographs using a variant of style transfer~\cite{gatys2016}. Unlike Deschaintre et al.~\shortcite{Deschaintre20}, our approach uses easy to find or capture photographs instead of SVBRDF maps as targets.

Using style transfer from photographs to material maps poses several challenges. First, we need to optimize the material maps to produce the desired fine-scale details in the rendered image, rather than directly optimizing the output image. Second, we need to make sure material properties remain realistic during this optimization, including the correlation between material maps.
In Fig.~\ref{fig:per-pixel}, we show that a naive adaptation of image style transfer fails to generate high-quality material maps due to these difficulties.
To address both of these challenges, we propose using material generation priors to guide the transfer of micro and meso-scale texture details while retaining realistic material properties. We use MaterialGAN, a generative model trained on material maps~\cite{Guo20}, as a prior and modify it to preserve \textit{tileability}, a particularly important feature to keep memory requirement low in production. Our transfer operation optimizes the latent space of our pre-trained tileable MaterialGAN using differentiable rendering to transfer the relevant details while preserving the structure of the original input material maps.

The idea of material appearance or style on a spatially-varying material can be ambiguous. There are different aspects of appearance that a designer may see in an exemplar that they would like to transfer to an existing material, requiring some control. To provide this control we introduce multi-example transfer with spatial guidance: our method allows transferring texture details to precise user-specified regions only. We propose using a sliced Wasserstein loss~\cite{Heitz2021} to guide our transfer and supports multi-target transfer thanks to a resampling strategy we introduce. We additionally describe a slightly slower but better grounded formulation to compare distributions with different sampling based on the Cramér loss~\cite{cramer1928}.

We demonstrate creating new materials with our image-guided editing operations in various applications. To summarize, our contributions are:
\begin{itemize}
    \item A material transfer method for controlling appearance of material maps using photo(s).
    \item MaterialGAN as a prior for tileable materials appearance transfer.
    \item A multi-target transfer option with fine region control.
\end{itemize}
\section{Related Work}
\subsection{Material Appearance Control}
Control and editing of material appearance is a long standing challenge in computer graphics. Different solutions have been proposed, depending on the targeted material representation. Tabulated materials are represented in a very high dimension, unintuitive space~\cite{Wills2009}, making their manipulation difficult. Lawrence et al. \cite{Lawrence2006} represent a spatially-varying (SV) measured materials as an inverse shade tree, decomposing it into spatial structure and basis BRDFs to facilitate editing through extracted "1D curves" representing physical directions. Ben-Artzi et al.~\shortcite{Artzi2006, Artzi2008} proposed a fast iteration editing framework of material "in situ", allowing to efficiently visualize global illumination effects. Lepage et al~\cite{Lepage2011} proposed material matting to decompose measured SV-BRDF into layers for spatial editing. More recently, Serrano et al.~\shortcite{Serrano16} and Shi et al~\shortcite{shi2021} proposed designing perceptual spaces for more intuitive BRDF editing and Hu et al.~\shortcite{Hu2020} proposed encoding the BRDF in a deep network, reducing its dimensionality and simplifying editing. These methods target measured BRDF editing and focus on extracting relevant dimensions (perceptual, spatial) along which globally uniform edits can be made. As opposed to these methods, our approach targets spatially varying analytical model editing, enabling complex spatial detail transfer.

Analytical BRDFs represent materials based on pre-determined models (e.g Cook-Torrance~\shortcite{CookTorrance}, Phong~\shortcite{Phong}, GGX~\shortcite{GGX}, etc.) and their parameters. To explore this parameter space, Ngan et al.~\shortcite{Ngan2006} propose an interface to navigate different BRDF properties with perceptually uniform steps and Talton et al.~\shortcite{Talton2009} leverage a collaborative space to define a good modeling space for users to explore. 
Image-space editing was also explored with gloss editing in lightfields~\cite{Gryaditskaya2016} and material properties modification~\cite{Zsolnai20}.
Recently, different methods were proposed to optimize or create procedural materials~\shortcite{hu2019, Shi20, hu2022}. While procedural materials are inherently editable, they are limited to the expressivity of their node graphs and are difficult to design.
More closely related to our approach are image-guided material properties and style transfer. Fiser et al.~\shortcite{fiser16} used drawing on a known shape (sphere) to transfer the style and texture to a more complex drawing.  Deschaintre et al.~\shortcite{Deschaintre20} proposed using a surface picture and material exemplars to create a material with the surface picture structure and exemplar properties. Recently, Rodriguez-Pardo et al.~\shortcite{rodriguezpardo2021transfer} proposed leveraging photometric inputs to transfer material properties annotated on a small region to larger samples.
As opposed to these methods our approach transfers the appearance and texture micro and meso structures from photo(s) to a pre-existing analytical SVBRDF.

\subsection{Style Transfer}
We formulate our by-example control on material maps as a material transfer problem. In recent years, neural style transfer~\cite{gatys2016} and neural texture synthesis~\cite{gatys2015} have been used in a variety of contexts (e.g. sketching~\cite{Texler2021, Sykora19}, video~\cite{Jamriska19,Texler20}, painting style~\cite{Texler20-CAG}). These methods are based on the matching of the statistics extracted by a pre-trained neural network between output and target images. For example, Gatys et al. \shortcite{gatys2015, gatys2016} leverage a pre-trained VGG neural network~\cite{vgg19} to guide style transfer, using the Gram Matrix of extracted deep features from the image as their statistical representation. Heitz et al.\shortcite{Heitz2021} described an alternative sliced Wasserstein loss as a more complete statistical description of extracted features. Different approaches proposed to train a neural network to transfer style of images or synthesize textures in a single forward operation~\cite{johnson2016, Ulyanov2016, huang2017, zhou2018}. These methods however focus on the transfer of the overall style of images and do not account for details. Domain specific (faces) and guided style transfer have been proposed~\cite{Kolkin19, Huang2019} to better control the process and transfer style between semantically compatible parts of the images.

In the context of materials, Nguyen et al. \shortcite{nguyen2012} proposed transferring the style or mood of an image to a 3D scene. They pose this problem as a combinatorial optimization of assigning discrete materials, extracted from the source image, to individual objects in the target 3D scene. Mazlov et al. 
\cite{Mazlov2019} proposed directly applying neural style transfer on material maps in a cloth dataset using a variant of RGB neural style transfer \cite{gatys2016}.

In contrast to this previous work, we focus on transferring micro- and meso-scale details from photos onto 2D material maps and enable transfer spatial control. In particular, we differ from Mazlov et al.~\shortcite{Mazlov2019} in our inputs. Our style examples are simple photographs, easily captured or found online. Our approach allows us to deal with the ambiguity of the material properties in the target photo through material priors and differentiable rendering while preserving the input material map structure and eventual tileability.
\subsection{Procedural Material Modeling}
As mentioned, procedural materials are inherently editable as they rely on parametrically controllable procedures. Recent work on inverse procedural material modeling~\cite{hu2019, GuoBayesian20, Shi20, hu2022} aim at reproducing material photographs, but are still limited to simple or existing procedural materials. While editable, these models remain challenging to create and their optimization relies on the expressiveness of the available procedural materials, with no guarantee that a realistic model exists~\cite{hu2019, Shi20}. Often procedural materials can look synthetic because they are not complex enough. In such cases, our method can augment the material appearance by transferring detailed real materials onto the synthetically generated material maps, while preserving their tileability. In this context, Hu et al. \shortcite{hu2019} proposed to improve the quality of fitted materials through a style augmentation step, but style was only transferred to the final rendered images and did not impact the material maps themselves.

\subsection{Material Acquisition}
As an alternative to material transfer, one could directly capture the target material. Classical material acquisition methods require dozens to thousands of material samples to be captured under controlled illuminations~\cite{Guarnera16}. Aittala et al.~\shortcite{aittala2016} proposed using a single flash image of a stationary material to reconstruct a patch of it through neural guided optimization. Recently, deep learning was used to improve single~\cite{Li17, Deschaintre18, henzler21neuralmaterial, Guo21, Zhou21} and few-images~\cite{Deschaintre19, Guo20, Gao19, Ye21} material acquisition. These methods recover 2D material maps based on an analytical BRDF model e.g. GGX~\cite{Walter2007}. As they mainly focus on acquisition, they do not enable control. As opposed to these methods, our approach provides a convenient way to control the appearance of material maps, leveraging the input material to retain high quality, allowing for multi-target control and preserving important properties such as tileability. Additionally, our method does not require a flash photograph as target, allowing the use of internet-searched references.

\section{Method}
Our method transfers fine-scale details from user-provided target material photos onto a set of material maps. The material maps contain diffuse albedo, normal, roughness and specular albedo properties, encoding parameters for a GGX \cite{Walter2007} shading model. 
Since fine-scale details are typically underdefined by single-view photos,
we choose to leverage a specialized prior that regularizes the transfer of materials details to ensure our material maps remain realistic and to help disambiguate unclear material properties in the target photo.

As prior, we train a tileable version of MaterialGAN~\cite{Guo20} using a large dataset of synthetic material maps. This prior is described in Section \ref{Sec:tileable-StyleGAN}. We transfer material appearance by optimizing our material maps in the latent space of this tileable MaterialGAN, rather than directly in pixel space. A differentiable GGX shading model renders the optimized material maps into an image where they are compared to the target photos. This appearance transfer approach is described in Section~\ref{Sec:Material-Transfer}.
In Section~\ref{Sec:Spatial-Control}, we describe how we add spatial control over the appearance transfer and enable transfer from multiple targets using label maps and
a resampled version of the sliced Wasserstein loss~\cite{Heitz2021}.
\subsection{Tileable MaterialGAN}
\label{Sec:tileable-StyleGAN}
MaterialGAN \cite{Guo20} has been shown to be a good prior for lightweight material acquisition. However, the original MaterialGAN, based on StyleGAN2~\cite{stylegan2}, is not designed to produce tileable outputs. It has slight artifacts on the borders, even if the training data is perfectly tileable. Moreover, material maps in most of publicly available material datasets \cite{Deschaintre18, Deschaintre20} are not tileable. To address this limitation, we modify StyleGAN's architecture following recent insights in GAN designs \cite{Karras2021, logan2022} to ensure tileability of the synthesized material maps, even if the training data is not tileable. Specifically, we prevent the network from processing the image borders differently from the rest of the image, by modifying all convolutional and upsampling operations with circular padding.

Once we enforce the generator network to always produce tileable outputs, we cannot show tileable synthesized and non-tileable real data to the discriminator, as it would have a clear signal to differentiate them. Instead, as suggested by \cite{logan2022}, we randomly crop both real and synthesized material maps. The discriminator cannot identify whether the crop comes from a tileable source or not, and instead has to identify whether the crop content looks like a real or fake material.

\begin{figure}
	\centering
	\renewcommand{\arraystretch}{0.6}
	\addtolength{\tabcolsep}{-4pt}
	\begin{tabular}{ccccc}
		Init. & Per-pixel & Ours & Target \\
		\includegraphics[width=0.11\textwidth]{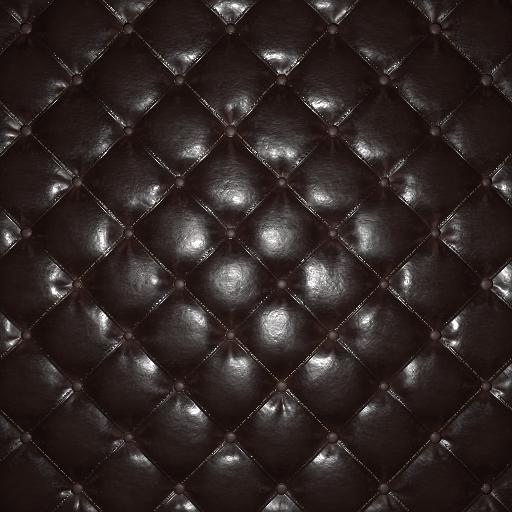}		\llap{\frame{\includegraphics[width=0.05\textwidth]{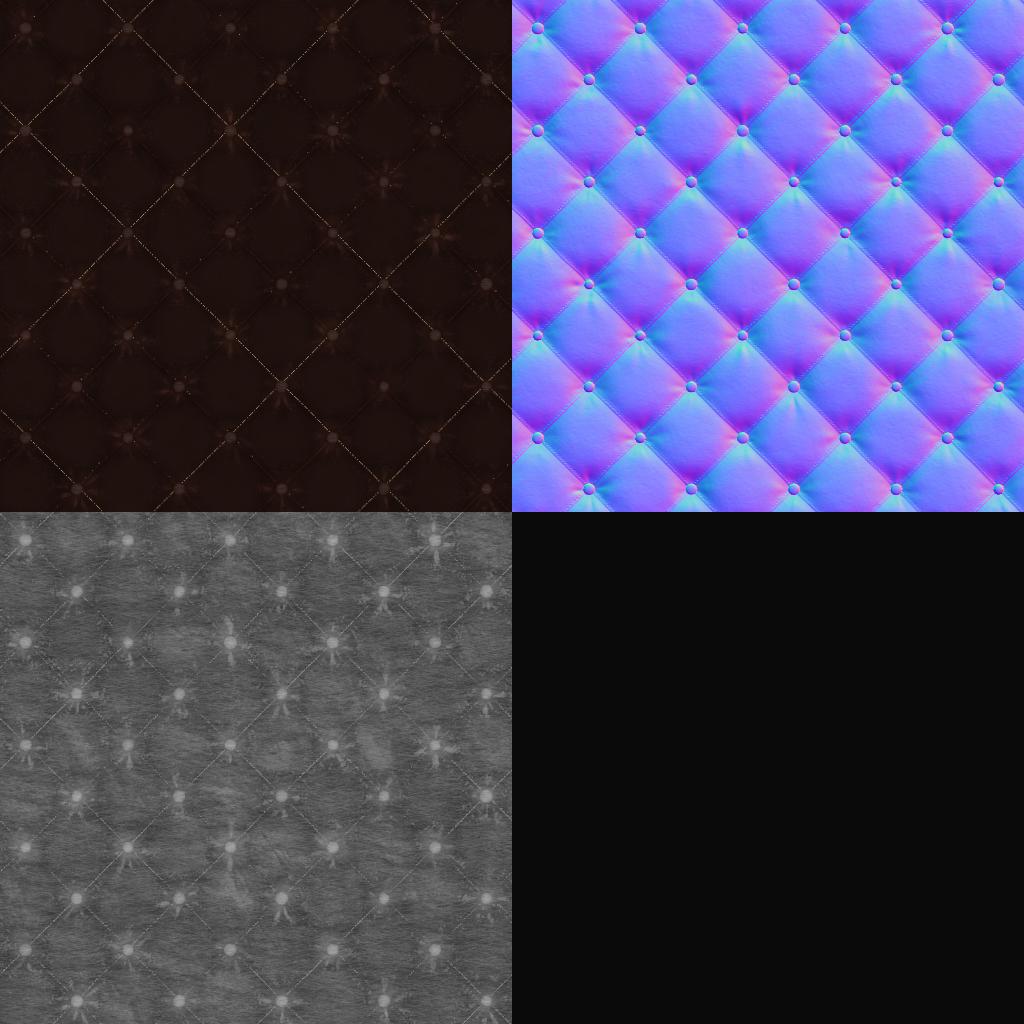}}} &
        \includegraphics[width=0.11\textwidth]{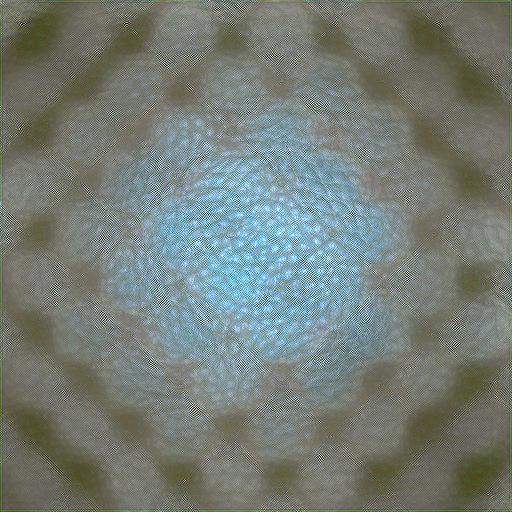}		\llap{\frame{\includegraphics[width=0.05\textwidth]{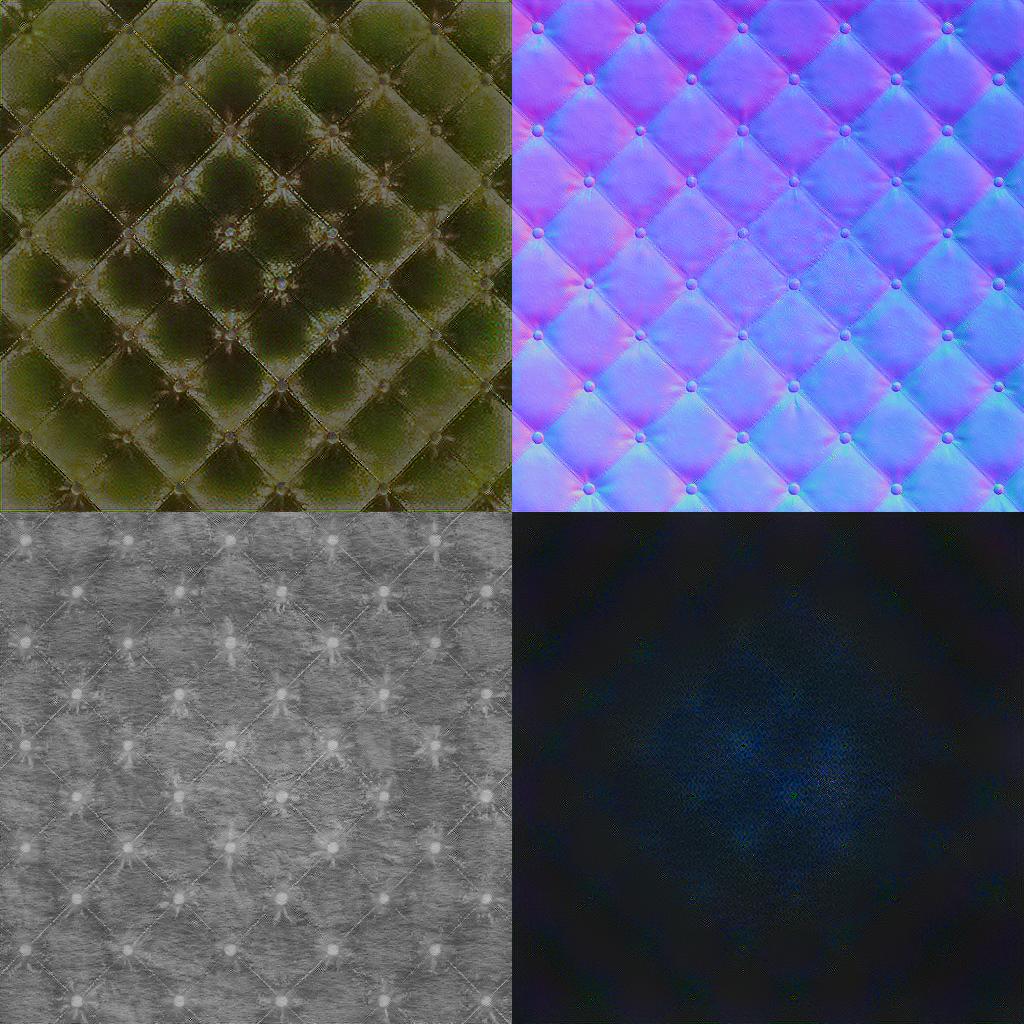}}} &
        \includegraphics[width=0.11\textwidth]{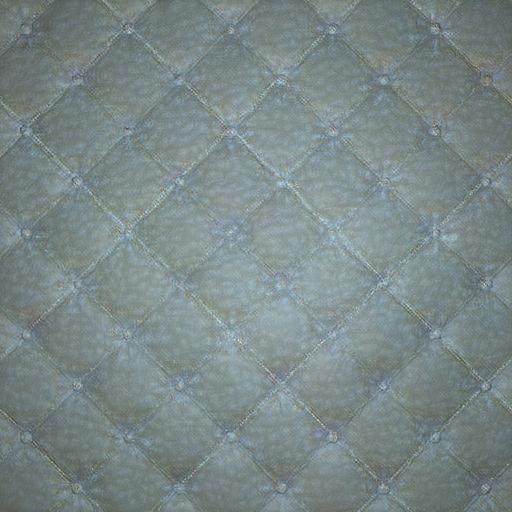} \llap{\frame{\includegraphics[width=0.05\textwidth]{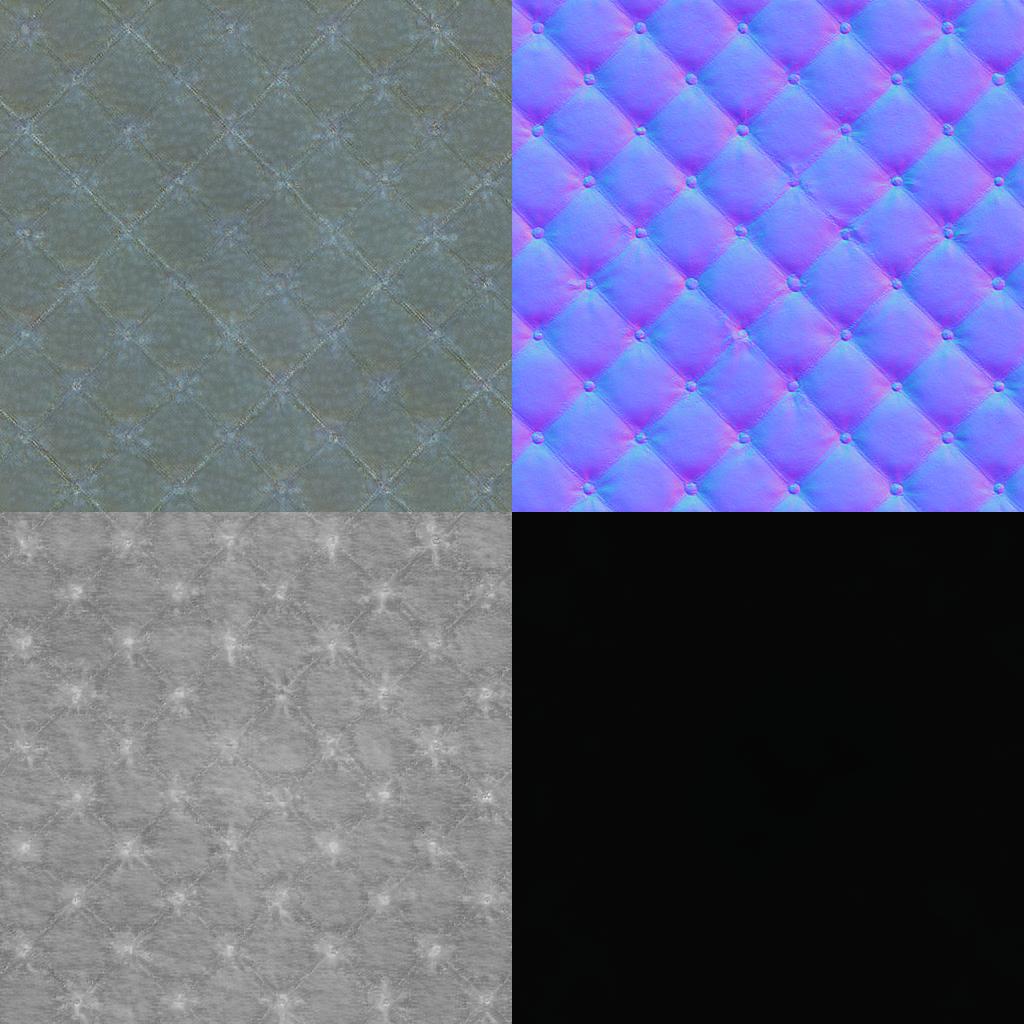}}} &
        \includegraphics[width=0.11\textwidth]{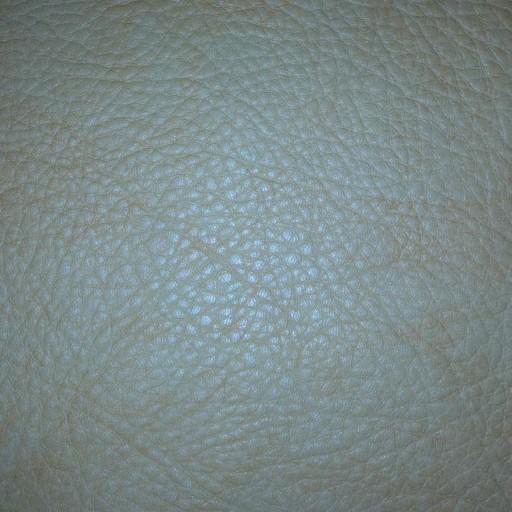} \\
		\includegraphics[width=0.11\textwidth]{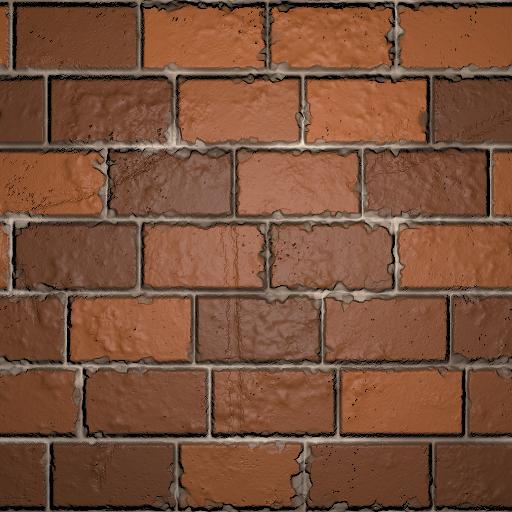}
		\llap{\frame{\includegraphics[width=0.05\textwidth]{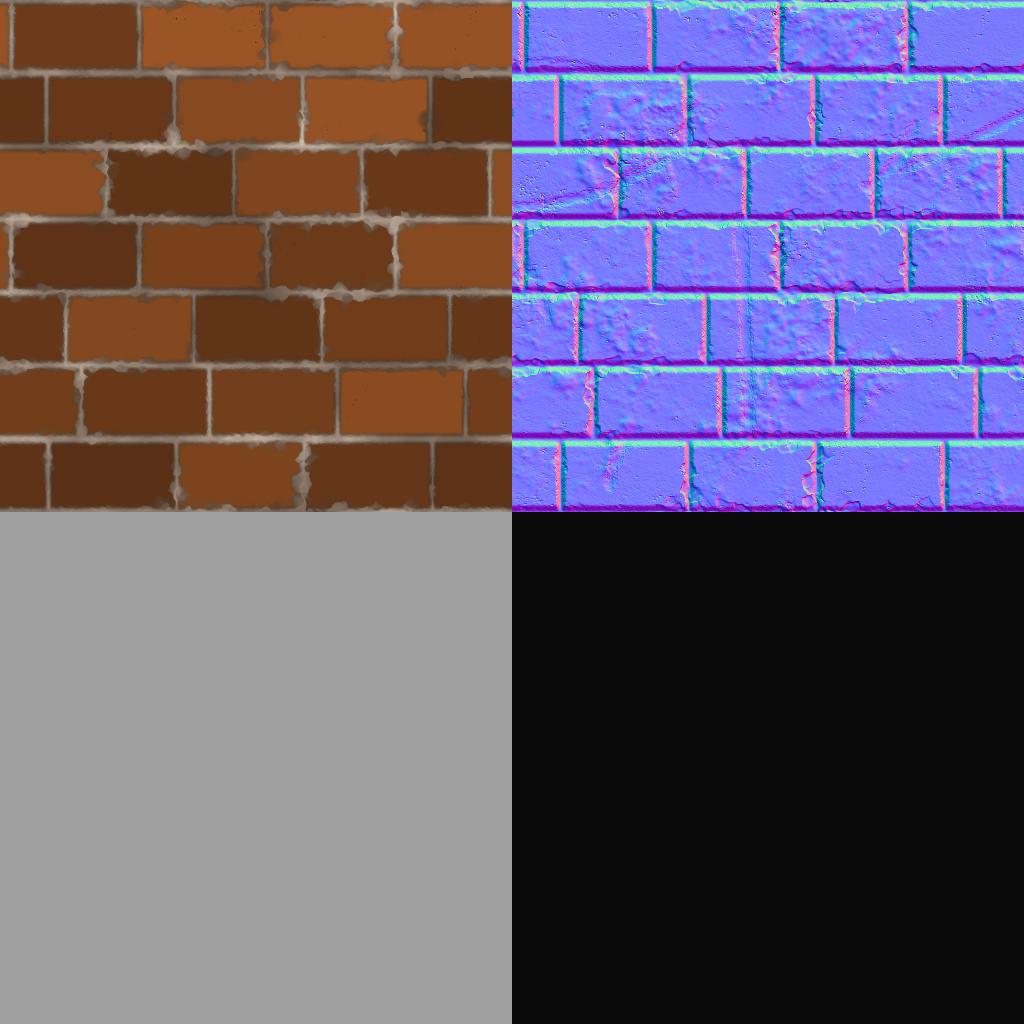}}}  &
        \includegraphics[width=0.11\textwidth]{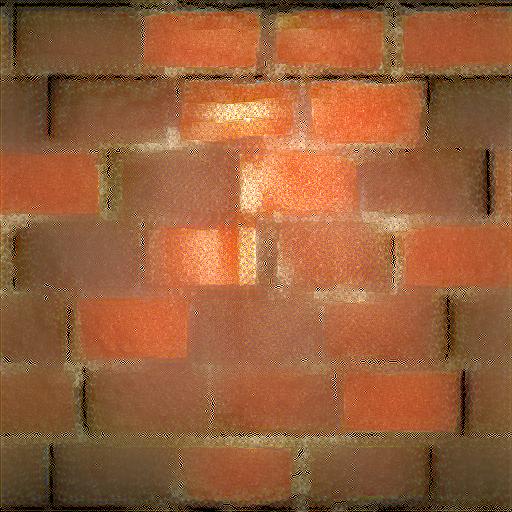}
		\llap{\frame{\includegraphics[width=0.05\textwidth]{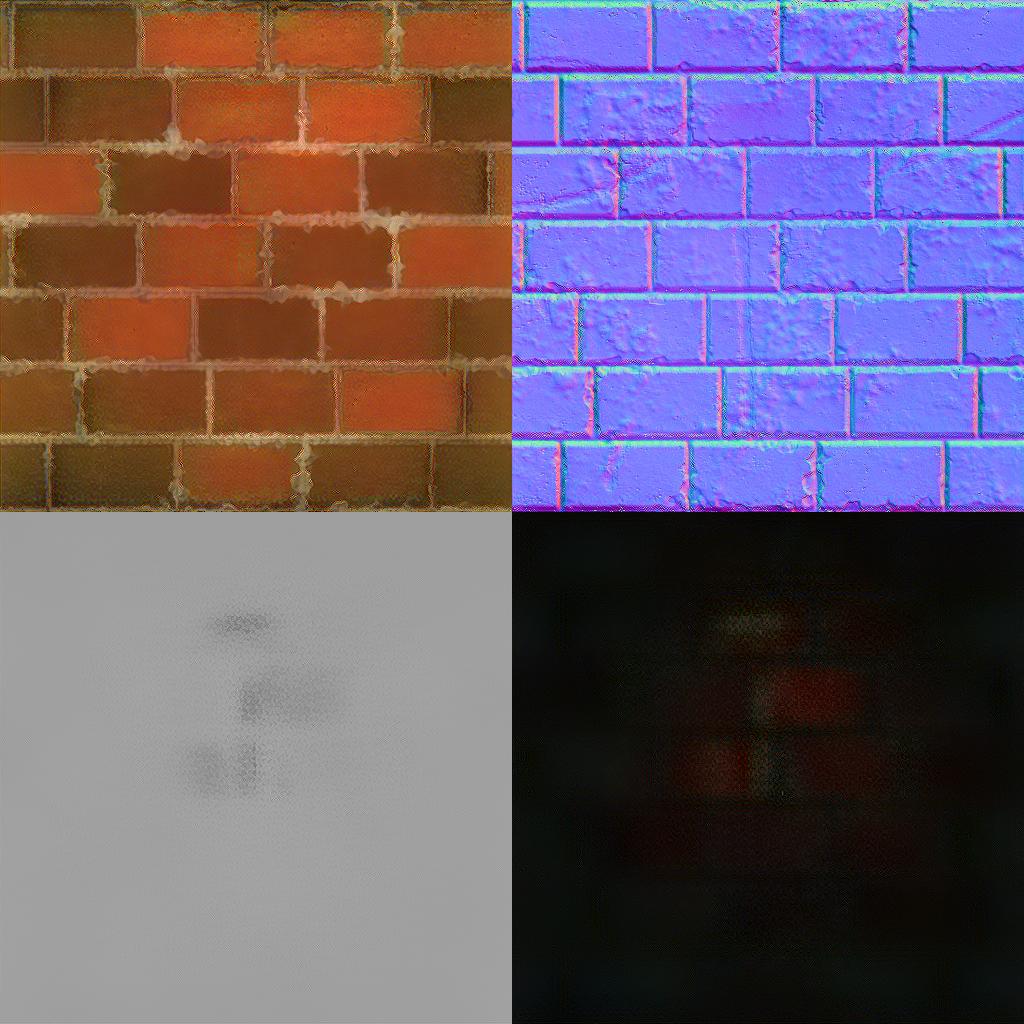}}} &
        \includegraphics[width=0.11\textwidth]{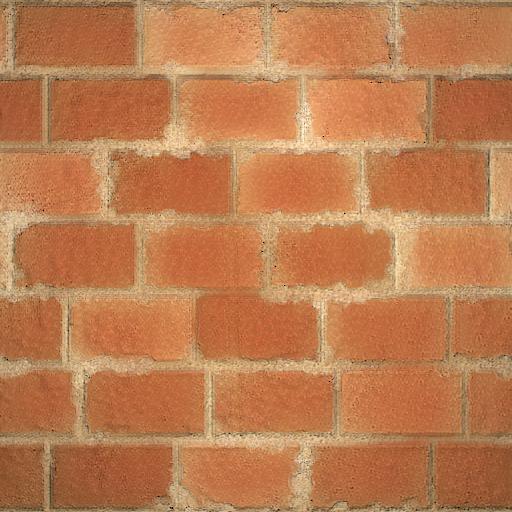}		\llap{\frame{\includegraphics[width=0.05\textwidth]{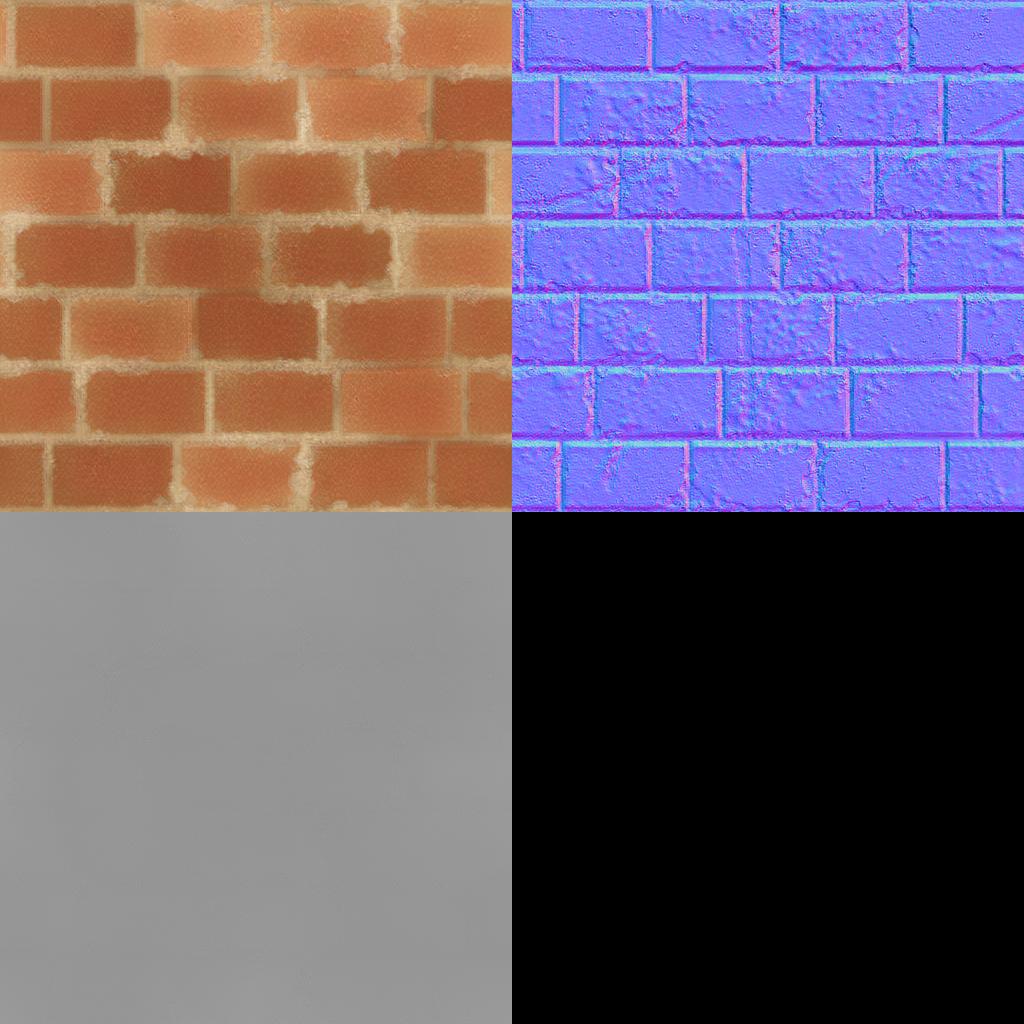}}} &
        \includegraphics[width=0.11\textwidth]{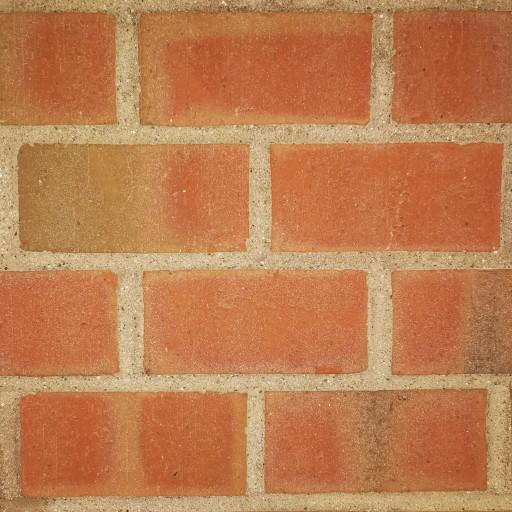} \\
	\end{tabular}
\vspace{-10pt}
\caption{Comparison between optimizing the material maps in pixel space and our approach, optimizing in the latent space of tileable MaterialGAN. Insets show the material maps used to render each image and \textit{Target} is a photograph representing the desired appearance.
Per-pixel optimization frequently gets trapped in local minima and suffers from artifacts and light-baking, while our use of a prior allows to reach the desired appearance.}
\label{fig:per-pixel}
\vspace{-3mm}
\end{figure}

\begin{figure}
	\centering
	\renewcommand{\arraystretch}{0.6}
	\addtolength{\tabcolsep}{-6pt}
	\begin{tabular}{cccc}
		Input/Target & Original MaterialGAN & Ours \\
		\includegraphics[width=0.09\textwidth]{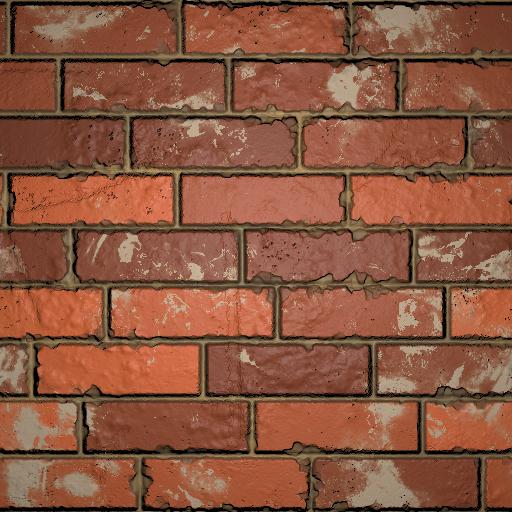} &
		\multirow{2}{*}[41pt]{
		\includegraphics[width=0.185\textwidth]{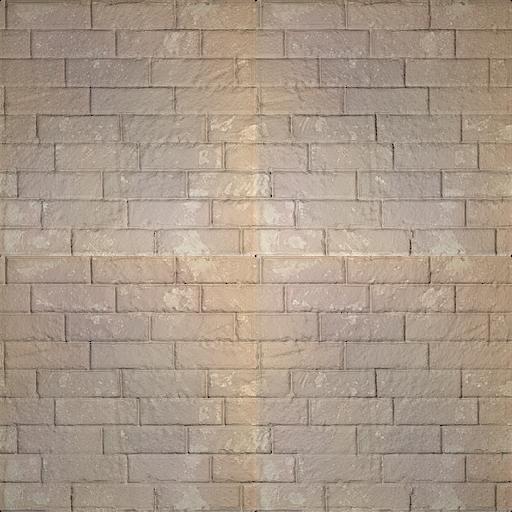}
		} &
		\multirow{2}{*}[41pt]{
		\includegraphics[width=0.185\textwidth]{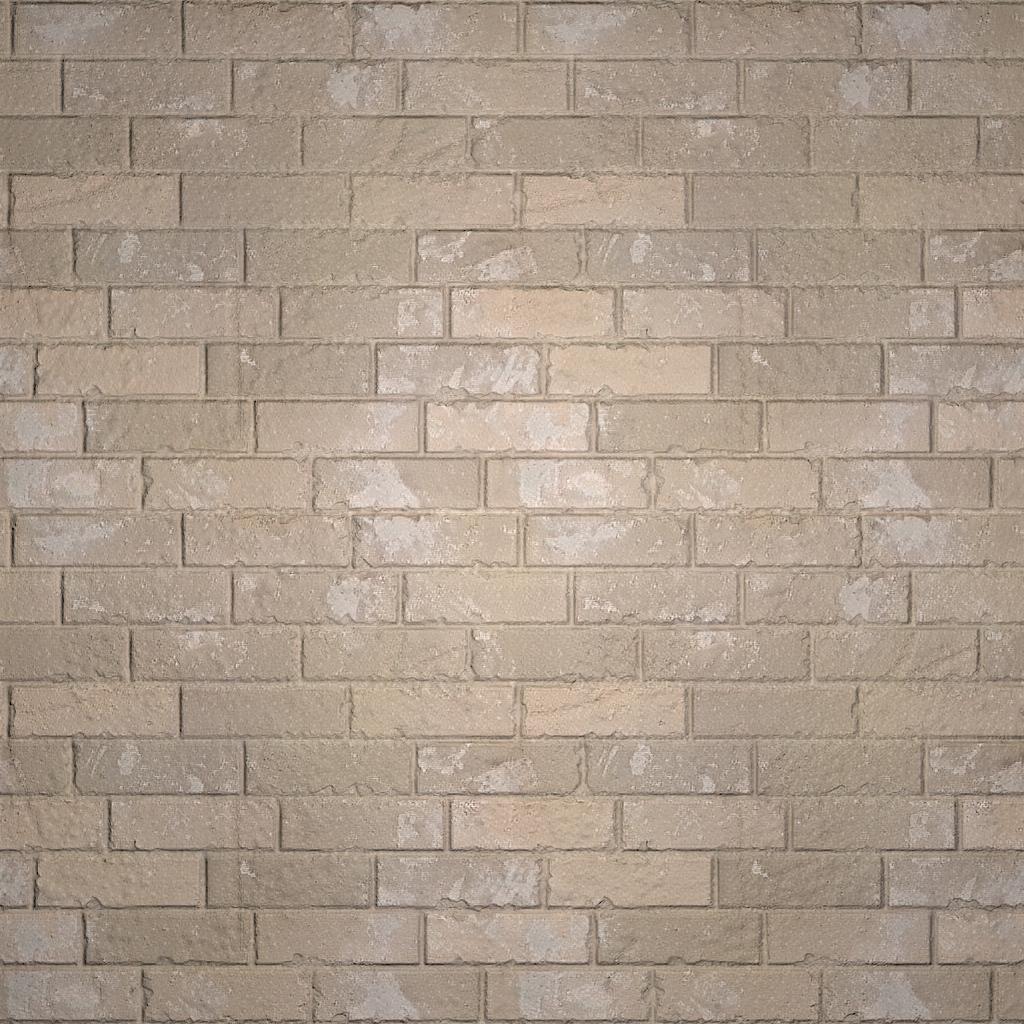}
		} \\
		\includegraphics[width=0.09\textwidth]{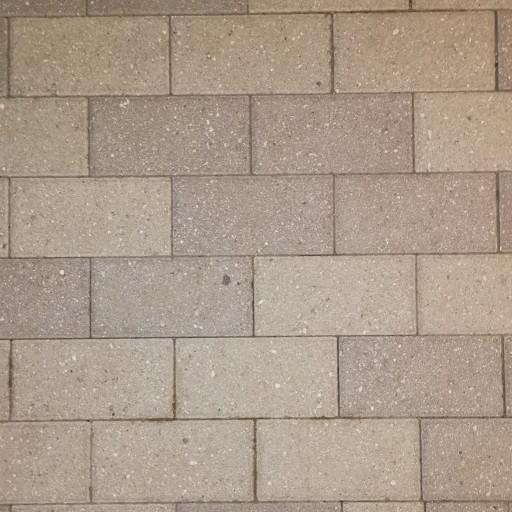} \\
		\includegraphics[width=0.09\textwidth]{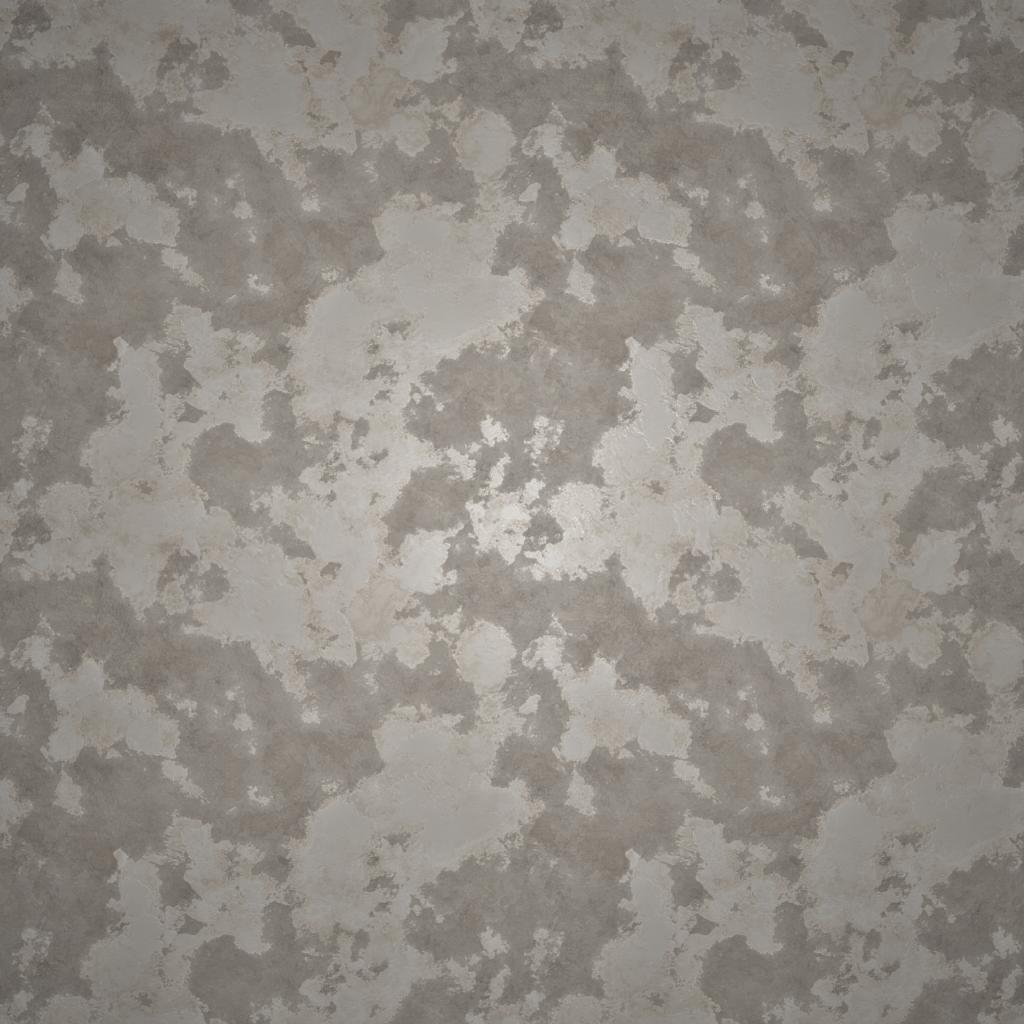} &
		\multirow{2}{*}[41pt]{
		\includegraphics[width=0.185\textwidth]{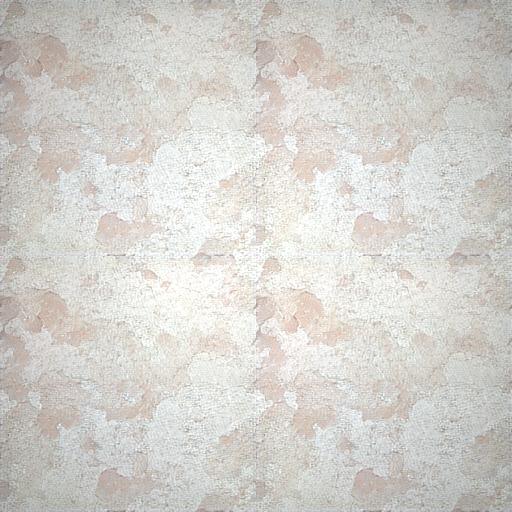}
		} &
		\multirow{2}{*}[41pt]{
		\includegraphics[width=0.185\textwidth]{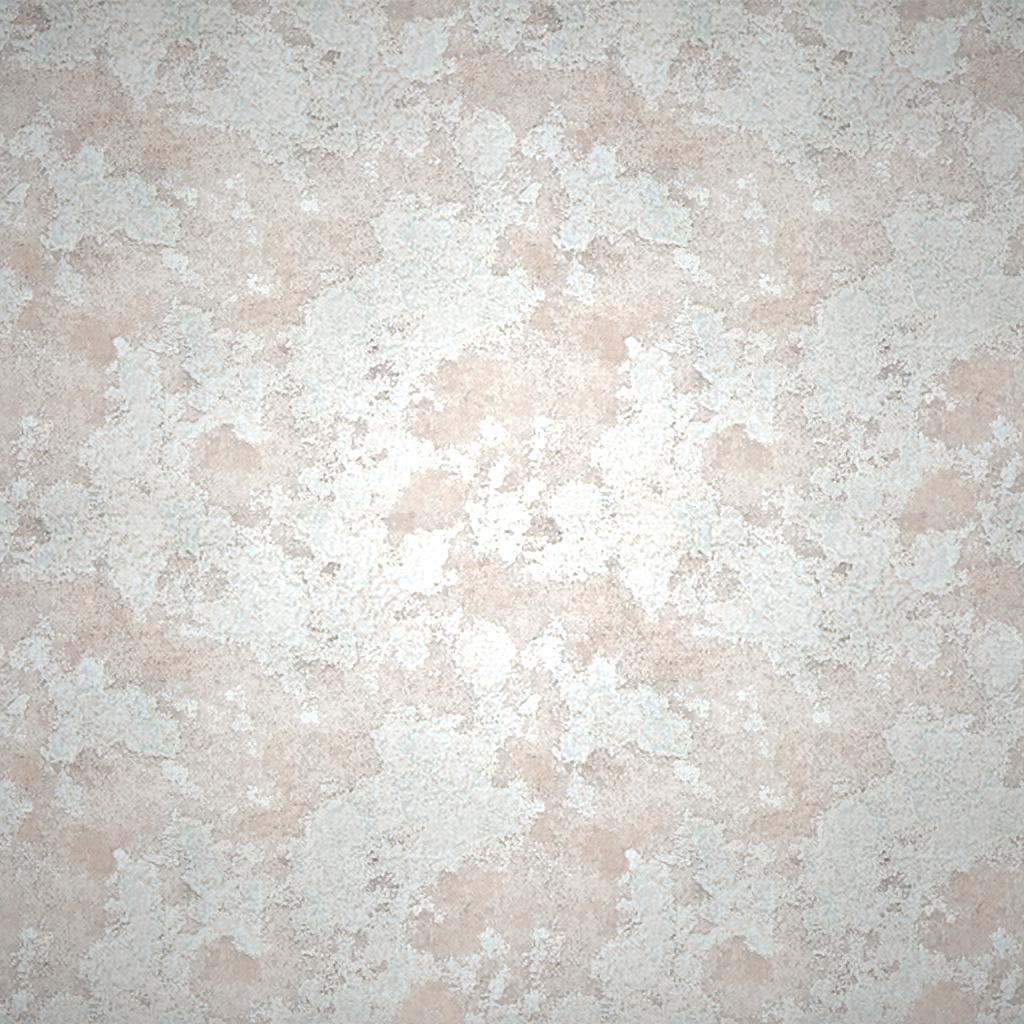}
		} \\
		\includegraphics[width=0.09\textwidth]{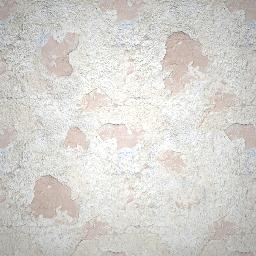} \\
	\end{tabular}
\vspace{-10pt}
\caption{The original MaterialGAN model does not produce tileable results, leading to visible seams on the boundary when tiled, while our tileable network preserves the input tileability without artifacts (Ours). Seams are particularly visible when zoomed in.}
\label{fig:tileability}
\end{figure}

We train our model with the same loss functions as StyleGAN2 \cite{stylegan2} including cross-entropy loss with R1 regularization and path regularization loss for generators. See Sec. \ref{Sec:implementation} for training details. With the trained tileable material prior, our material transfer method overcomes the local minimum problems caused by simple per-pixel optimization (Fig. \ref{fig:per-pixel}) and reconstructs high-quality material maps compared to an adaption of Deep Image Prior (Fig. \ref{fig:deep_image_prior}). In Fig. \ref{fig:tileability}, we show our modified network architecture can successfully preserve tileability after transfer compared to the original MaterialGAN. Importantly, the preserved tileability allows us to directly apply our transfer materials on different objects seamlessly in a large-scale scene as shown in Fig.~\ref{fig:teaser}.

\subsection{Neural Material Transfer}\label{Sec:Material-Transfer}
In previous work, Hu et al.~\shortcite{hu2019} proposed an extra style augmentation step to add realistic details on their fitted procedural materials given a photo as target. However, their transfer is only applied to the rendered images and does not modify the material maps, limiting the impact of their transfer step (e.g. the material can not be relit). Different from their method, we directly transfer the target photo(s) material appearance onto material maps, allowing the use of the new material in a traditional rendering pipeline.

Given a set of input material maps $M_0$ and a user-provided target image $I$, we compute the transferred material maps $M$ as follows:
\begin{equation}
\arg\min_M\quad d_0(R(M), I) + d_1(M, M_0)
\label{Eq:material-transfer}
\end{equation}
where $R$ is a differentiable rendering operator, rendering material maps $M$ into an image. $d_0(R(M), I)$ measures the statistical similarity between the synthetic image $R(M)$ and target image $I$. $d_1(M, M_0)$ is a regularization term that penalizes the structure difference between transferred material maps $M$ and the original input $M_0$. The lighting used in $R$ can easily be adapted if needed to roughly match the lighting conditions of target images. However, we found that a simple co-located point light works well in the examples we tested. Similar to neural style transfer and neural texture synthesis, we apply a statistics-based method to measure the similarity for $d_0$ and $d_1$. Common choices for $d_0$ and $d_1$ are style loss and feature loss~\cite{gatys2015}. However, simply performing per-pixel optimization on material maps $M$ (i) fails to reach the appearance of the target photograph. This is caused by challenging local minima in the optimization and a high sensitivity to the learning rate, requiring careful tuning (see Figure~\ref{fig:per-pixel}). And (ii) the optimization may result in departure from the manifold of realistic materials.
Instead, we take advantage of the learned latent space of our pre-trained tileable MaterialGAN to regularize our material transfer, effectively solving these problems.

We tackle the material transfer challenges in two steps. First, we project the input material maps $M_0$ into the latent space of the pre-trained MaterialGAN model $f$ by optimizing the latent code $\theta$. The optimization is guided by $L_1$ loss and feature loss:
\begin{equation}
L_\theta=||f(\theta) - M_0||_1 + ||F(f(\theta)) -  F(M_0)||_1
\label{Eq:project}
\end{equation}
where $F$ is a feature extractor using a pre-trained VGG network~\cite{vgg19}. With the projected latent vector, we perform material transfer by optimizing $\theta$ to minimize the statistical difference between rendered material $R(f(\theta))$ and the material target $I$
\begin{equation}
L_\theta=||S(R(f(\theta))) - S(I)||_1 + ||F(f(\theta)) -  F(M_0)||_1
\label{Eq:style-transfer}
\end{equation}
The statistical descriptor $S$, the style loss, can be implemented in different ways. The Gram Matrix~\cite{gatys2015} is a commonly used statistical loss. Recently, Heitz et al. ~\shortcite{Heitz2021} proposed a sliced Wasserstein loss to measure a more complete statistical difference, taking additional statistical moments into consideration. Their implementation however relies on comparing signals (here images) with the same number of samples (here pixels). To remove this limitation, we propose using uniform resampling to balance the number of samples in both distributions to facilitate spatial control and multi-target transfer (Sec. \ref{Sec:Spatial-Control}). We derive in supplemental material a slightly slower but better grounded sliced Cramér loss, comparing the distribution CDFs rather than PDFs.

\begin{figure}
	\centering
    \includegraphics[width=1.0\columnwidth]{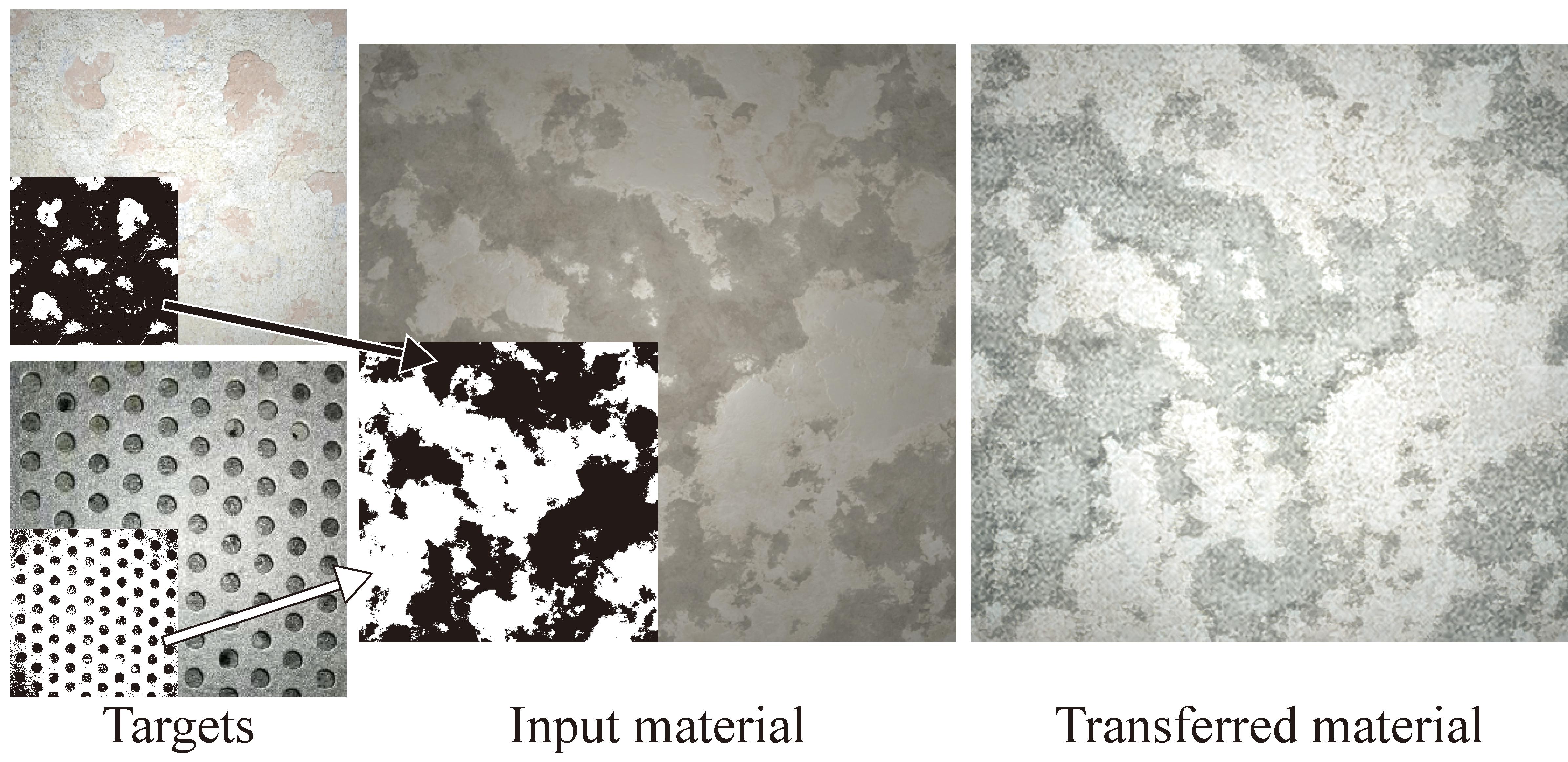}
\vspace{-10pt}
\caption{For our multi-target transfer, we transfer the appearance of two targets to the same input material. A binary map on the input material defines which of the two targets should be used for each pixel. Additionally, a binary map can be defined on each target defines which regions should be included in the appearance transfer.
Here we transfer the appearance from the black region of the mask of the top target to the black regions on the input mask. The white regions are similarly transferred from the bottom target. This binary system is only used for visualization. There is no limit to the number of correspondences that can be made.}
\label{fig:label-define}
\end{figure}
\subsection{Spatial Control and Multi-target Transfer}\label{Sec:Spatial-Control}
The sliced Wasserstein loss has been shown to be a good statistical metric~\cite{Heitz2021} to compare deep feature maps, but does not allow comparing feature maps with different numbers of samples (pixels) trivially. The "tag" trick introduced in the original paper cannot be applied to our goal for multi-target transfer. The sliced Wasserstein loss compares two images by projecting per-pixel VGG feature vectors onto randomly sampled directions in feature space, giving two sets of 1D sample points $u$ and $v$, one for each image. These are compared by taking the difference between the sorted sample points. To allow for different sample counts $|u|<|v|$, we introduce the \emph{resampled} sliced Wasserstein loss as:
\begin{equation}
    L_{SW1D}(u, v)=\frac{1}{|u|}||\textrm{sort}(u) - \textrm{sort}(U(v))||_1
\label{Eq:Wasserstein}
\end{equation}
where $U(v)$ is an operator that uniformly random subsamples a vector to obtain $|u|$ samples from $v$. Note here we compute $L_1$ error as opposed to squared error suggested in the original paper as we found it to perform better. 

Using this resampling approach, we can compute statistical differences between labeled regions of different size. Assume we have label maps associated with material maps and each target photo, we define our transfer rule as \textbf{Label X: Target Y, Z} which means to transfer material appearance from regions labeled by Z
in target photo Y onto regions labeled by X in the input material maps. We show examples of how to define the label in Fig. \ref{fig:label-define}. The sliced Wasserstein loss will be computed between deep features on each labeled regions. Without loss of generality, we extract a deep feature $p^l$ and $\hat{p}^l$ from layer $l$ on the rendered image $R(\theta)$ and one of the material target $\hat{I}$. Similar to \cite{Heitz2021}, we randomly sample $N$ directions $V\in S^{N_l}$ and project features $p^l$ and $\hat{p}^l$ onto the sampled directions to get projected 1D features $p^l_V$ and $\hat{p}^l_V$.

\begin{figure}
	\centering
	\renewcommand{\arraystretch}{0.6}
	\addtolength{\tabcolsep}{-5.5pt}
	\begin{tabular}{cccccc}
	    \multicolumn{2}{c}{
	    \includegraphics[width=0.15\textwidth]{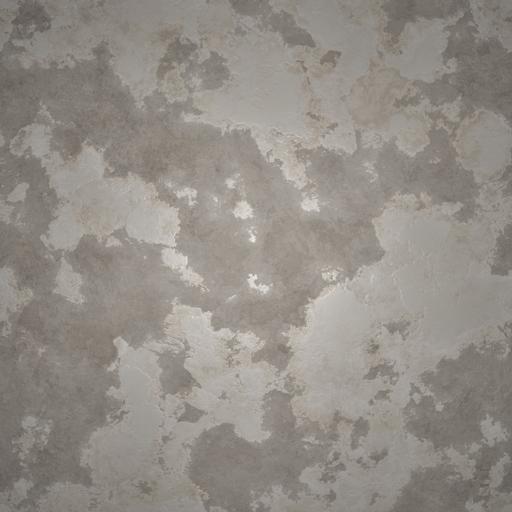}
	    \llap{\frame{\includegraphics[width=0.05\textwidth]{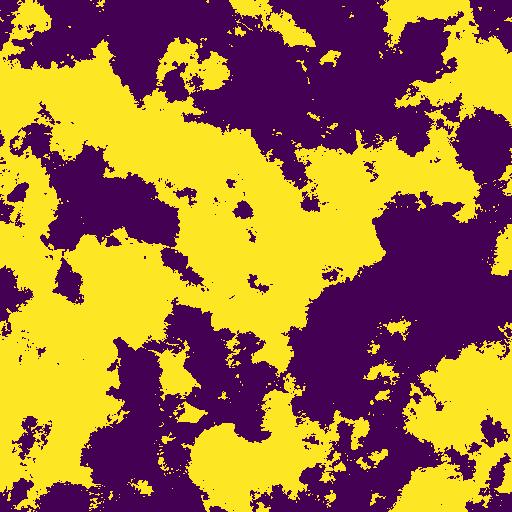}}}} & 
		\multicolumn{2}{c}{
	    \includegraphics[width=0.15\textwidth]{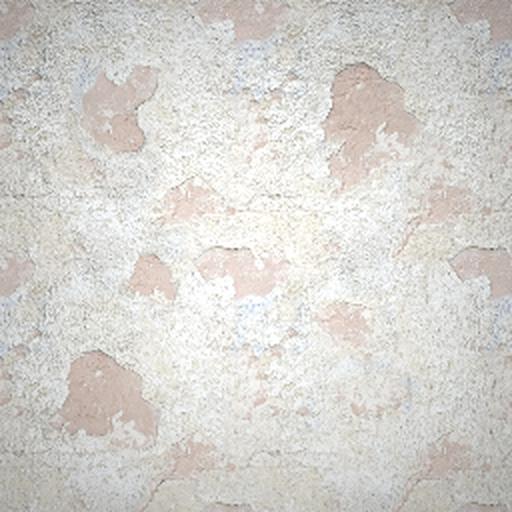}
	    \llap{\frame{\includegraphics[width=0.05\textwidth]{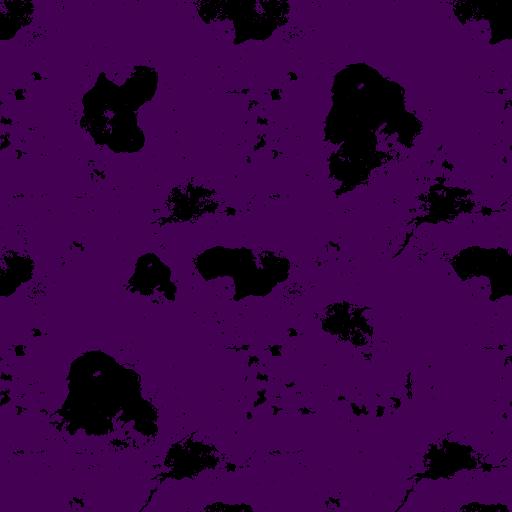}}}} &
	    \multicolumn{2}{c}{
	    \includegraphics[width=0.15\textwidth]{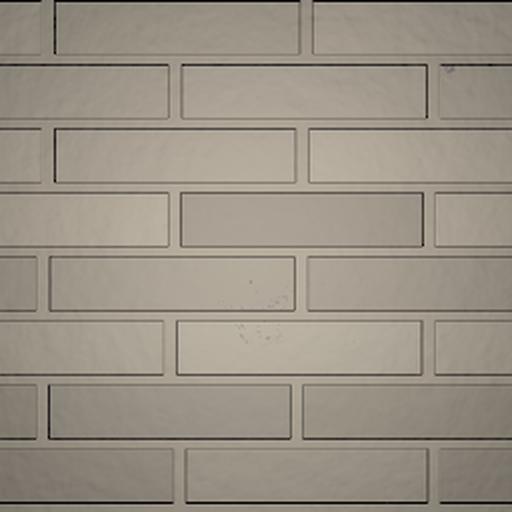}
	    \llap{\frame{\includegraphics[width=0.05\textwidth]{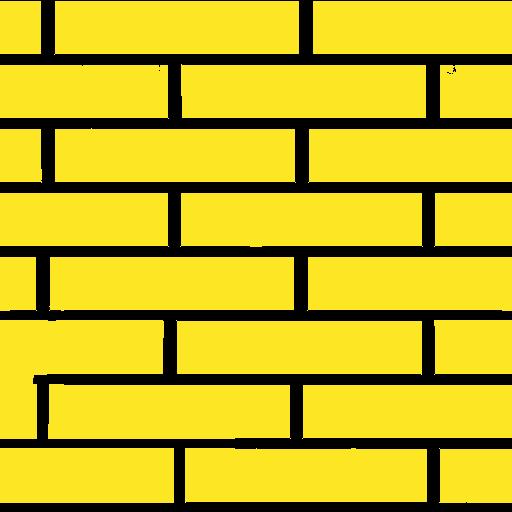}}}} \\
	    \multicolumn{2}{c}{Input} & \multicolumn{2}{c}{Target0} & \multicolumn{2}{c}{Target1} \\
	    \multicolumn{3}{c}{
	    \includegraphics[width=0.23\textwidth]{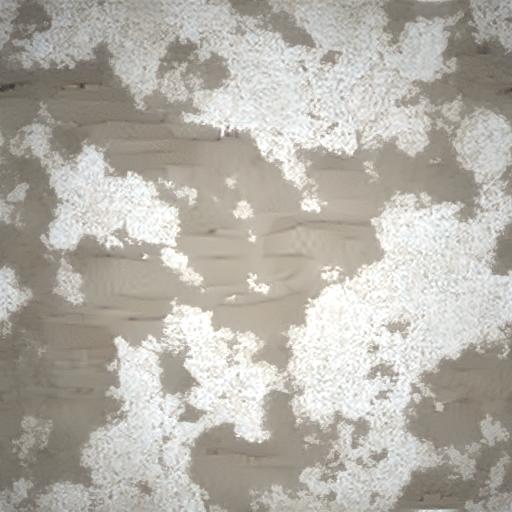}
	    \llap{\frame{\includegraphics[width=0.1\textwidth]{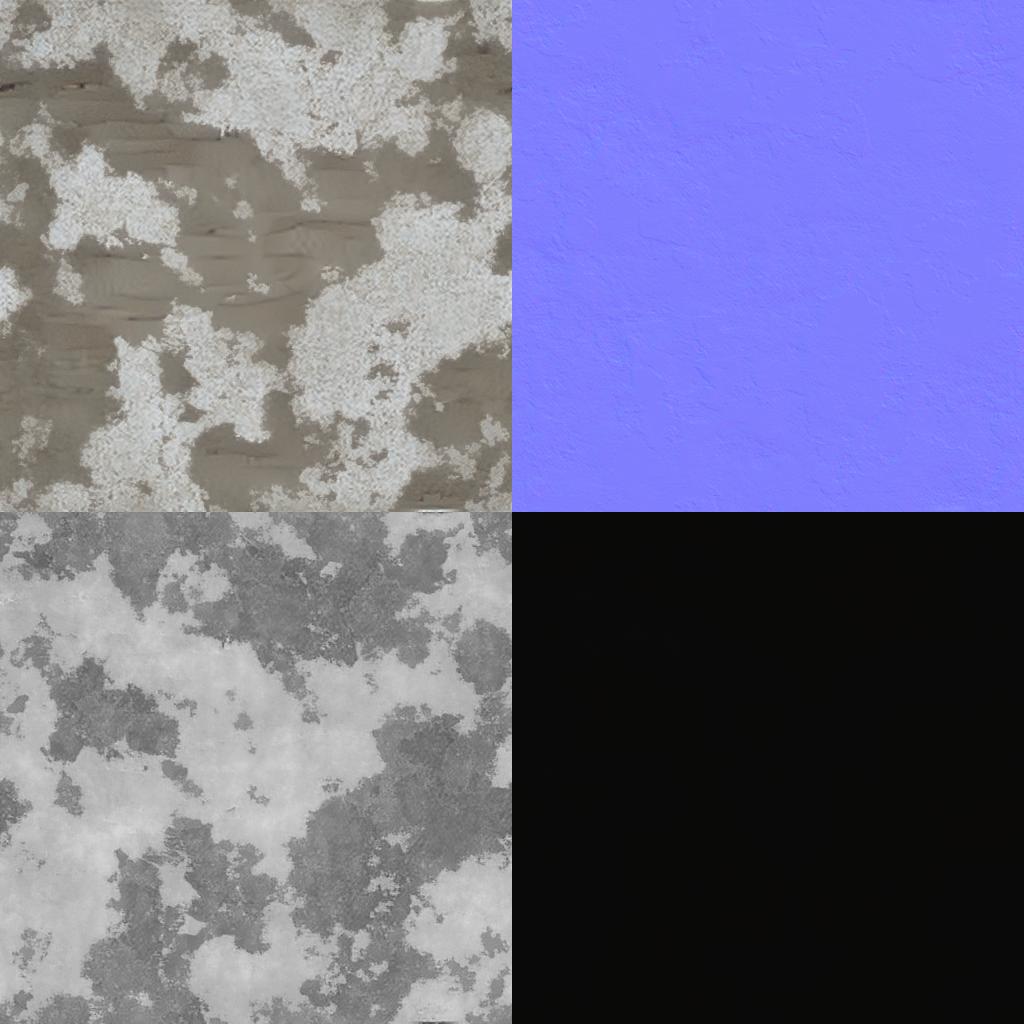}}}} & 
	    \multicolumn{3}{c}{
	    \includegraphics[width=0.23\textwidth]{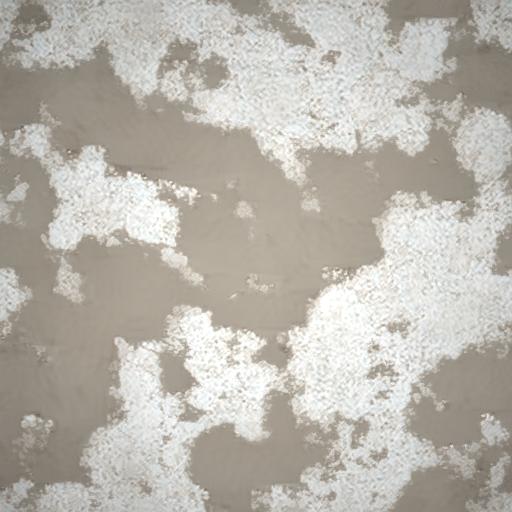}
	    \llap{\frame{\includegraphics[width=0.1\textwidth]{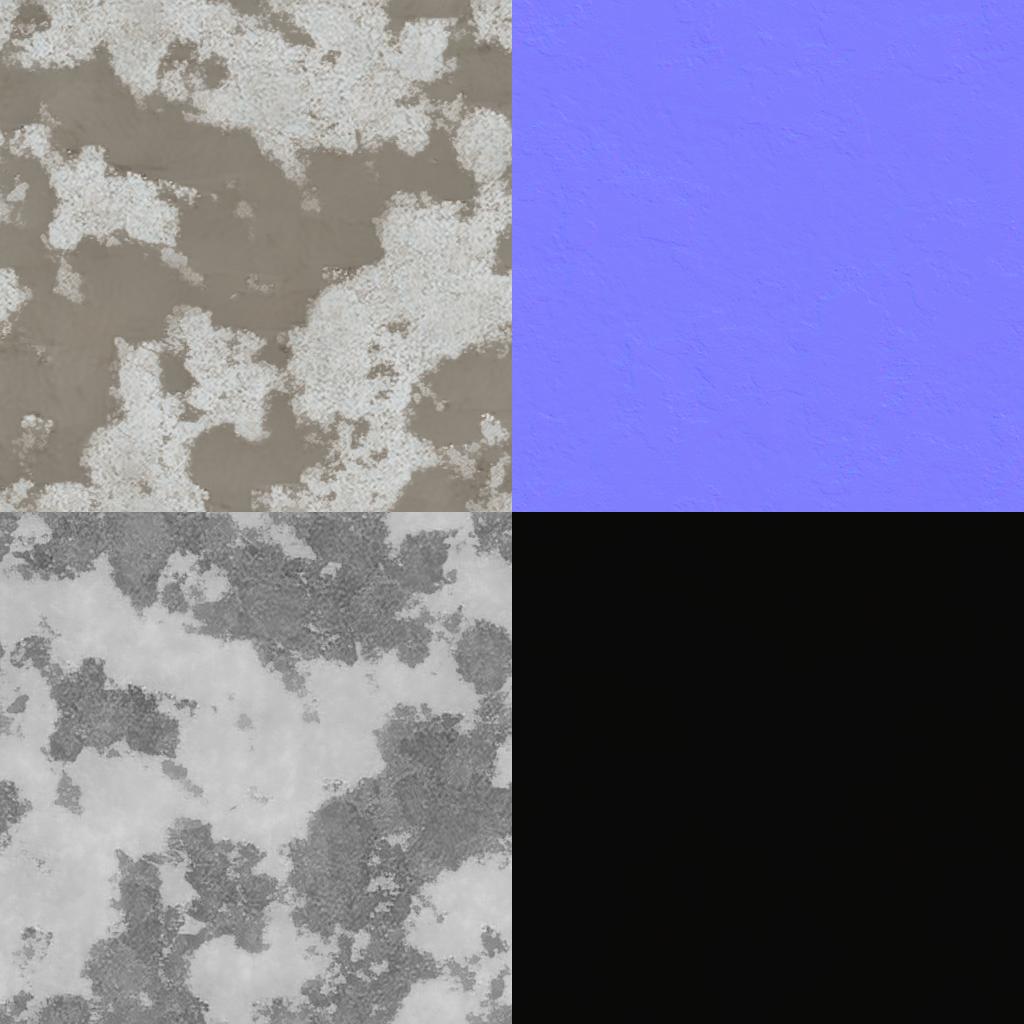}}}} \\
	    \multicolumn{3}{c}{Guided Gram Matrix} & \multicolumn{3}{c}{Sliced Wasserstein Loss}\\
	\end{tabular}
\vspace{-10pt}
\caption{For our multi-target transfer, we evaluate a Guided Gram Matrix which leads to artifacts which do not appear with our resampled Wasserstein loss. We hypothesize that these artifacts appear because the Guided Gram Matrix remains impacted by the statistics of target bricks borders despite the erosion we perform.}
\label{fig:gram}
\end{figure}

Now suppose we have a transfer rule \textbf{Label i: Target $\hat{I}$, j}, instructing us to transfer materials from regions labeled by $j$ in $\hat{I}$ to regions labeled by $i$ in the rendered image $R(\theta)$. We take samples labeled with $i$ from $p^l_V$ and samples labeled with $j$ from $\hat{p}^l_V$ as $p^l_V\{i\}$ and $\hat{p}^l_V\{j\}$. Note here $p^l_V\{i\}$ and $\hat{p}^l_V\{j\}$ usually contain different number of samples, therefore we compute the sliced Wasserstein loss using our proposed resampling technique (Eq. \ref{Eq:Wasserstein}). We compute this loss for each transfer rule separately and take their average as our final loss. For completeness, we also evaluate the Gram Matrix \cite{gatys2017} for partial transfer and show that it can lead to artifacts as shown in Fig.~\ref{fig:gram}. We therefore adopt sliced Wasserstein loss with resampling in all our experiments.

\begin{figure}
	\centering
	\renewcommand{\arraystretch}{0.6}
	\addtolength{\tabcolsep}{-4pt}
	\begin{tabular}{ccc}
		\includegraphics[width=0.225\textwidth]{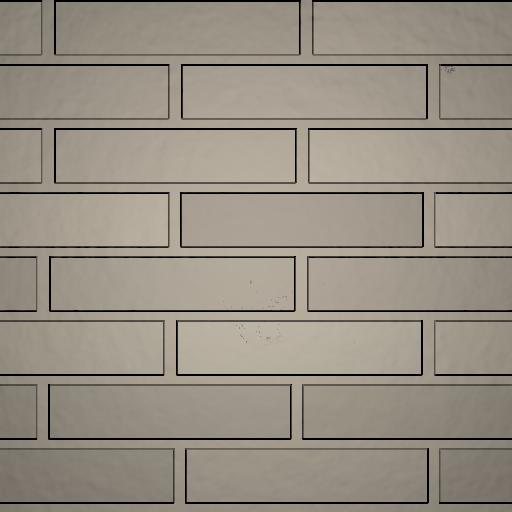}\llap{\frame{\includegraphics[width=0.07\textwidth]{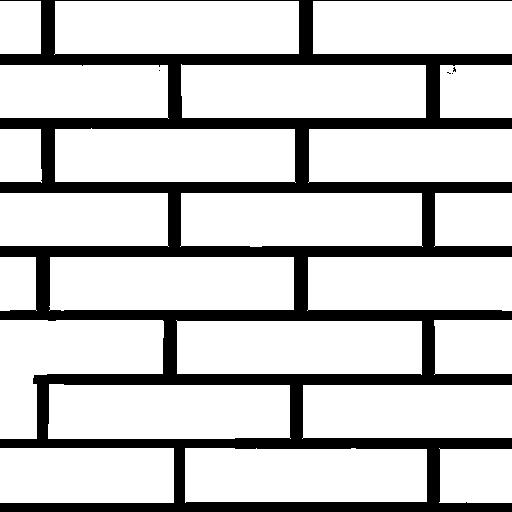}}} &
	    \includegraphics[width=0.225\textwidth]{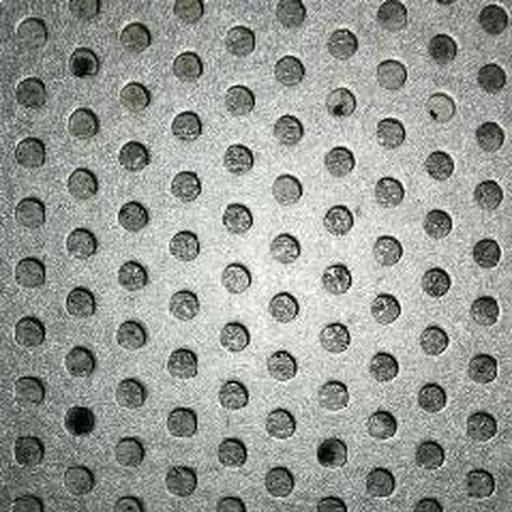}\llap{\frame{\includegraphics[width=0.07\textwidth]{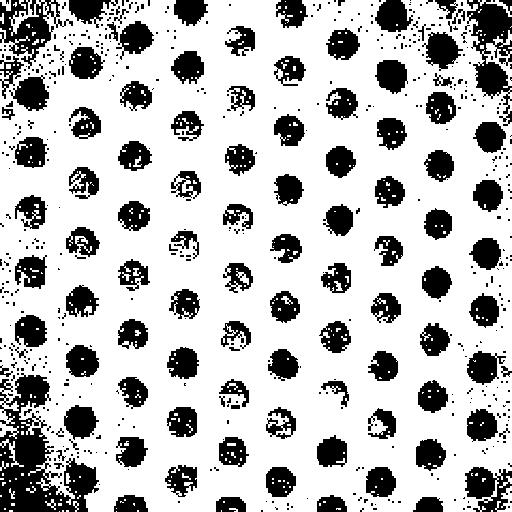}}} \\
	    Init. & Target \\
        \includegraphics[width=0.225\textwidth]{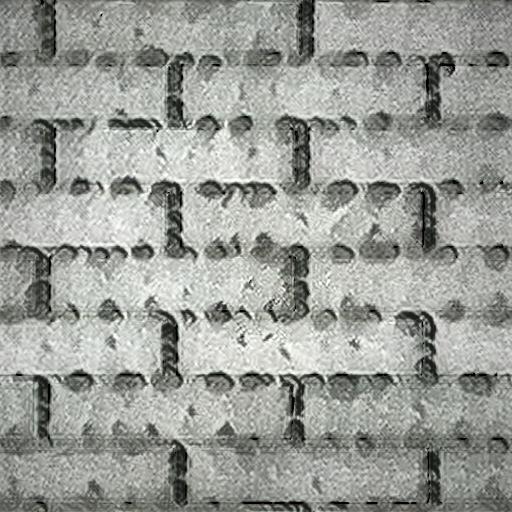} &
        \includegraphics[width=0.225\textwidth]{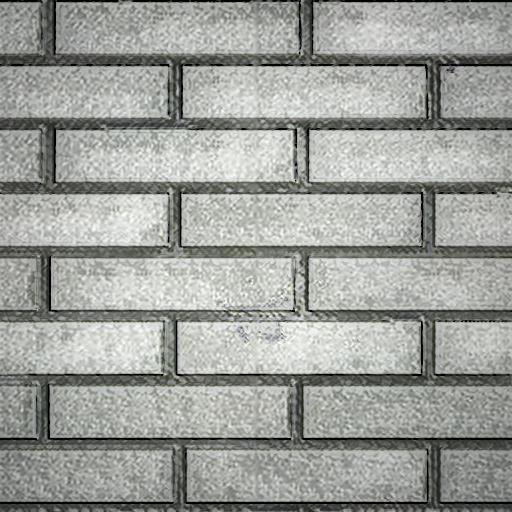} \\
        W/o Erosion & With Erosion \\
	\end{tabular}
\vspace{-10pt}
\caption{We show the importance of applying erosion on label maps. When computing appearance features at boundary pixels of the label map, the receptive field extends beyond the boundary. Therefore without erosion, the transferred appearance would include regions outside the label mask.
Erosion of the binary masks makes sure the appearance features only include information inside the masked region, 
preserving the desired appearance.
}
\label{fig:erosion}
\end{figure}

A particular case is made of the boundary features, as neurons on the labeled boundary will have a receptive field which crosses the boundary due to the footprint of the deep convolutions, forcing them to consider statistics from irrelevant nearby regions. To prevent transferring unrelated material statistics, similar to Gatys et al~\shortcite{gatys2017}, we perform an erosion operation on the labeled regions, and only evaluate the sliced Wasserstein loss on the eroded regions. Fig. \ref{fig:erosion} shows an example with and without this erosion step. We note that while an erosion step reduces irrelevant texture transfer, too large an erosion may remove all samples from the distribution at deeper layers. In such case, we do not compute the loss for deeper layers with no valid pixels.
\begin{figure}
	\centering
	\renewcommand{\arraystretch}{0.6}
	\addtolength{\tabcolsep}{-5pt}
	\begin{tabular}{cccccc}
		& \scalebox{0.8}{Render} & \scalebox{0.8}{Albedo} & \scalebox{0.8}{Normal} & \scalebox{0.8}{Roughness} & \scalebox{0.8}{Specular} \\
		\raisebox{20pt}{\rotatebox[origin=c]{90}{\scalebox{0.8}{GT}}} &
		\includegraphics[width=0.09\textwidth]{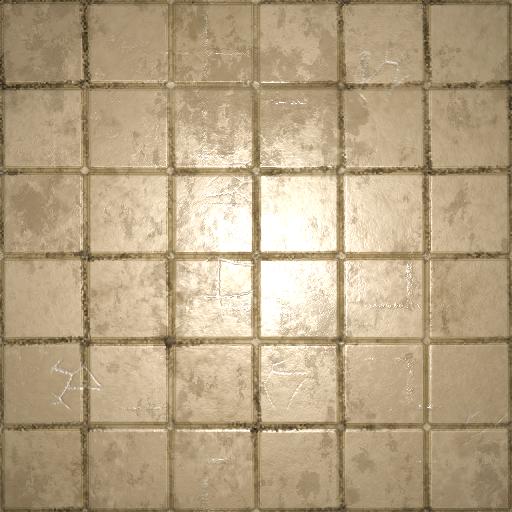} &
		\includegraphics[width=0.09\textwidth]{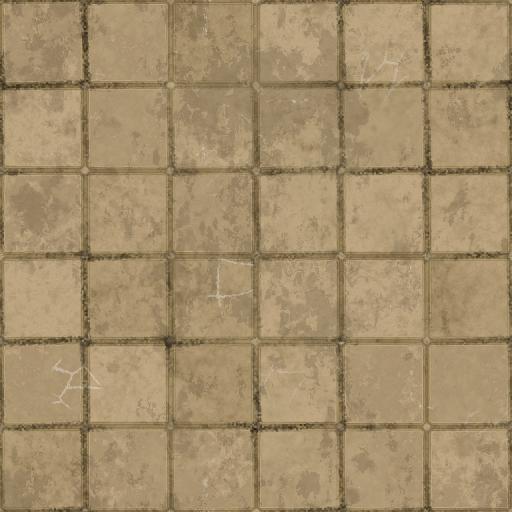} &
		\includegraphics[width=0.09\textwidth]{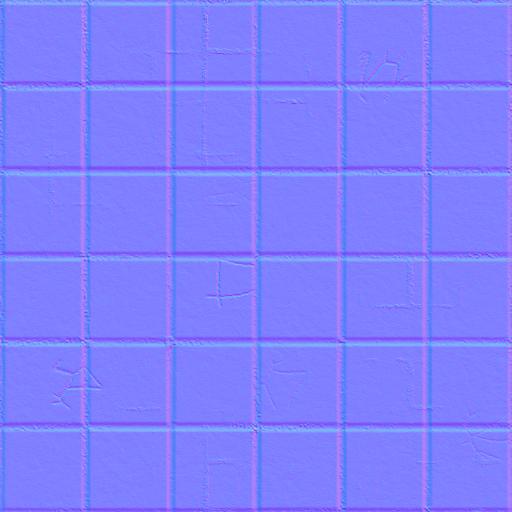} &
		\includegraphics[width=0.09\textwidth]{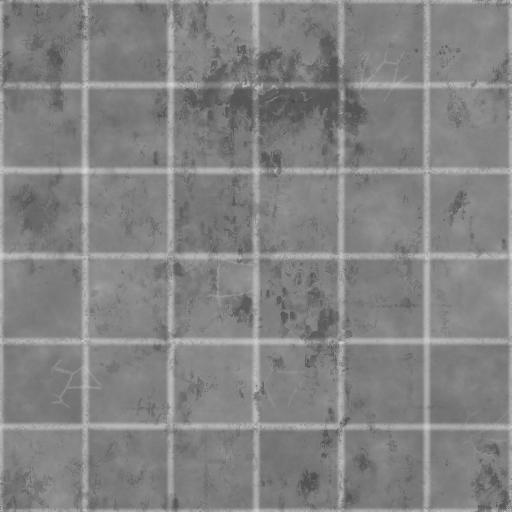} &
		\includegraphics[width=0.09\textwidth]{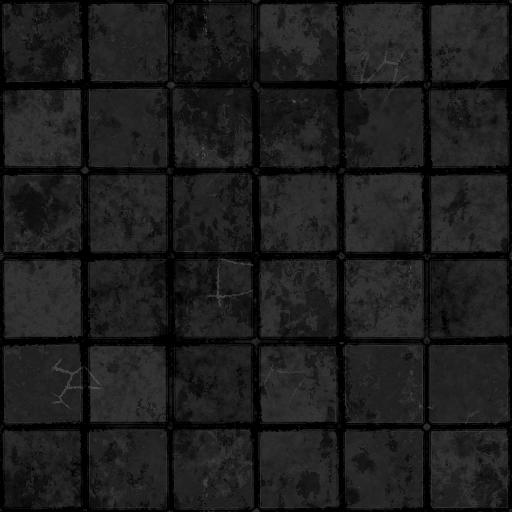} \\
		
		\raisebox{20pt}{\rotatebox[origin=c]{90}{\scalebox{0.8}{$W$}}} &
		\includegraphics[width=0.09\textwidth]{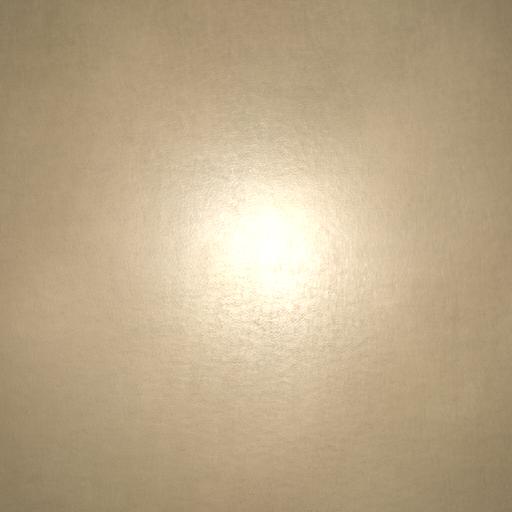} &
		\includegraphics[width=0.09\textwidth]{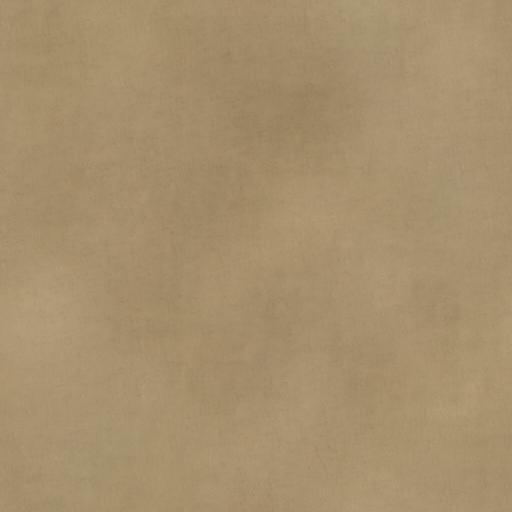} &
		\includegraphics[width=0.09\textwidth]{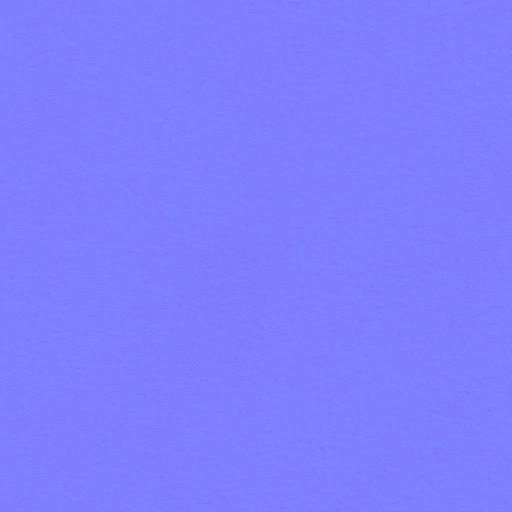} &
		\includegraphics[width=0.09\textwidth]{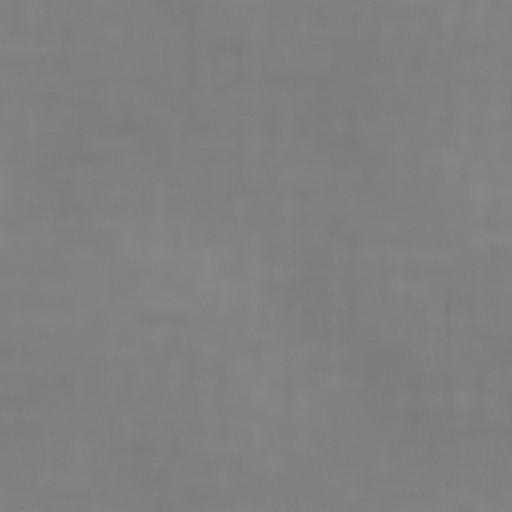} &
		\includegraphics[width=0.09\textwidth]{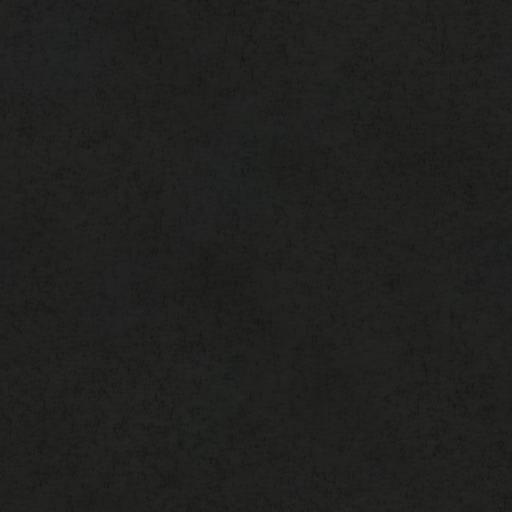} \\
		
		\raisebox{20pt}{\rotatebox[origin=c]{90}{\scalebox{0.8}{$W^+$}}} &
		\includegraphics[width=0.09\textwidth]{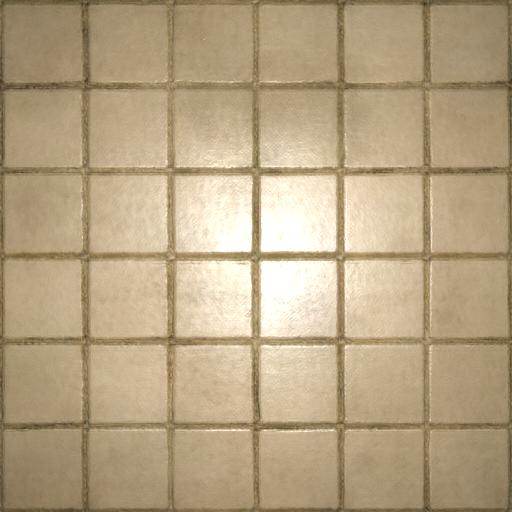} &
		\includegraphics[width=0.09\textwidth]{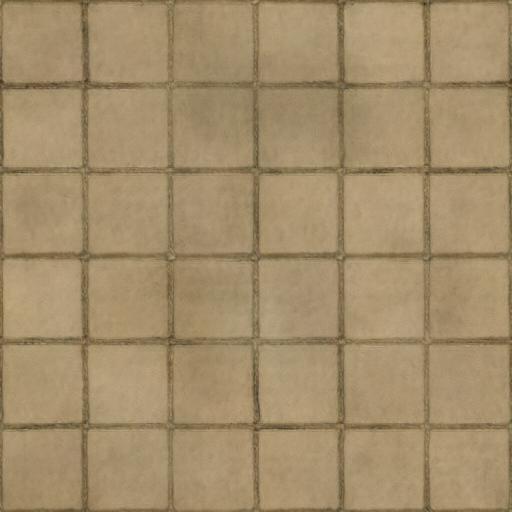} &
		\includegraphics[width=0.09\textwidth]{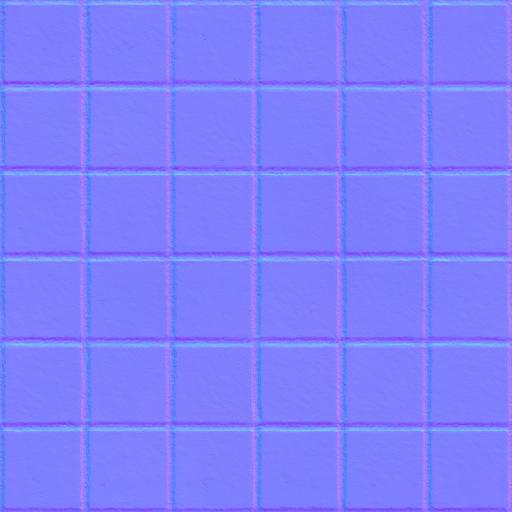} &
		\includegraphics[width=0.09\textwidth]{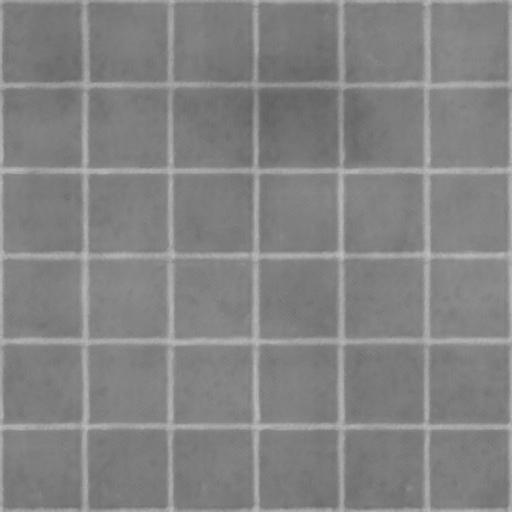} &
		\includegraphics[width=0.09\textwidth]{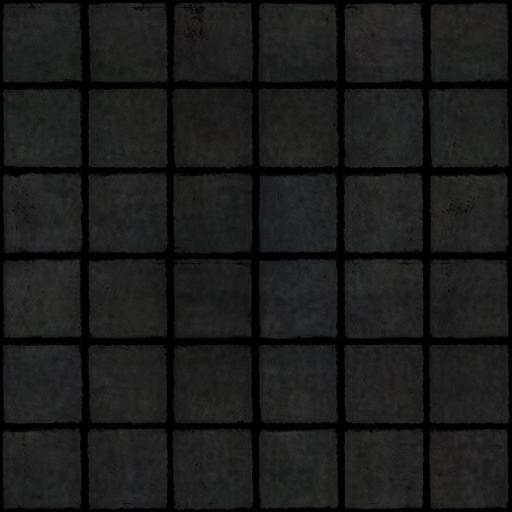} \\
		
		\raisebox{20pt}{\rotatebox[origin=c]{90}{\scalebox{0.8}{$W^+N$}}} &
		\includegraphics[width=0.09\textwidth]{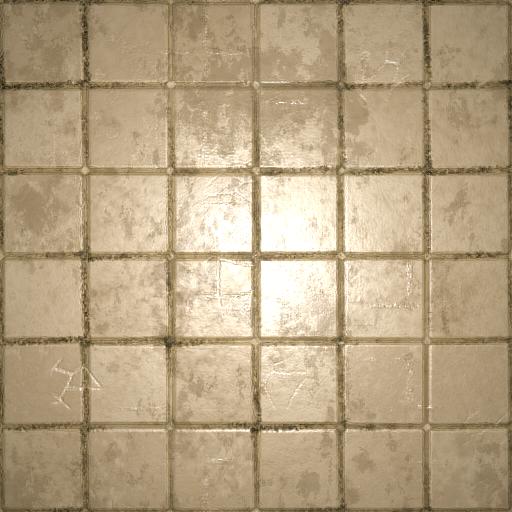} &
		\includegraphics[width=0.09\textwidth]{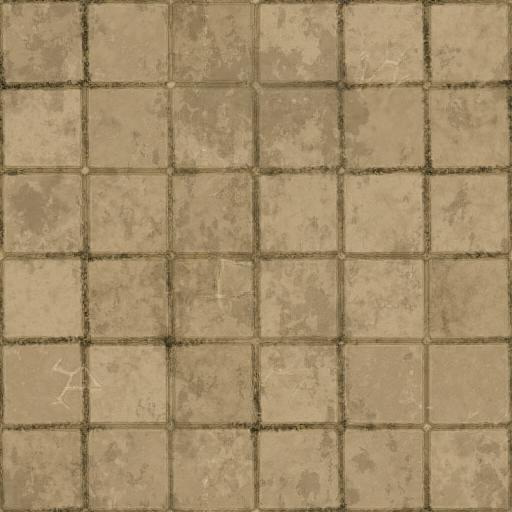} &
		\includegraphics[width=0.09\textwidth]{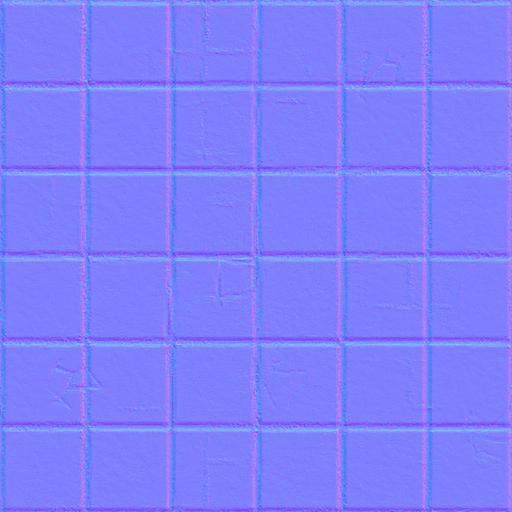} &
		\includegraphics[width=0.09\textwidth]{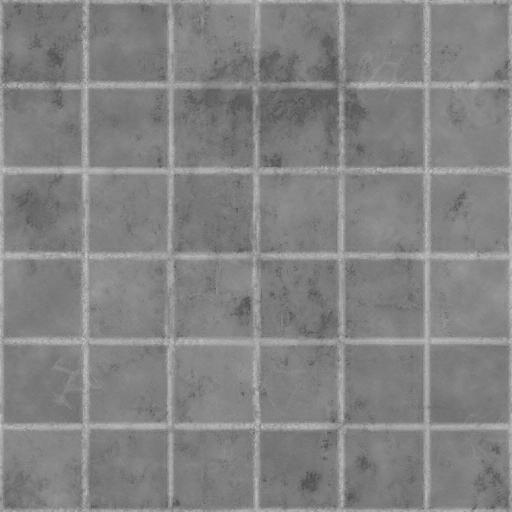} &
		\includegraphics[width=0.09\textwidth]{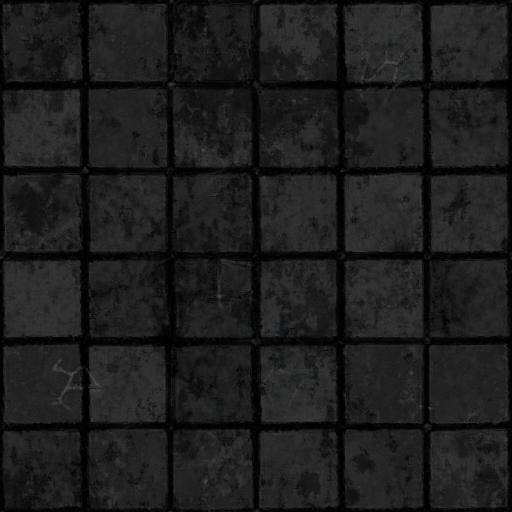} \\
		
	\end{tabular}
\vspace{-10pt}
\caption{We show projection results when optimizing Eq. \ref{Eq:project} with three different latent space of tileable MaterialGAN: $W$, $W^+$ and $W^+N$. We see that we can only capture both large and small scale structures using $W^+N$, while $W$, $W^+$ either fail to approximate the detail appearance or miss fine scale structures.}
\label{fig:projection}
\end{figure}
\section{Implementation Details} \label{Sec:implementation}
We implement our algorithm in PyTorch for both training and optimization. We train our tileable MaterialGAN model using a synthetic material dataset containing 199,068 images with a resolution of 288x288~\cite{Deschaintre18}. The material maps are encoded as 9-channel 2D images (3 for albedo, 2 for normal, 1 for roughness and 3 for specular). The full model is trained by crops: we train the generator to synthesize 512x512 material maps and we make a 2x2 tile (1024x1024) and randomly crop to 256x256 to compare with the randomly cropped 256x256 ground truth material maps. The architecture of the GAN model ensure tileability (Sec. \ref{Sec:tileable-StyleGAN}), despite the crops being not tileable. For important hyperparameters, we empirically set $\gamma=10$ for R1 regularization and weight of path length regularization as $1$ \cite{stylegan2}. We train the network using an Adam optimizer ($\beta=(0.9, 0.999)$) with a learning rate of 0.002 on 8 Nvidia Tesla V100 GPUs. The full training takes 2$\sim$3 days with a batch size of 32. 

We run experiments on our optimization-based material transfer on an Intel i9-10850K machine with Nvidia RTX 3090. The optimization is built on the pre-trained tileable MaterialGAN model as a material prior. Specifically, MaterialGAN has multiple latent spaces: $z\in Z$, the input latent code; $w\in W$, the intermediate latent code after linear mapping; per-layer style code $w^+\in W^+$; and noise inputs for each blocks $n\in N$. In previous work, different latent codes have been proposed for various applications \cite{stylegan2ada, Guo20}. In our experiments, we optimize both $W^+$ and $N$, enabling our optimization to capture both large-scale structure and fine-scale details. In Fig. \ref{fig:projection}, we show projection results (Eq. \ref{Eq:project}) only by optimizing $W$ or $W^+$. Results show that $W$ is less expressive, while $W^+$ can capture large scale structures, but misses fine scale structures.

For the projection step, we run 1000 iterations with an Adam optimizer with a learning rate of 0.08, taking around 5 minutes. We extract deep features from [relu1\_2, relu2\_2, relu3\_2, relu4\_2] in a pre-trained VGG19 neural network to evaluate the feature loss in Eq. \ref{Eq:project}. As we find the projection of the details in the normal map to be harder than for other maps, we assign a weight of 5 to the normal maps loss while 1 for other material maps.

After projection, we optimize the embedded latent code $\theta$ to minimize the loss function in Eq. \ref{Eq:style-transfer}. Similar to style transfer, we take deep features from relu4\_2 to compute feature loss, and extract deep features from layers [relu1\_1, relu2\_1, relu3\_1, relu4\_1] to compute the sliced Wasserstein loss. We weigh style losses from different layers by [5, 5, 5, 0.5] respectively, emphasizing local features. To compute the sliced Wasserstein loss, we sample a number of random projection directions equal to the number of channels of the compared deep features as suggested in the original paper~\cite{Heitz2021}. We run 500 iterations using an Adam optimizer with a learning rate of 0.02, taking about 2.5 minutes.

For spatial control, if the input material maps come from a procedural material \cite{SubstanceDes}, the label maps can be extracted from the graph. For target photos, we compute the label maps with a scribble-based segmentation method~\cite{hu2022}, which takes less than a minute overall. However, a precise and full segmentation of the example photos is not always necessary: users only need to indicate a few example regions of the material they want to transfer. We apply an erosion operation on each resampled label map with a kernel size of 5 pixels for 512x512 images as described in Sec. \ref{Sec:Spatial-Control}. This erosion size may need to be adapted for different image resolutions.
\begin{figure*} %
	\centering
	\addtolength{\tabcolsep}{-5.5pt}
	\setlength{\extrarowheight}{1.5pt}
 	\def\arraystretch{1.0}
	\begin{tabular}{ccccccccc}
		Target & & Albedo & Normal & Roughness & Specular & Rendered & Our 2x2 Tiled\\

		\includegraphics[width=0.12\textwidth]{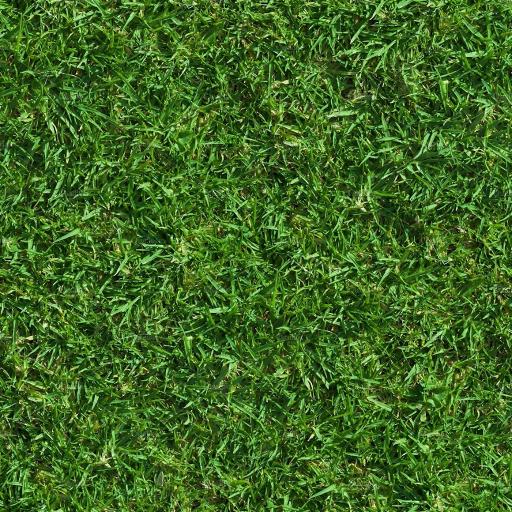} &
		\raisebox{30pt}{\scalebox{1.0}{\rotatebox[origin=c]{90}{Input}}} &
		\includegraphics[width=0.12\textwidth]{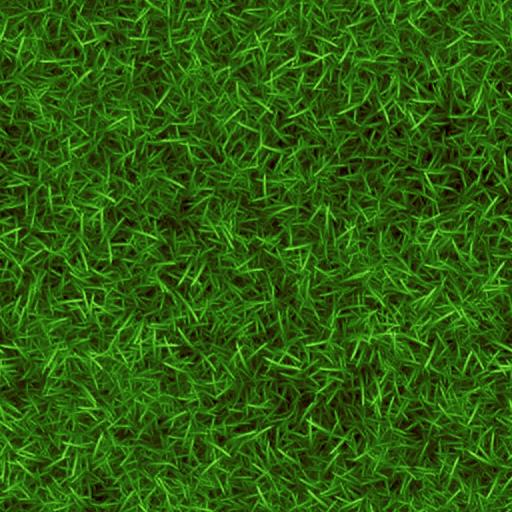} &
		\includegraphics[width=0.12\textwidth]{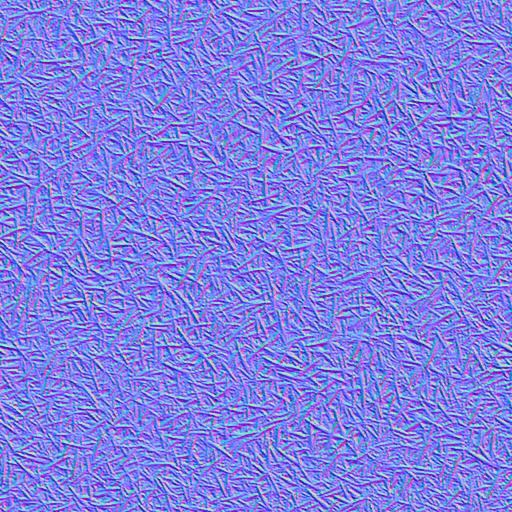} &
		\includegraphics[width=0.12\textwidth]{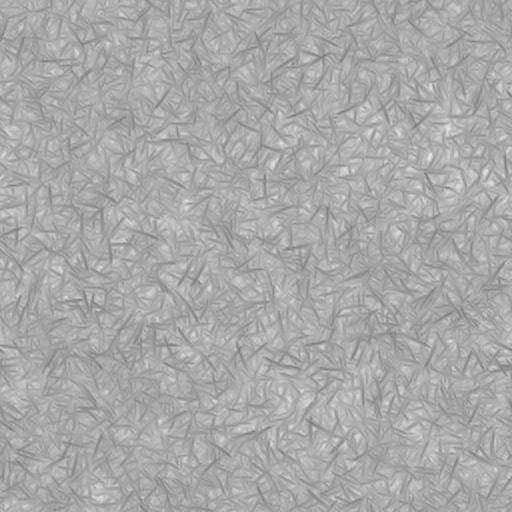} &
		\includegraphics[width=0.12\textwidth]{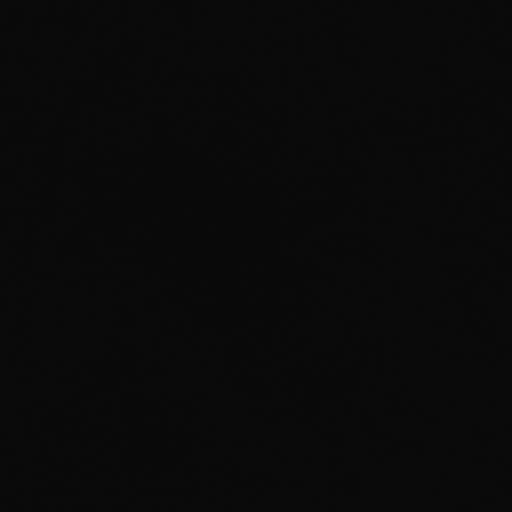} &
        \includegraphics[width=0.12\textwidth]{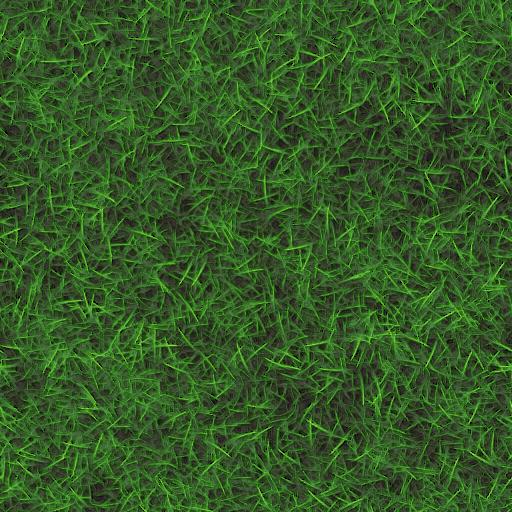} &
        \multirow{2}{*}[51pt]{\includegraphics[width=0.246\textwidth]{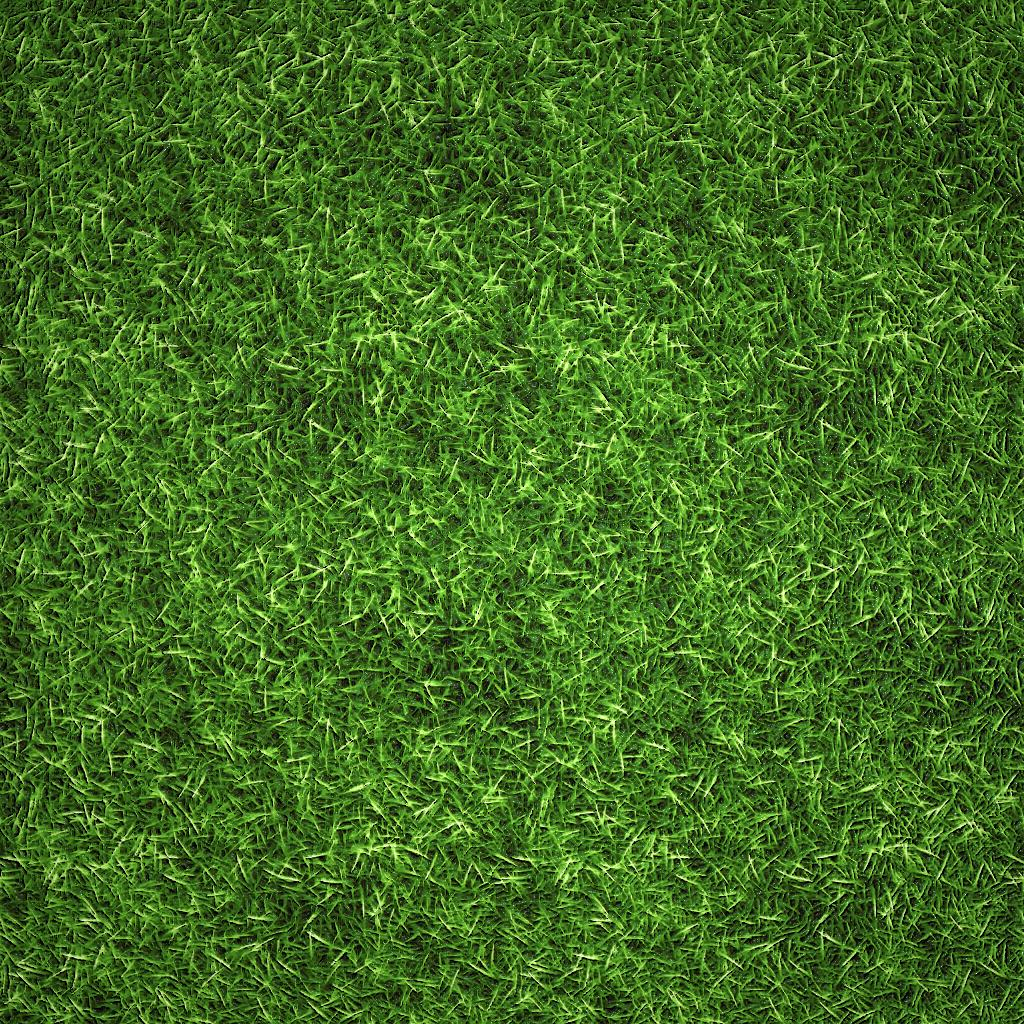}} \\
        &
		\raisebox{30pt}{\scalebox{1.0}{\rotatebox[origin=c]{90}{Transferred}}} &
		\includegraphics[width=0.12\textwidth]{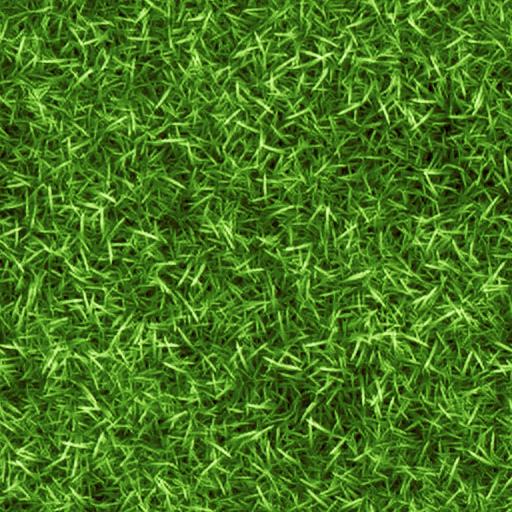} &
		\includegraphics[width=0.12\textwidth]{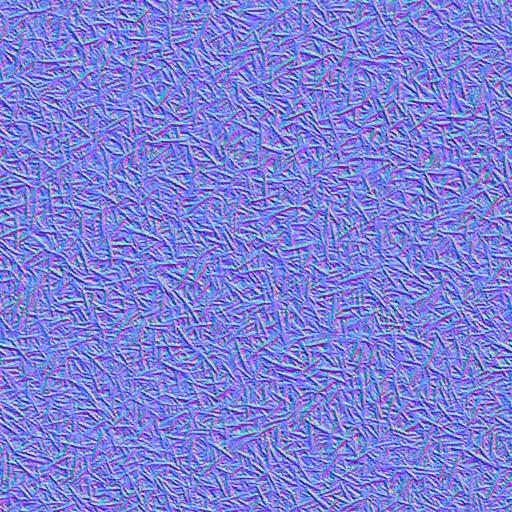} &
		\includegraphics[width=0.12\textwidth]{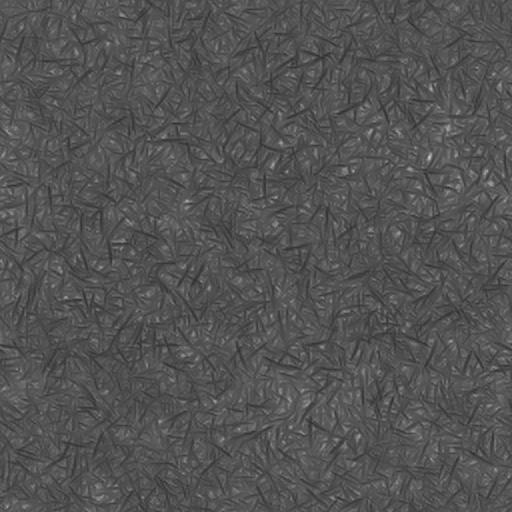} &
		\includegraphics[width=0.12\textwidth]{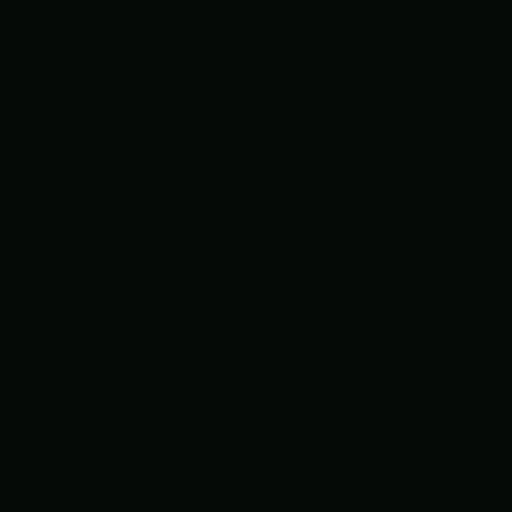} &
        \includegraphics[width=0.12\textwidth]{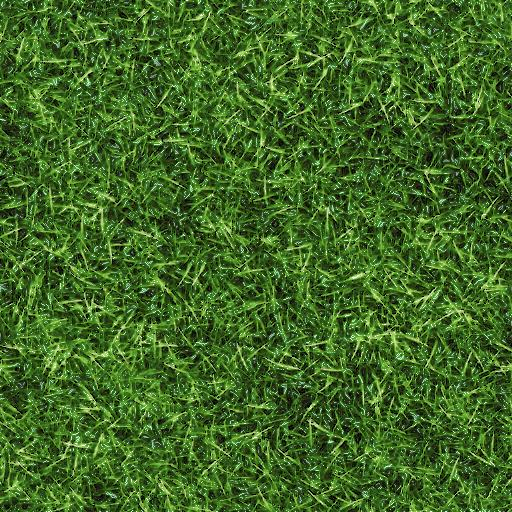} & & \\
        
        \includegraphics[width=0.12\textwidth]{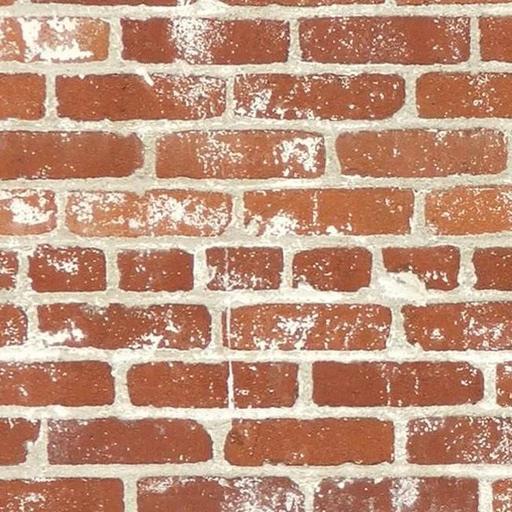} &
		\raisebox{30pt}{\scalebox{1.0}{\rotatebox[origin=c]{90}{Input}}} &
		\includegraphics[width=0.12\textwidth]{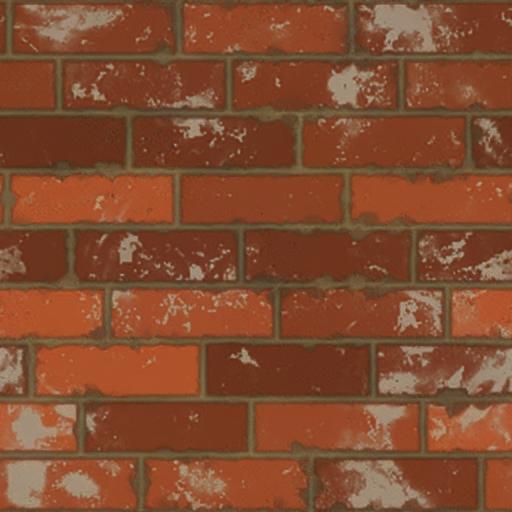} &
		\includegraphics[width=0.12\textwidth]{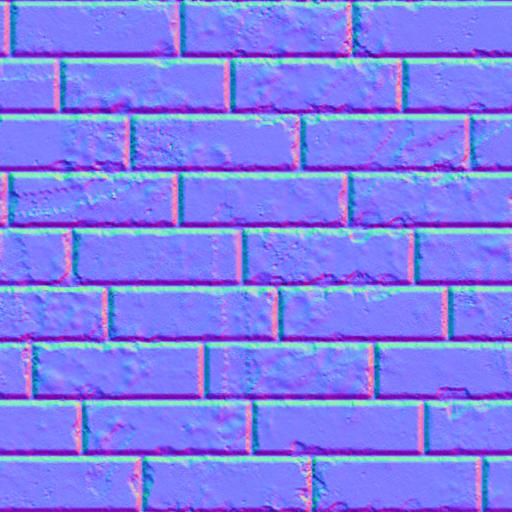} &
		\includegraphics[width=0.12\textwidth]{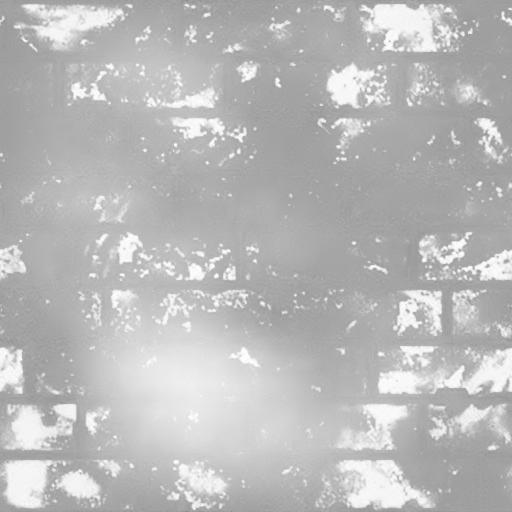} &
		\includegraphics[width=0.12\textwidth]{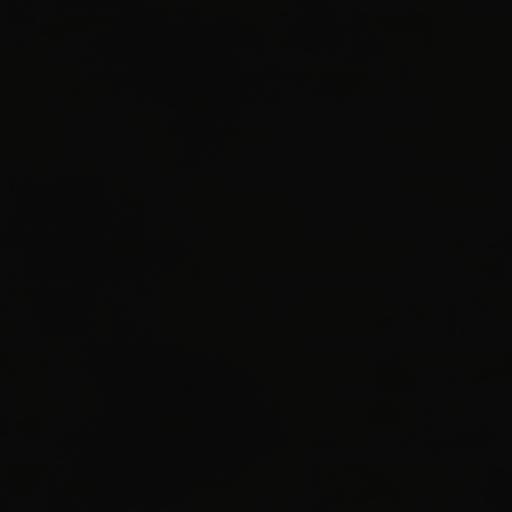} &
        \includegraphics[width=0.12\textwidth]{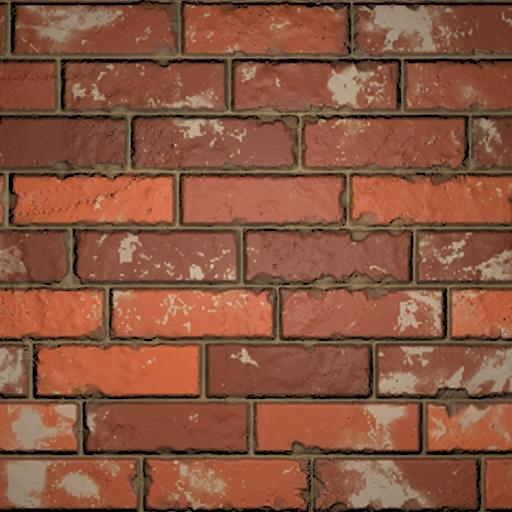} &
        \multirow{2}{*}[51pt]{\includegraphics[width=0.246\textwidth]{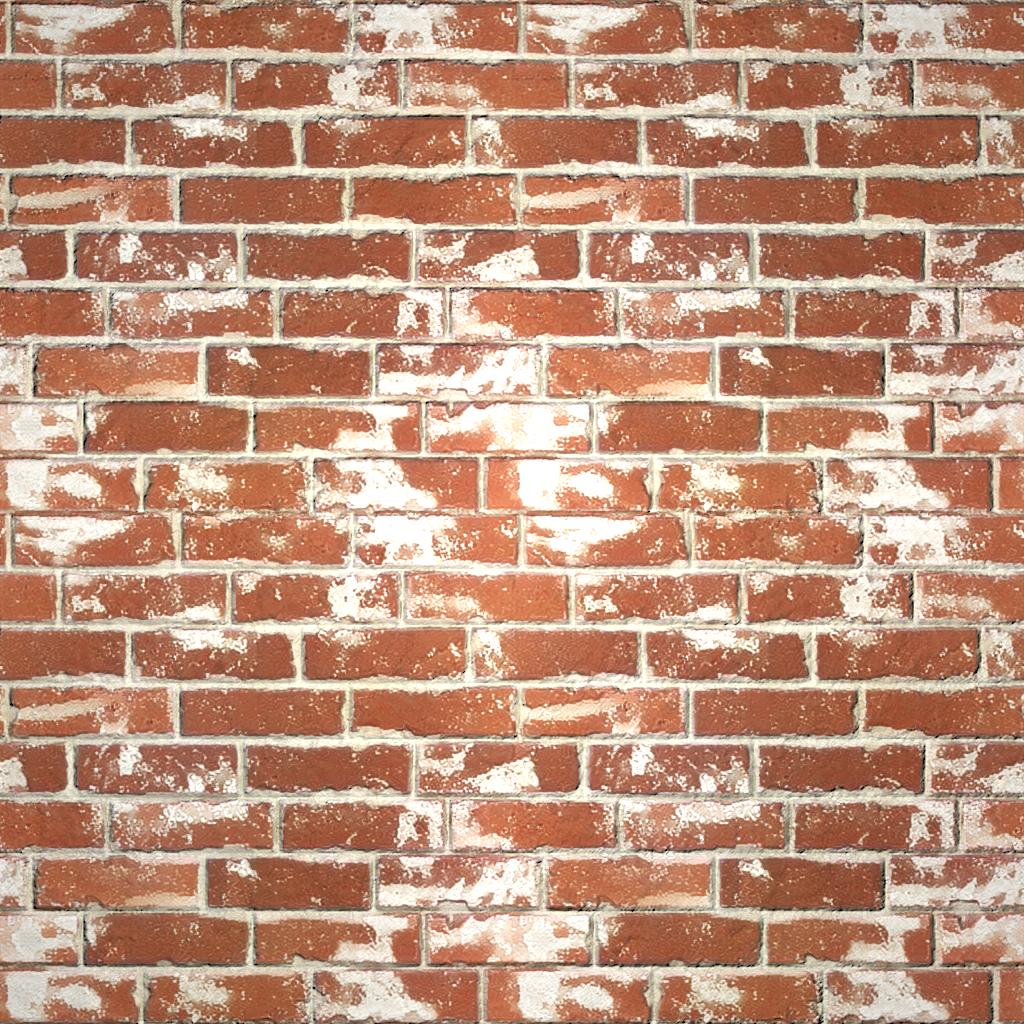}} \\
        &
		\raisebox{30pt}{\scalebox{1.0}{\rotatebox[origin=c]{90}{Transferred}}} &
		\includegraphics[width=0.12\textwidth]{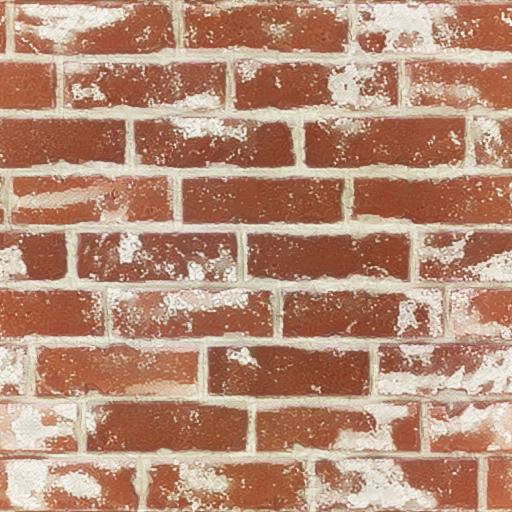} &
		\includegraphics[width=0.12\textwidth]{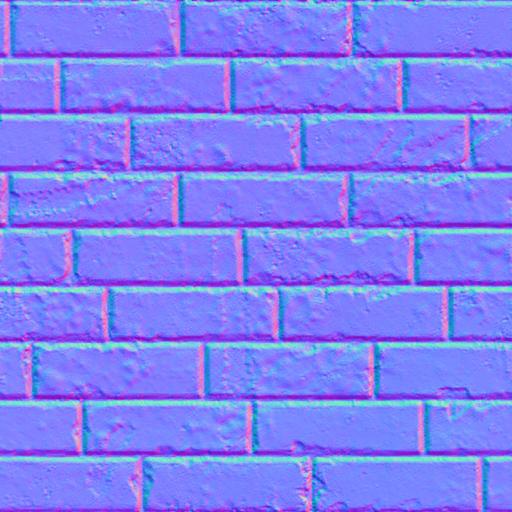} &
		\includegraphics[width=0.12\textwidth]{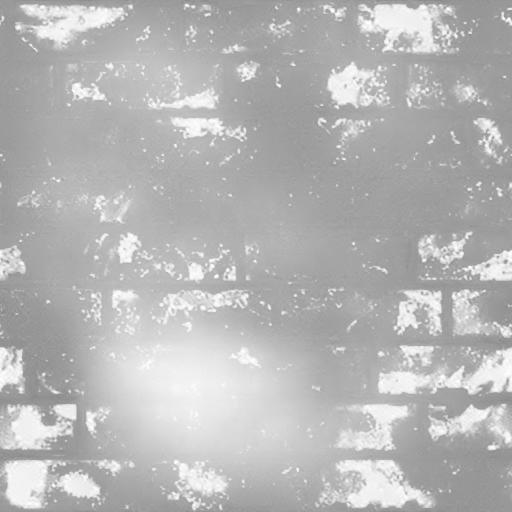} &
		\includegraphics[width=0.12\textwidth]{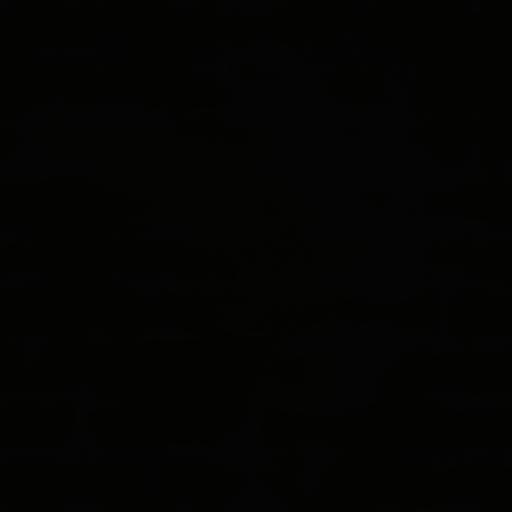} &
        \includegraphics[width=0.12\textwidth]{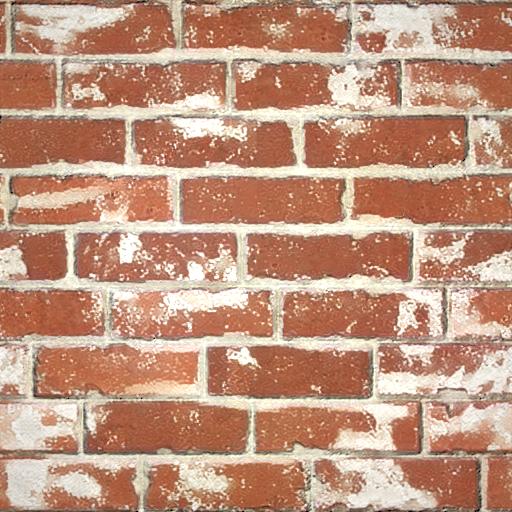} & & \\
        
        \includegraphics[width=0.12\textwidth]{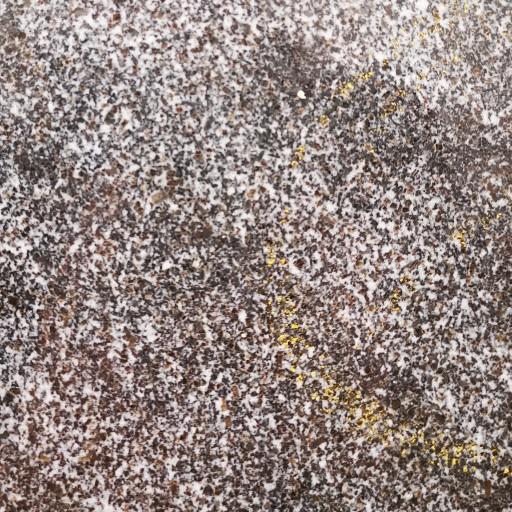} &
		\raisebox{30pt}{\scalebox{1.0}{\rotatebox[origin=c]{90}{Input}}} &
		\includegraphics[width=0.12\textwidth]{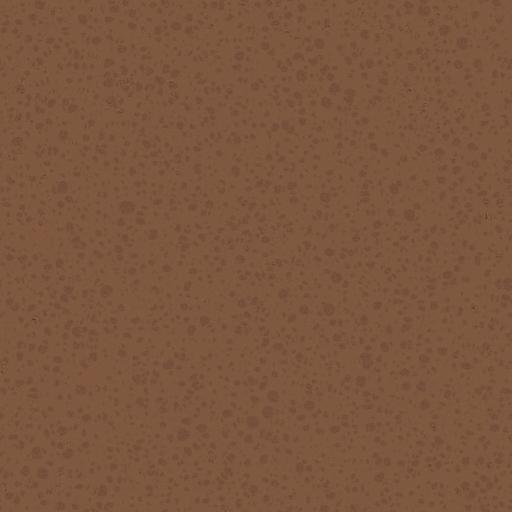} &
		\includegraphics[width=0.12\textwidth]{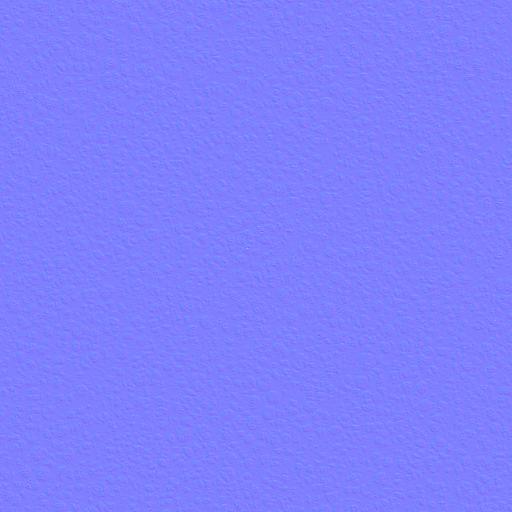} &
		\includegraphics[width=0.12\textwidth]{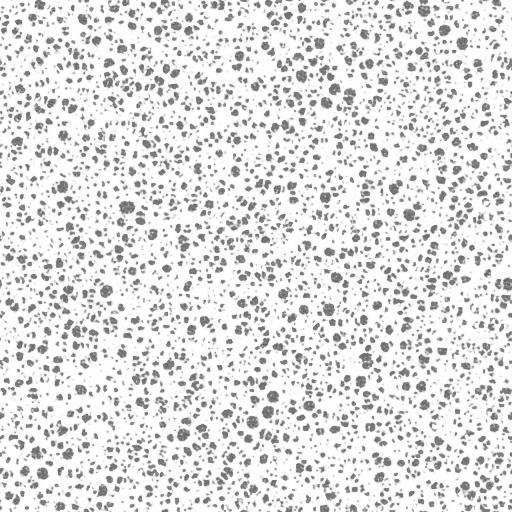} &
		\includegraphics[width=0.12\textwidth]{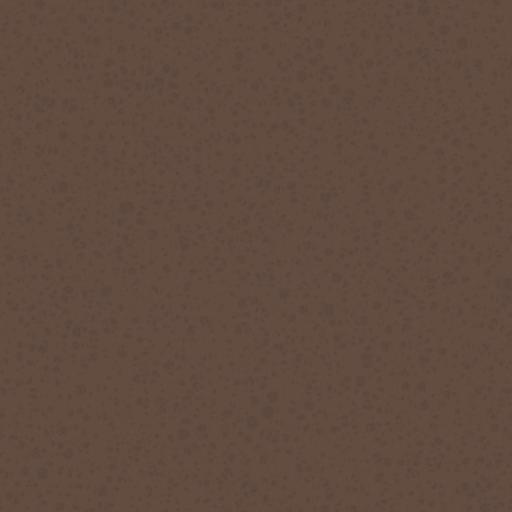} &
        \includegraphics[width=0.12\textwidth]{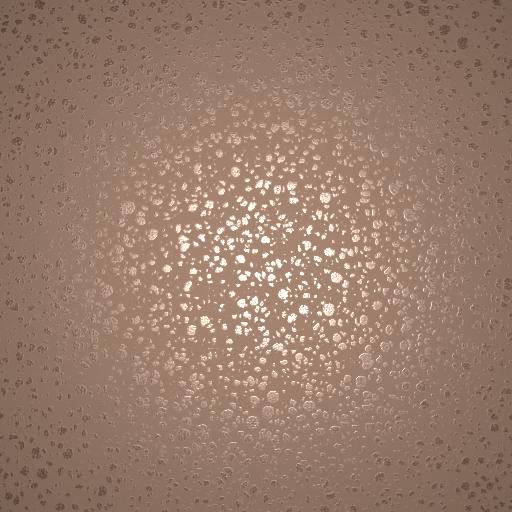} &
        \multirow{2}{*}[51pt]{\includegraphics[width=0.246\textwidth]{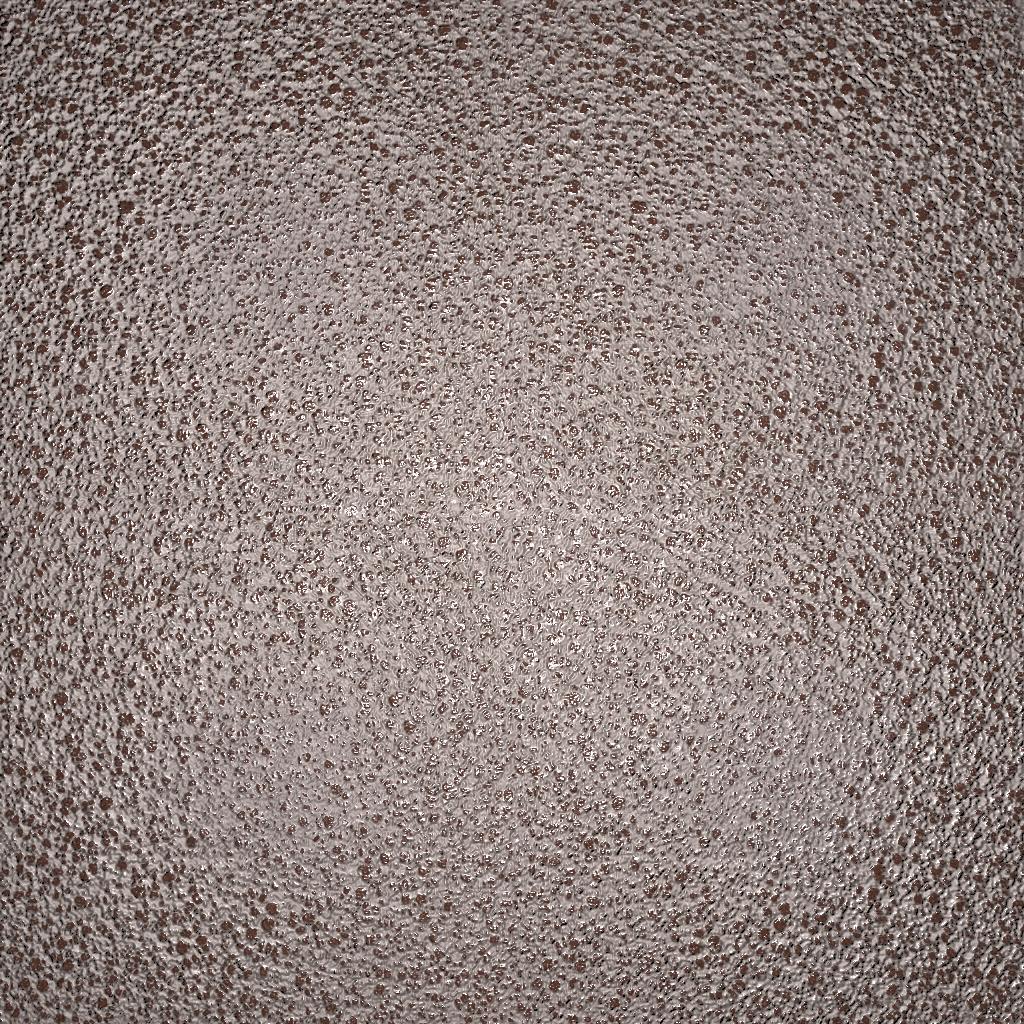}} \\
        &
		\raisebox{30pt}{\scalebox{1.0}{\rotatebox[origin=c]{90}{Transferred}}} &
		\includegraphics[width=0.12\textwidth]{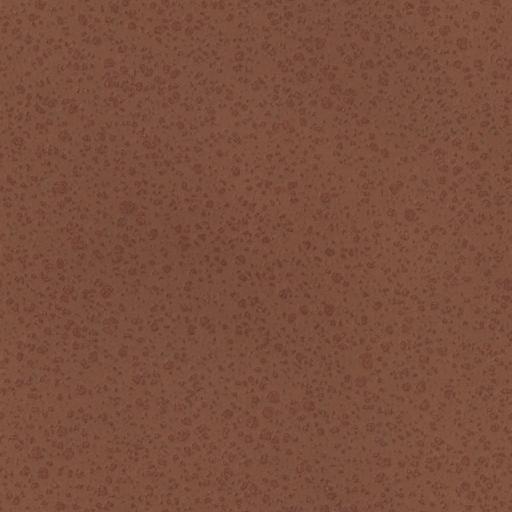} &
		\includegraphics[width=0.12\textwidth]{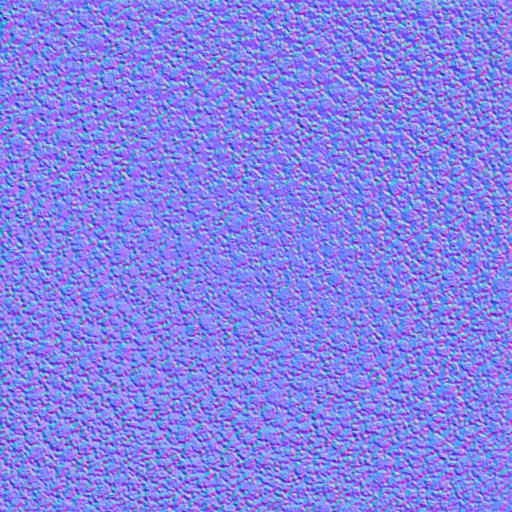} &
		\includegraphics[width=0.12\textwidth]{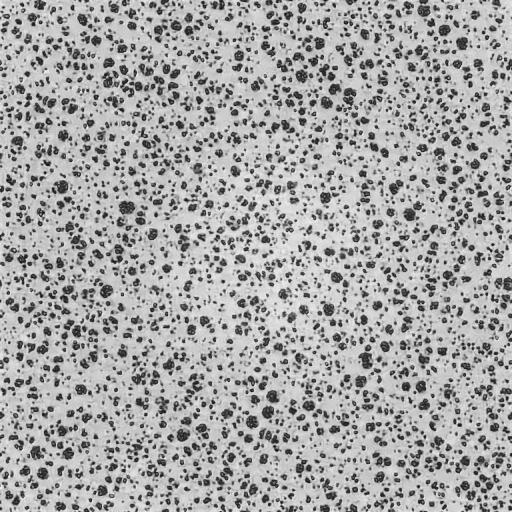} &
		\includegraphics[width=0.12\textwidth]{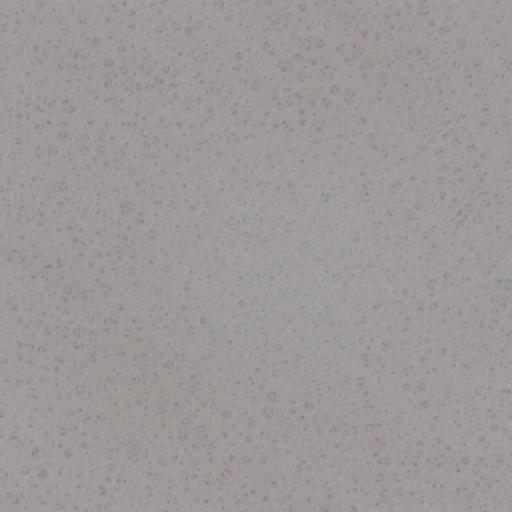} &
        \includegraphics[width=0.12\textwidth]{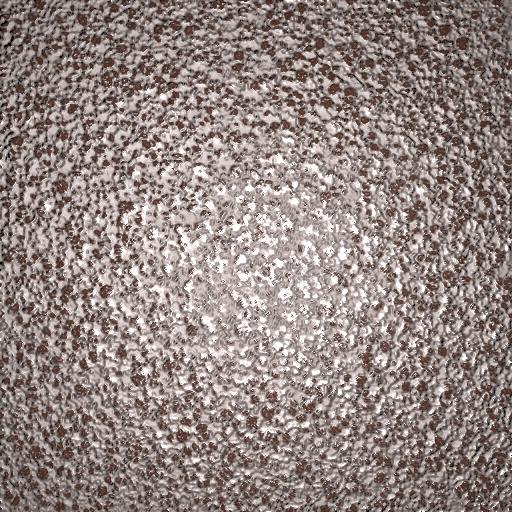} & & \\
        
        \includegraphics[width=0.12\textwidth]{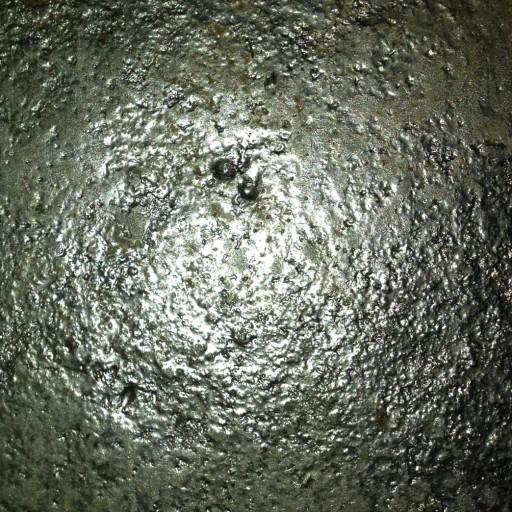} &
		\raisebox{30pt}{\scalebox{1.0}{\rotatebox[origin=c]{90}{Input}}} &
		\includegraphics[width=0.12\textwidth]{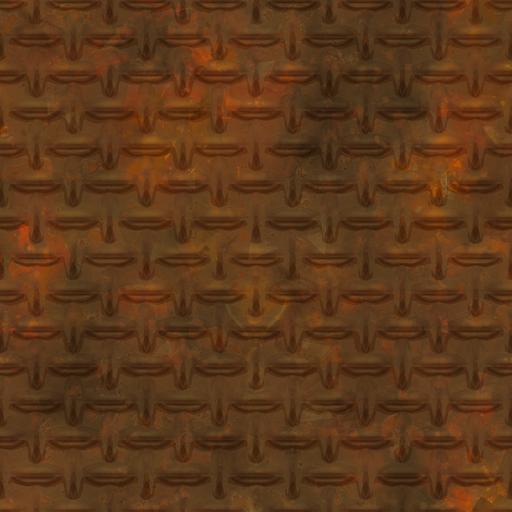} &
		\includegraphics[width=0.12\textwidth]{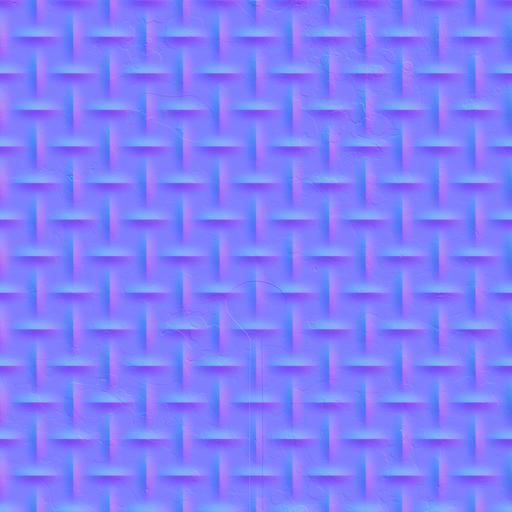} &
		\includegraphics[width=0.12\textwidth]{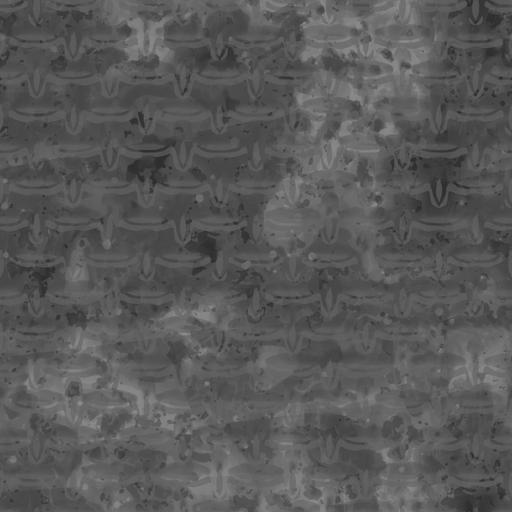} &
		\includegraphics[width=0.12\textwidth]{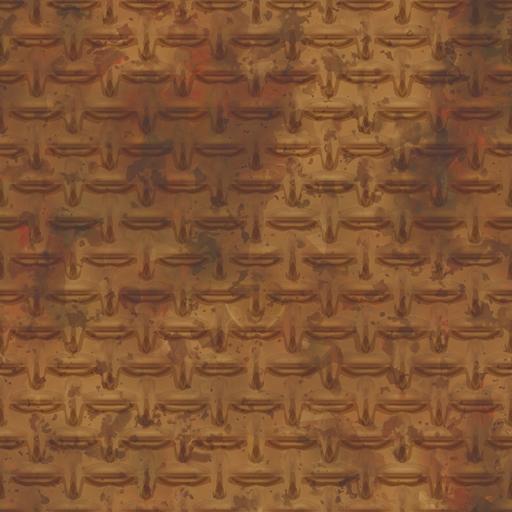} &
        \includegraphics[width=0.12\textwidth]{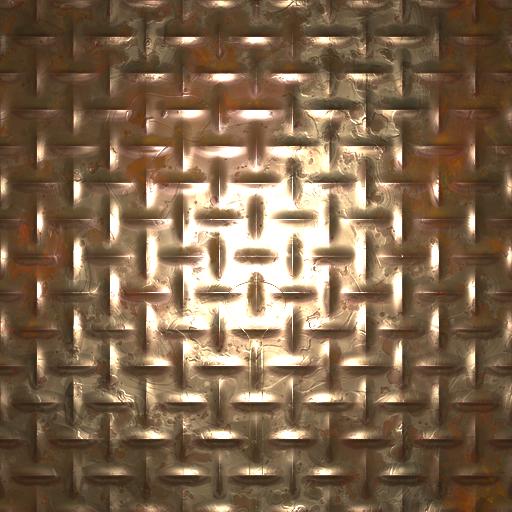} &
        \multirow{2}{*}[51pt]{\includegraphics[width=0.246\textwidth]{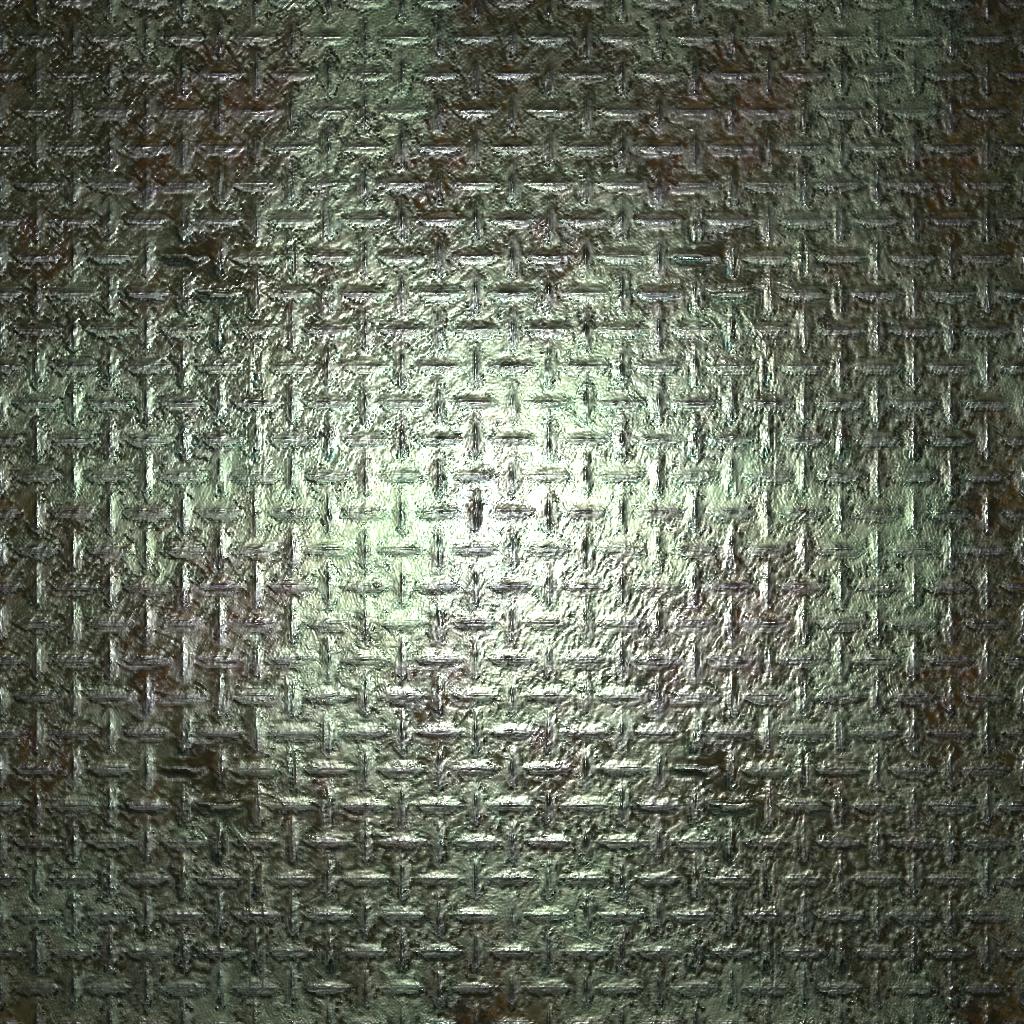}} \\
        &
		\raisebox{30pt}{\scalebox{1.0}{\rotatebox[origin=c]{90}{Transferred}}} &
		\includegraphics[width=0.12\textwidth]{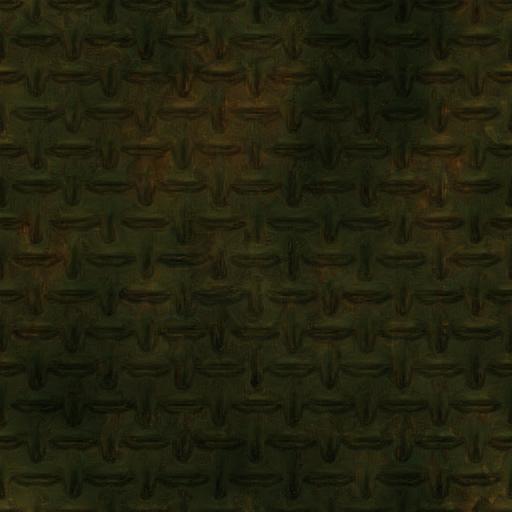} &
		\includegraphics[width=0.12\textwidth]{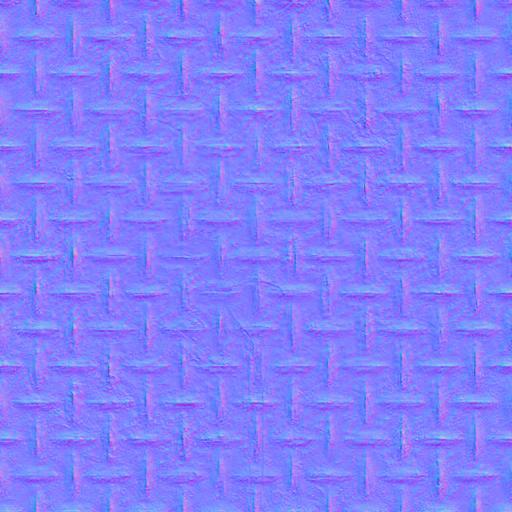} &
		\includegraphics[width=0.12\textwidth]{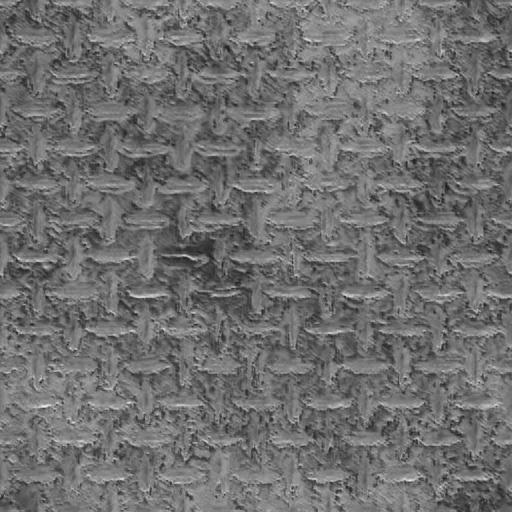} &
		\includegraphics[width=0.12\textwidth]{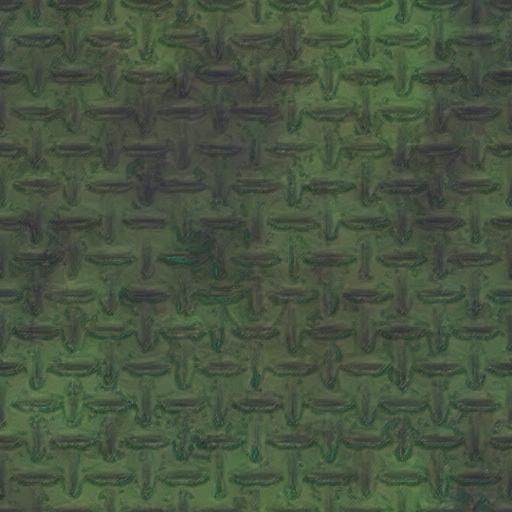} &
        \includegraphics[width=0.12\textwidth]{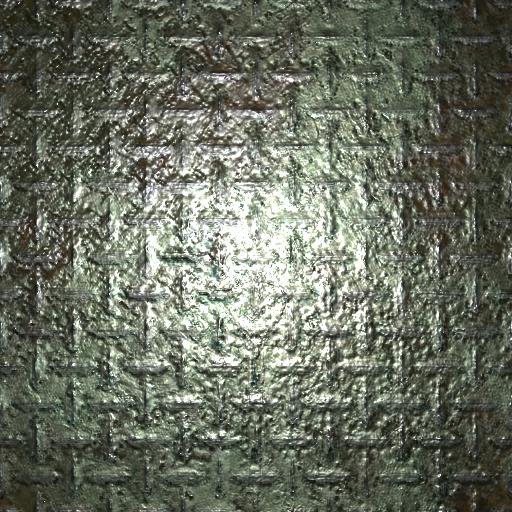} & & \\
	\end{tabular}
\vspace{-10pt}
\caption{Material Transfer Results. All target images are real photographs. We show each material map (512x512) before and after optimization. We also show a 2x2 tiled rendered image (1024x1024) using our transferred material maps. The optimized material maps have the same large-scale structure as the input material maps but now share the material appearance of the target photo. We see that our transfer does not only modify the diffuse albedo, but can modify various properties when required.}
\label{fig:results}
\vspace{-1mm}
\end{figure*}
\begin{figure*} %
	\centering
	\addtolength{\tabcolsep}{-5.25pt}
	\setlength{\extrarowheight}{3pt}
 	\def\arraystretch{1.0}
	\begin{tabular}{ccccc|ccccc}
		\scalebox{0.9}{Target(s)} & & \scalebox{0.9}{Material Map} & \scalebox{0.9}{Render} & \scalebox{0.9}{Tiled} & \scalebox{0.9}{Target(s)} & &  \scalebox{0.9}{Material Map} & \scalebox{0.9}{Render} & \scalebox{0.9}{Tiled} \\

	    \includegraphics[width=0.09\textwidth]{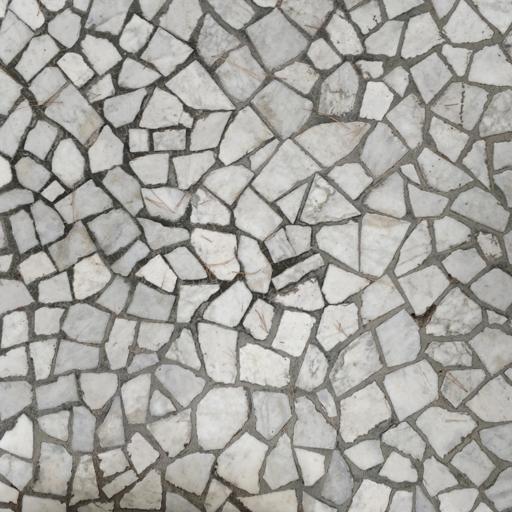}
	    \llap{\frame{\includegraphics[width=0.03\textwidth]{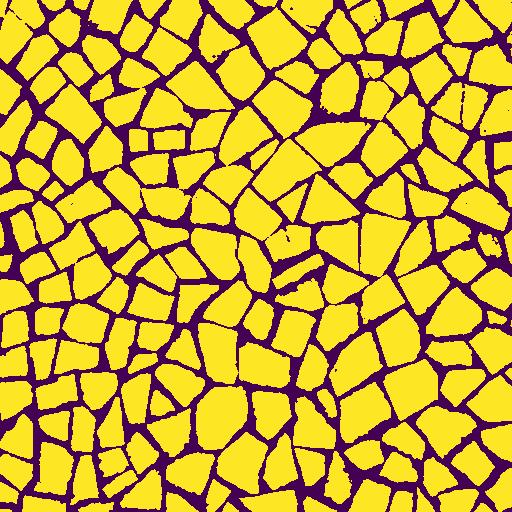}}} &
    	\raisebox{20pt}{\scalebox{0.8}{\rotatebox[origin=c]{90}{Input}}} &
	    \includegraphics[width=0.09\textwidth]{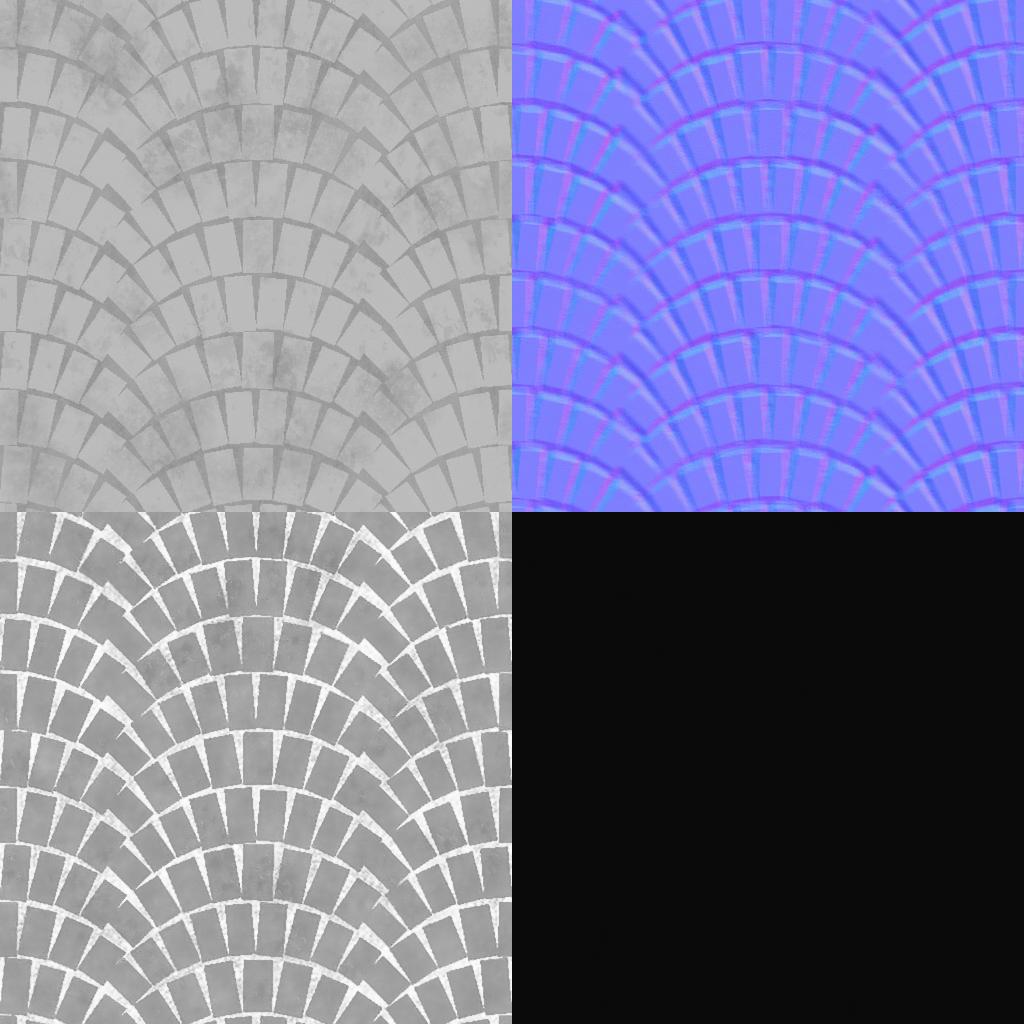} &
        \includegraphics[width=0.09\textwidth]{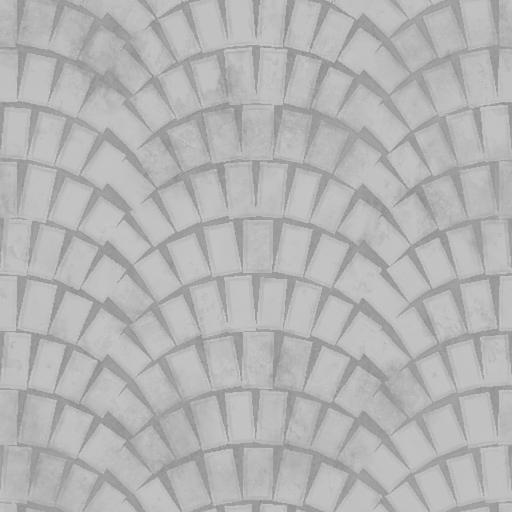}	   \llap{\frame{\includegraphics[width=0.03\textwidth]{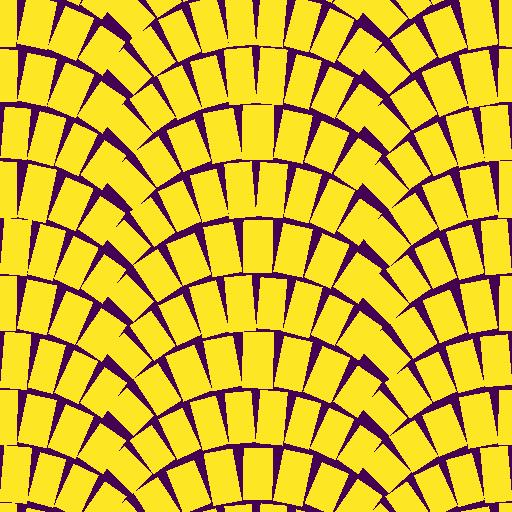}}} &
        \multirow{2}{*}[35pt]{        \includegraphics[width=0.19\textwidth]{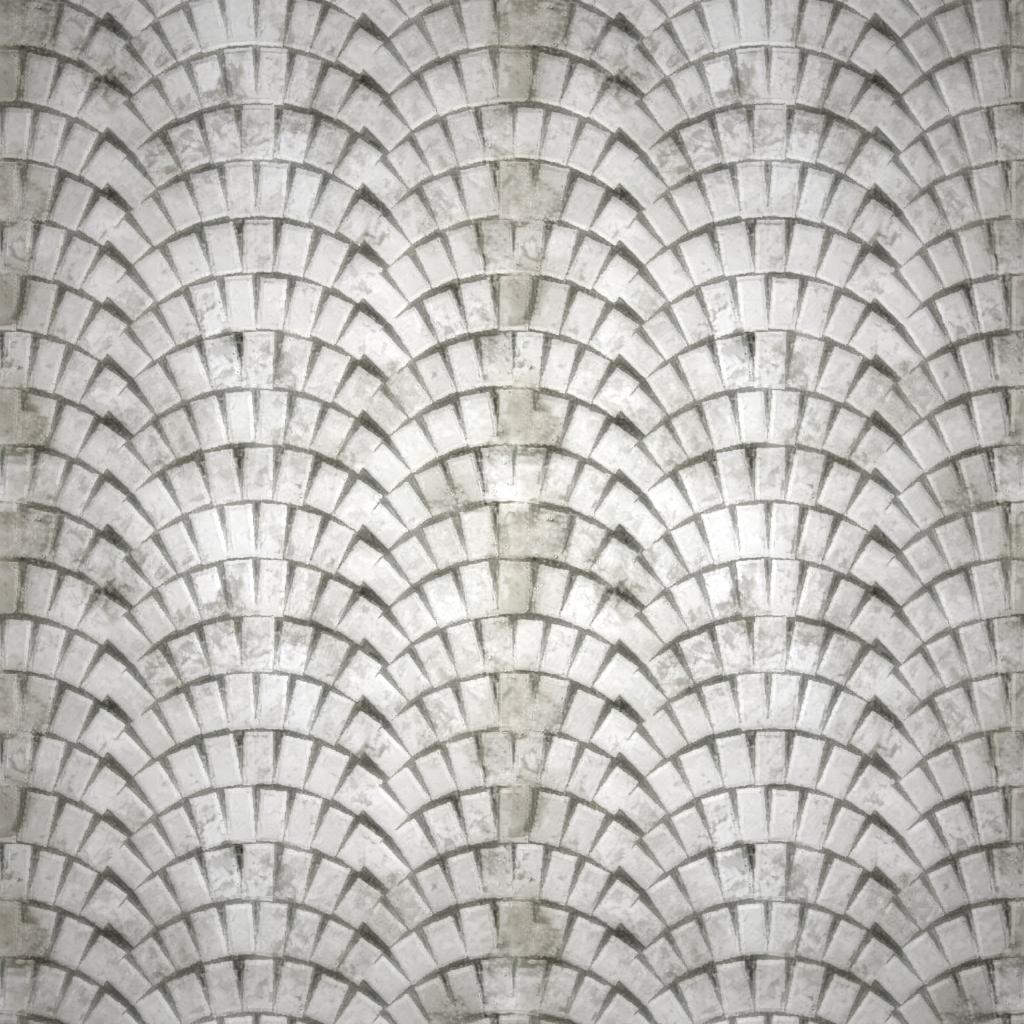}} & 
        
        \includegraphics[width=0.09\textwidth]{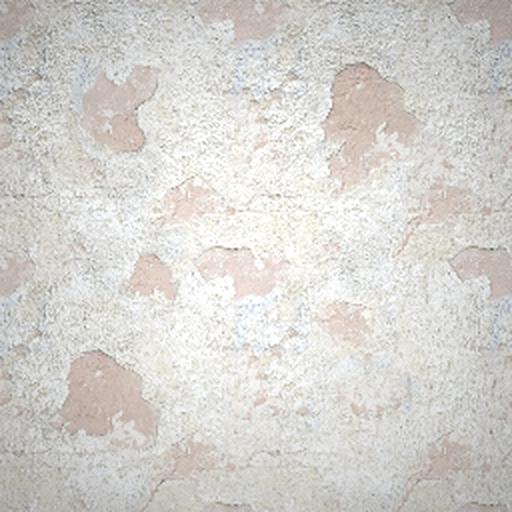}
	    \llap{\frame{\includegraphics[width=0.03\textwidth]{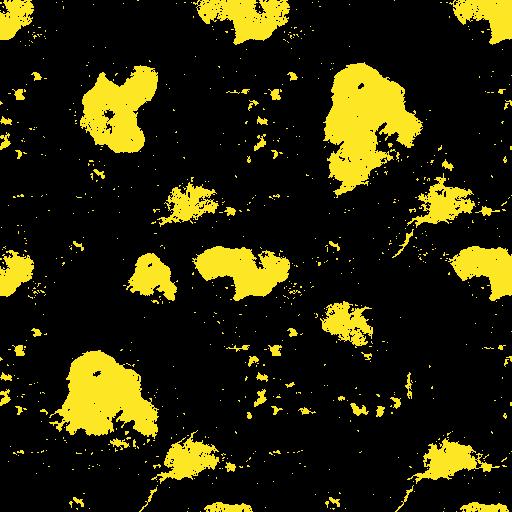}}} &
	    \raisebox{20pt}{\scalebox{0.8}{\rotatebox[origin=c]{90}{Input}}} &
	    \includegraphics[width=0.09\textwidth]{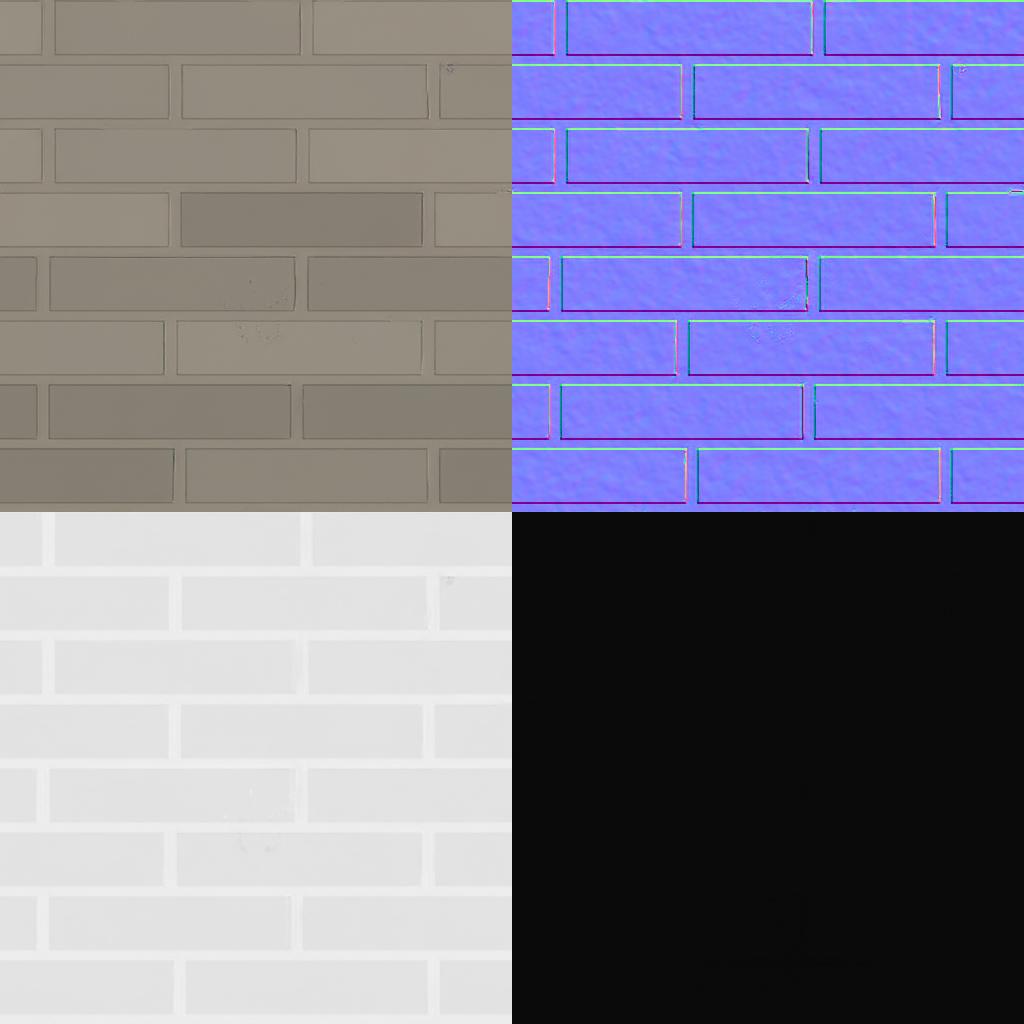} &
        \includegraphics[width=0.09\textwidth]{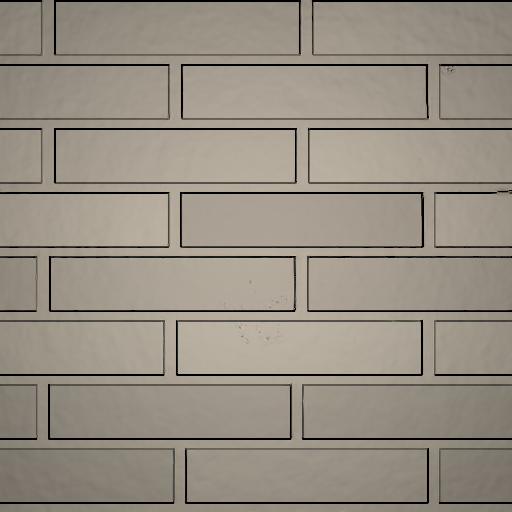}
        \llap{\frame{\includegraphics[width=0.03\textwidth]{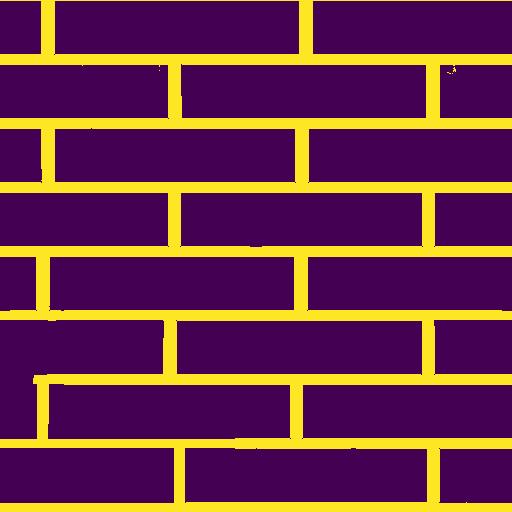}}} &
        \multirow{2}{*}[35pt]{        \includegraphics[width=0.19\textwidth]{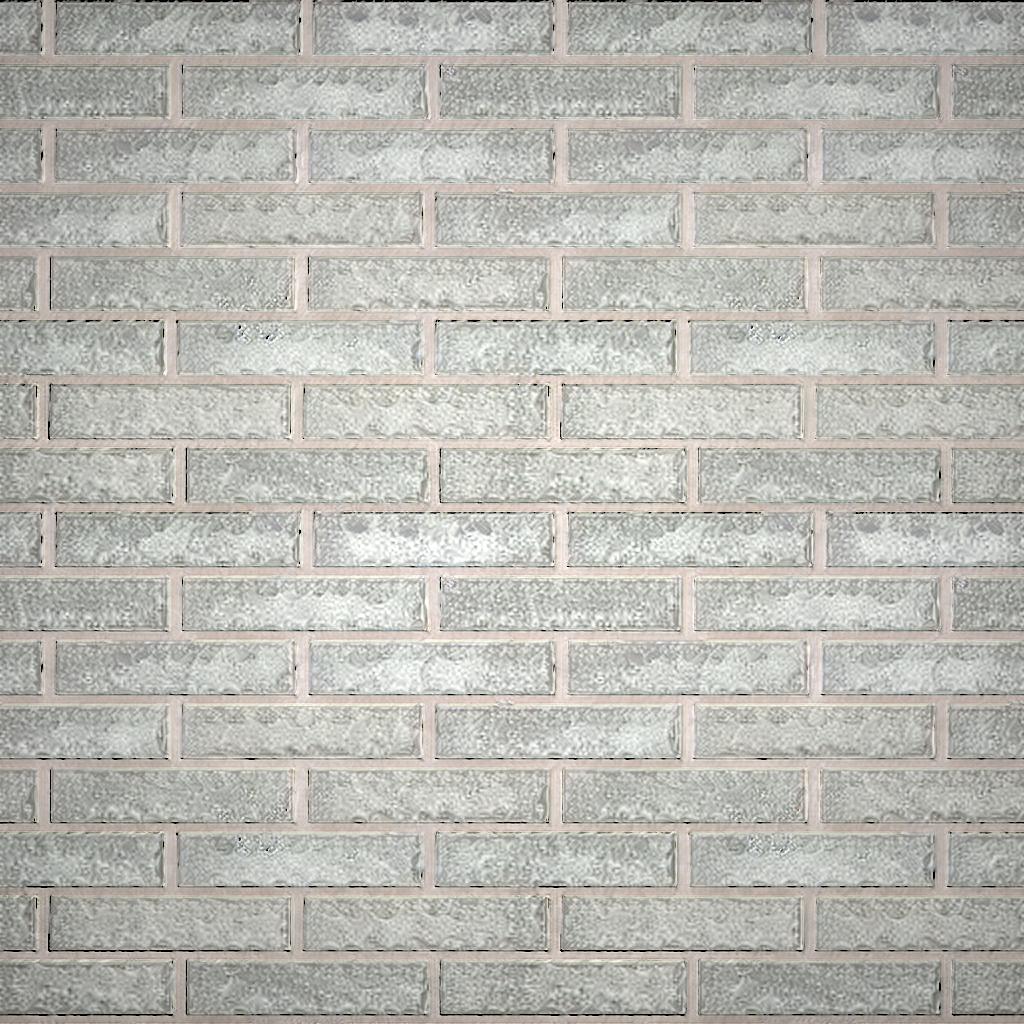}} \\
        
        &
        \raisebox{20pt}{\scalebox{0.8}{\rotatebox[origin=c]{90}{Transferred}}} &
		\includegraphics[width=0.09\textwidth]{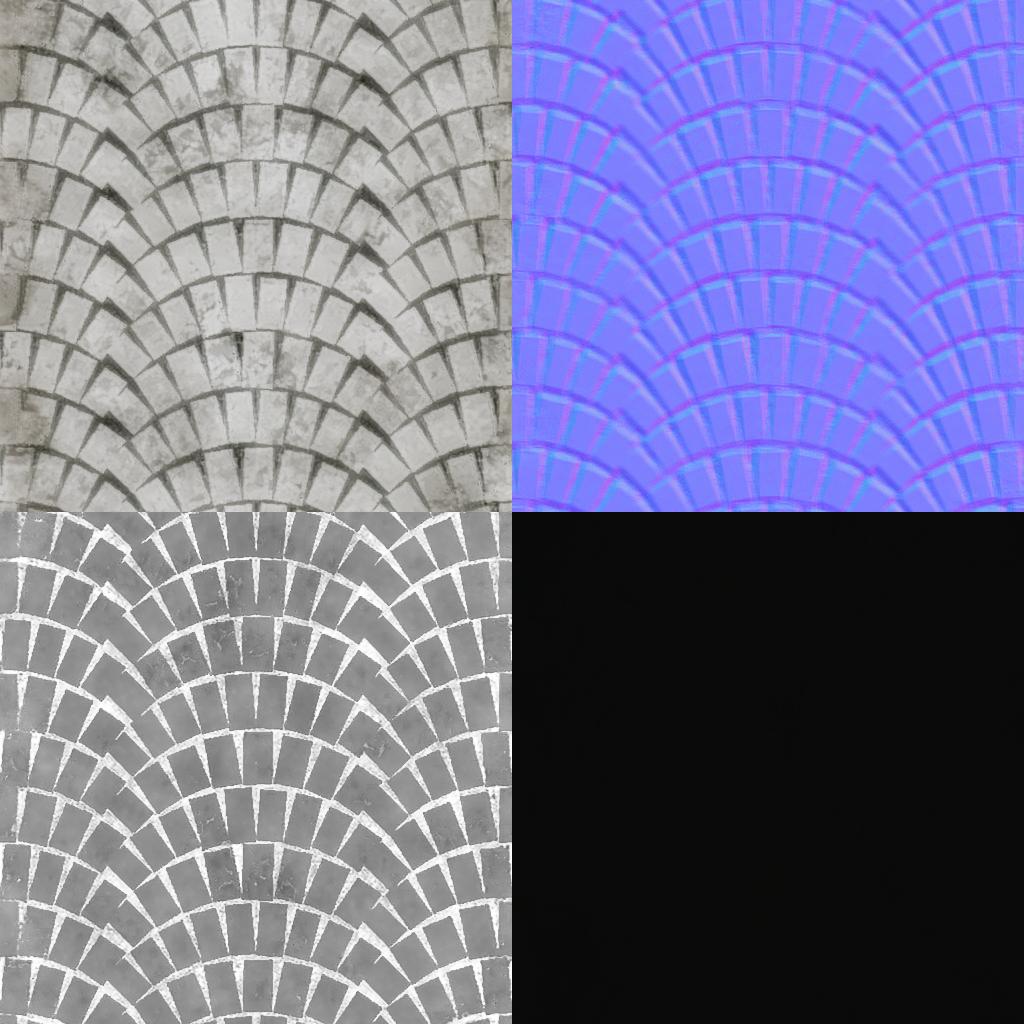} &
        \includegraphics[width=0.09\textwidth]{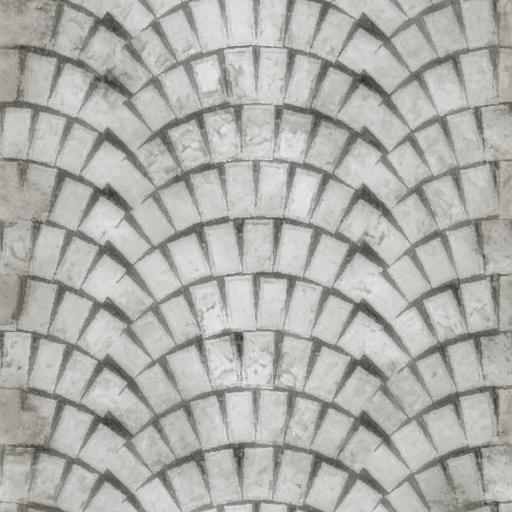} & &
        \includegraphics[width=0.09\textwidth]{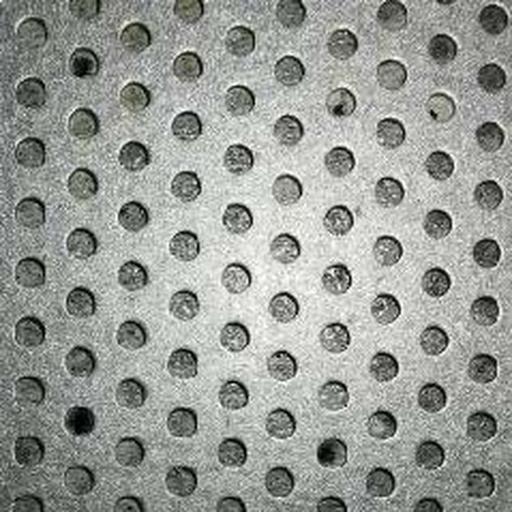}
	    \llap{\frame{\includegraphics[width=0.03\textwidth]{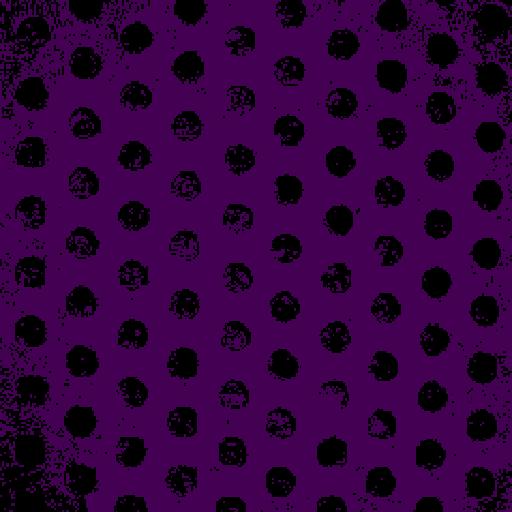}}} &
        \raisebox{20pt}{\scalebox{0.8}{\rotatebox[origin=c]{90}{Transferred}}} &
		\includegraphics[width=0.09\textwidth]{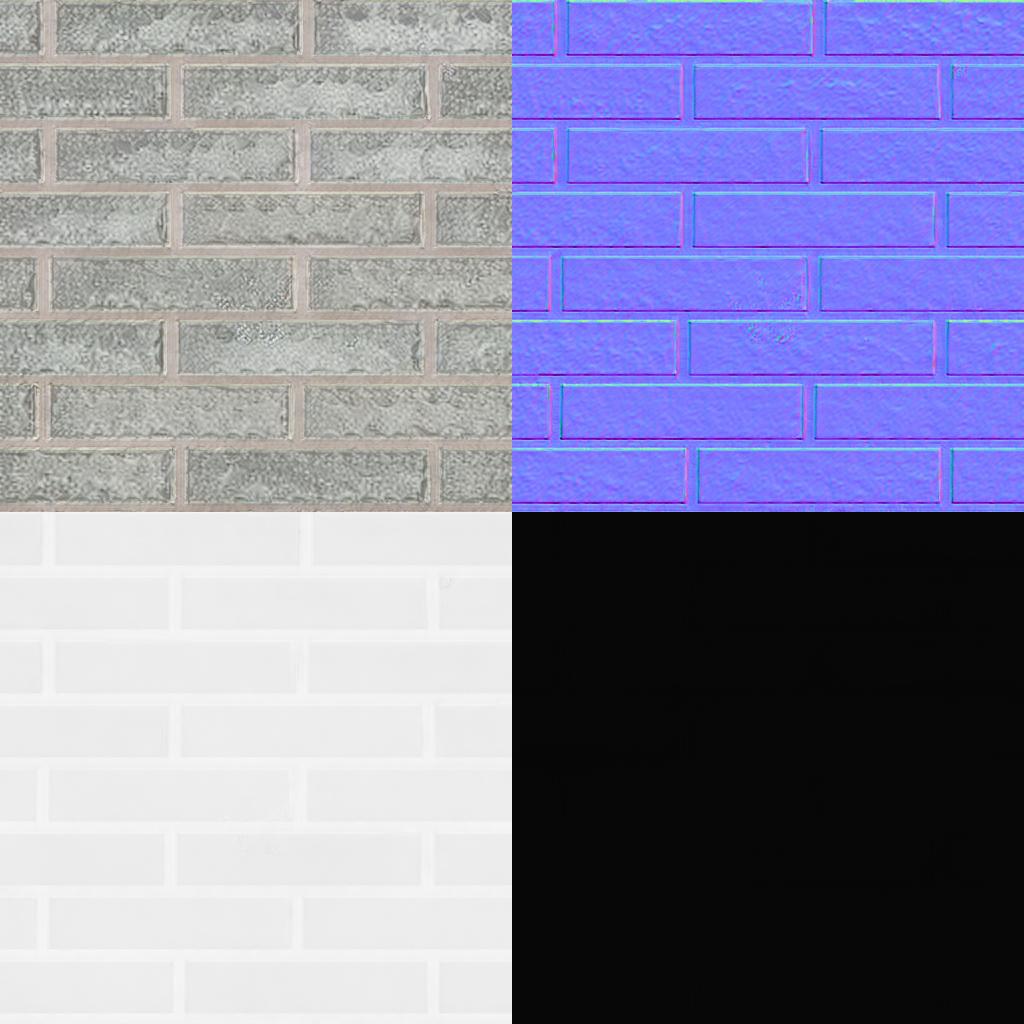} &
        \includegraphics[width=0.09\textwidth]{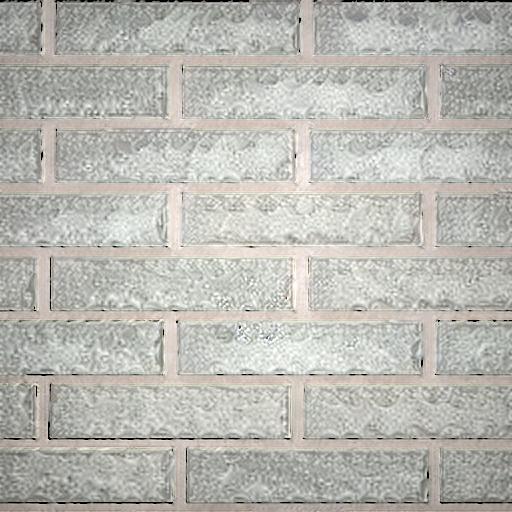} & \\

	    \includegraphics[width=0.09\textwidth]{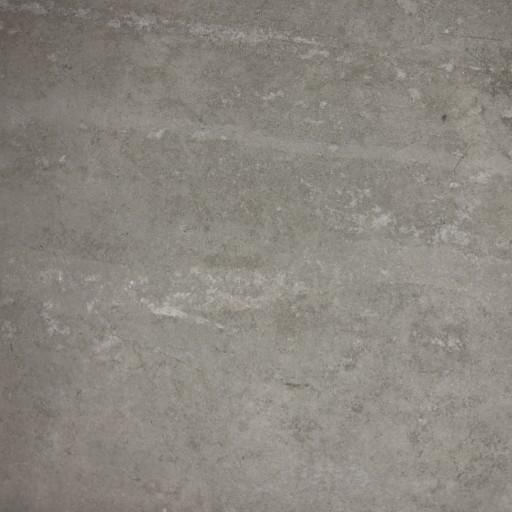}
	    \llap{\frame{\includegraphics[width=0.03\textwidth]{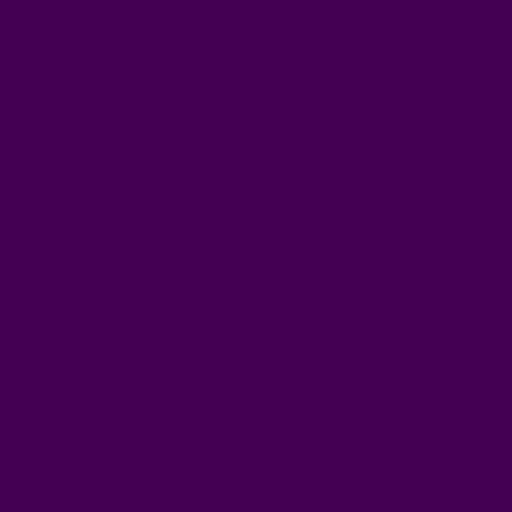}}} &
    	\raisebox{20pt}{\scalebox{0.8}{\rotatebox[origin=c]{90}{Input}}} &
	    \includegraphics[width=0.09\textwidth]{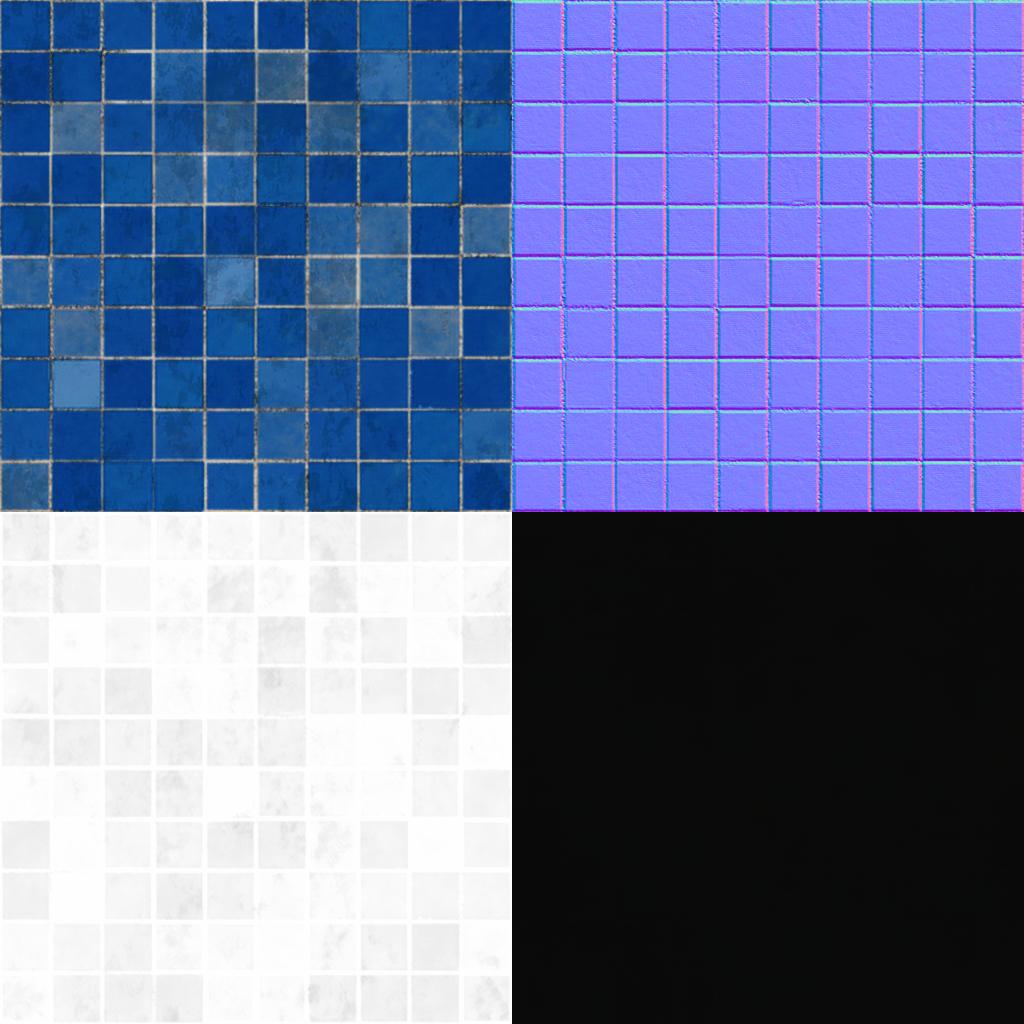} &
        \includegraphics[width=0.09\textwidth]{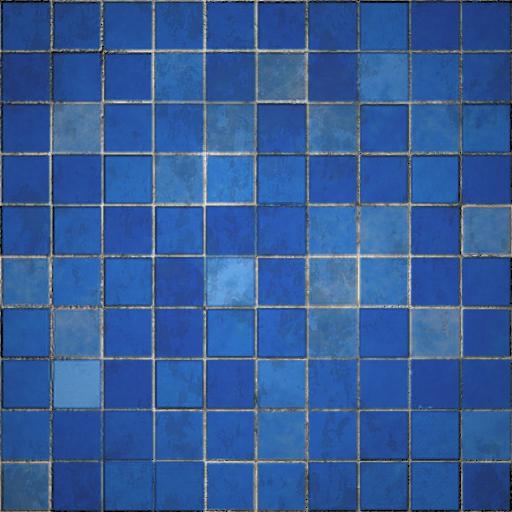}	   \llap{\frame{\includegraphics[width=0.03\textwidth]{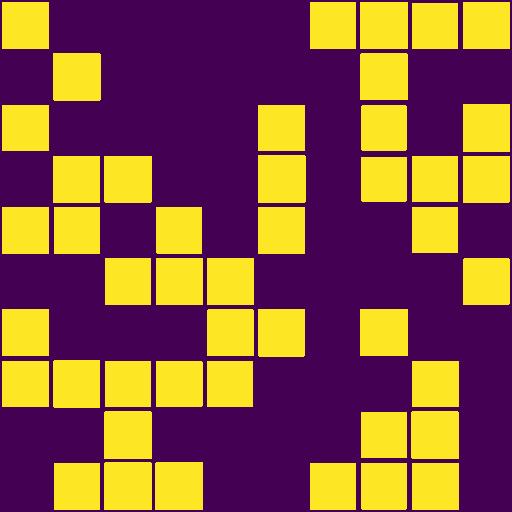}}} &
        \multirow{2}{*}[35pt]{        \includegraphics[width=0.19\textwidth]{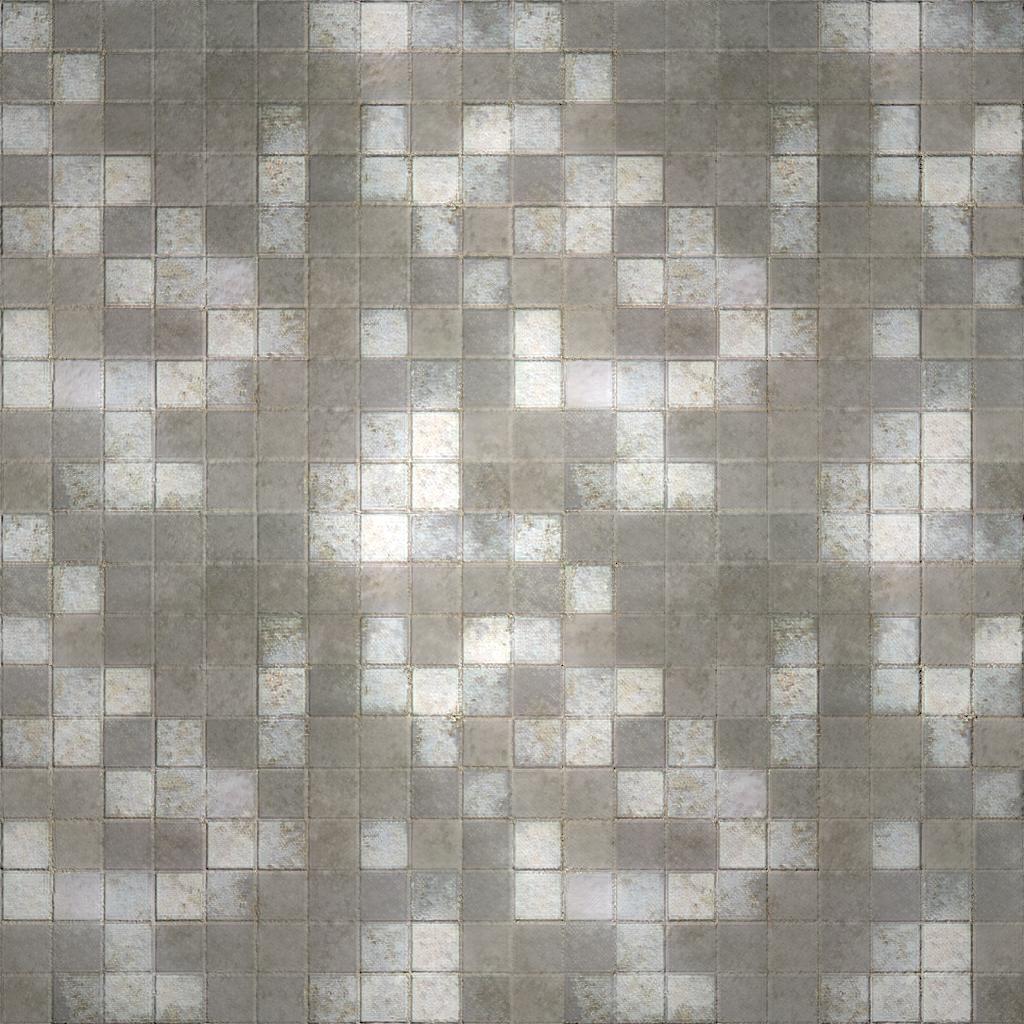}} & 
        
        \includegraphics[width=0.09\textwidth]{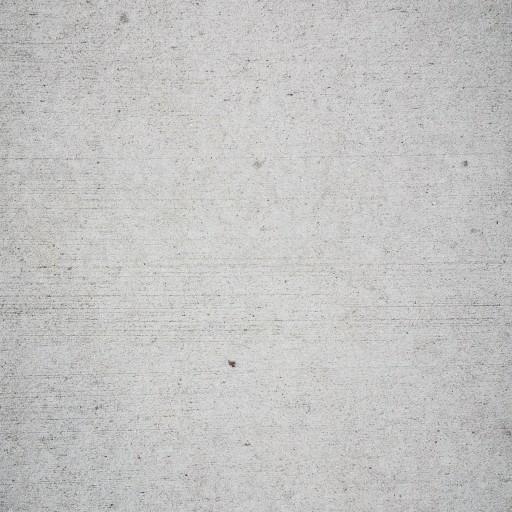}
	    \llap{\frame{\includegraphics[width=0.03\textwidth]{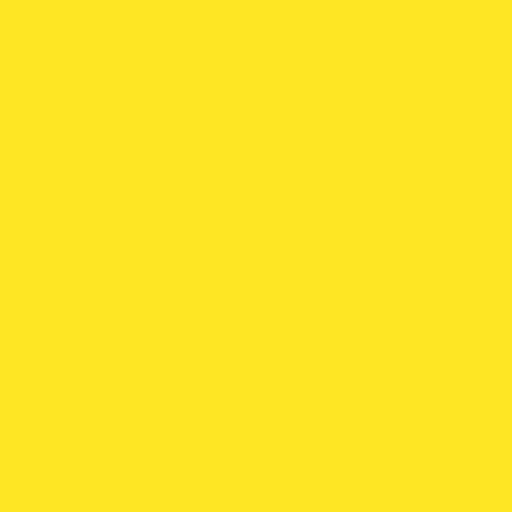}}} &
	    \raisebox{20pt}{\scalebox{0.8}{\rotatebox[origin=c]{90}{Input}}} &
	    \includegraphics[width=0.09\textwidth]{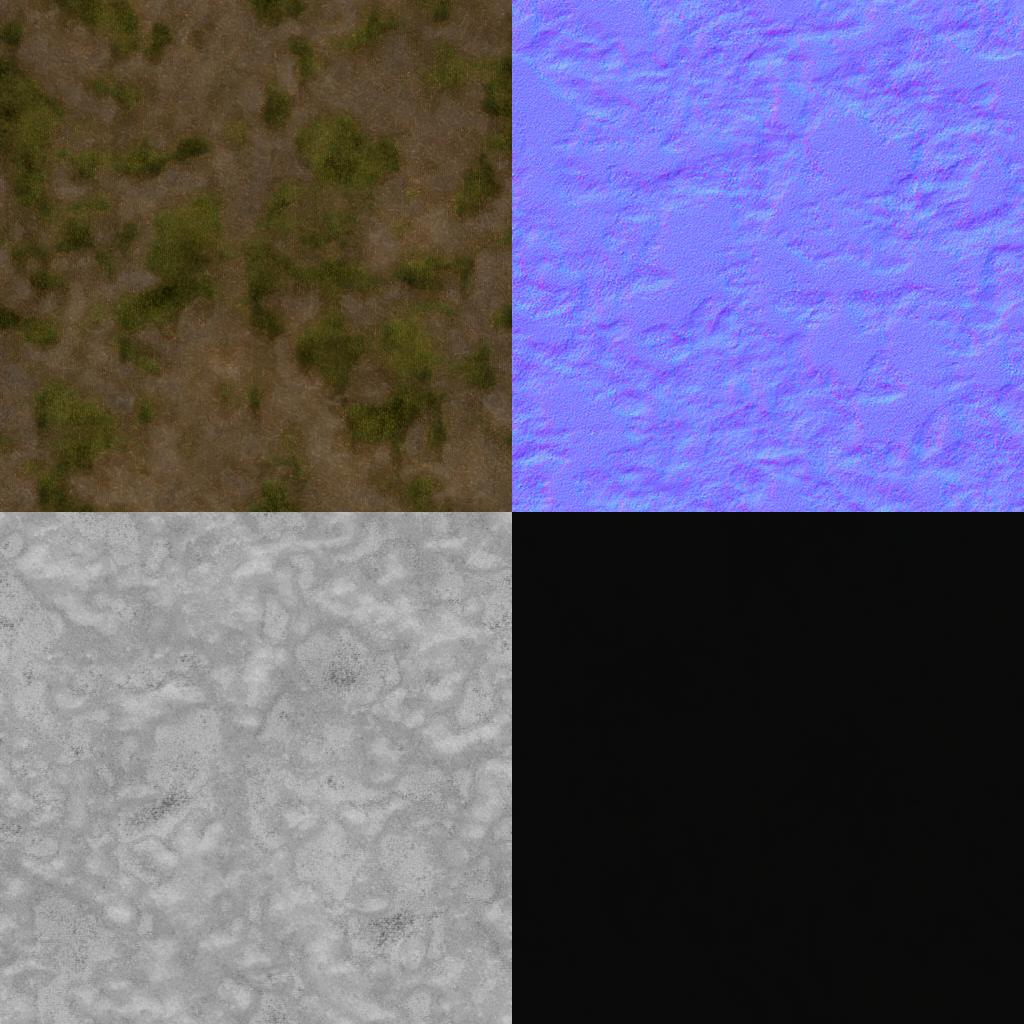} &
        \includegraphics[width=0.09\textwidth]{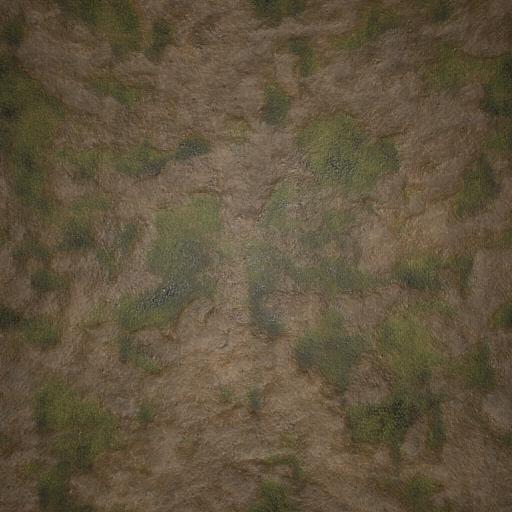}
        \llap{\frame{\includegraphics[width=0.03\textwidth]{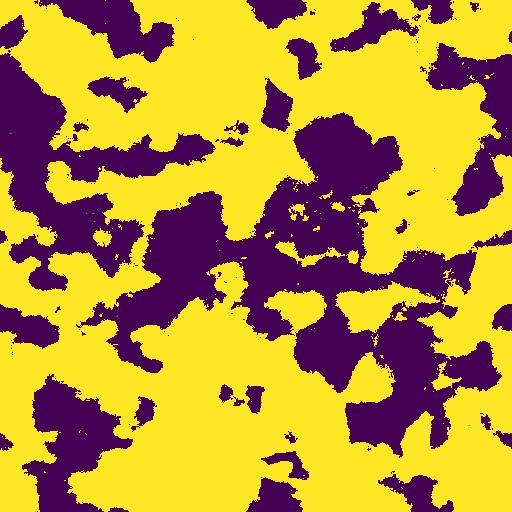}}} &
        \multirow{2}{*}[35pt]{        \includegraphics[width=0.19\textwidth]{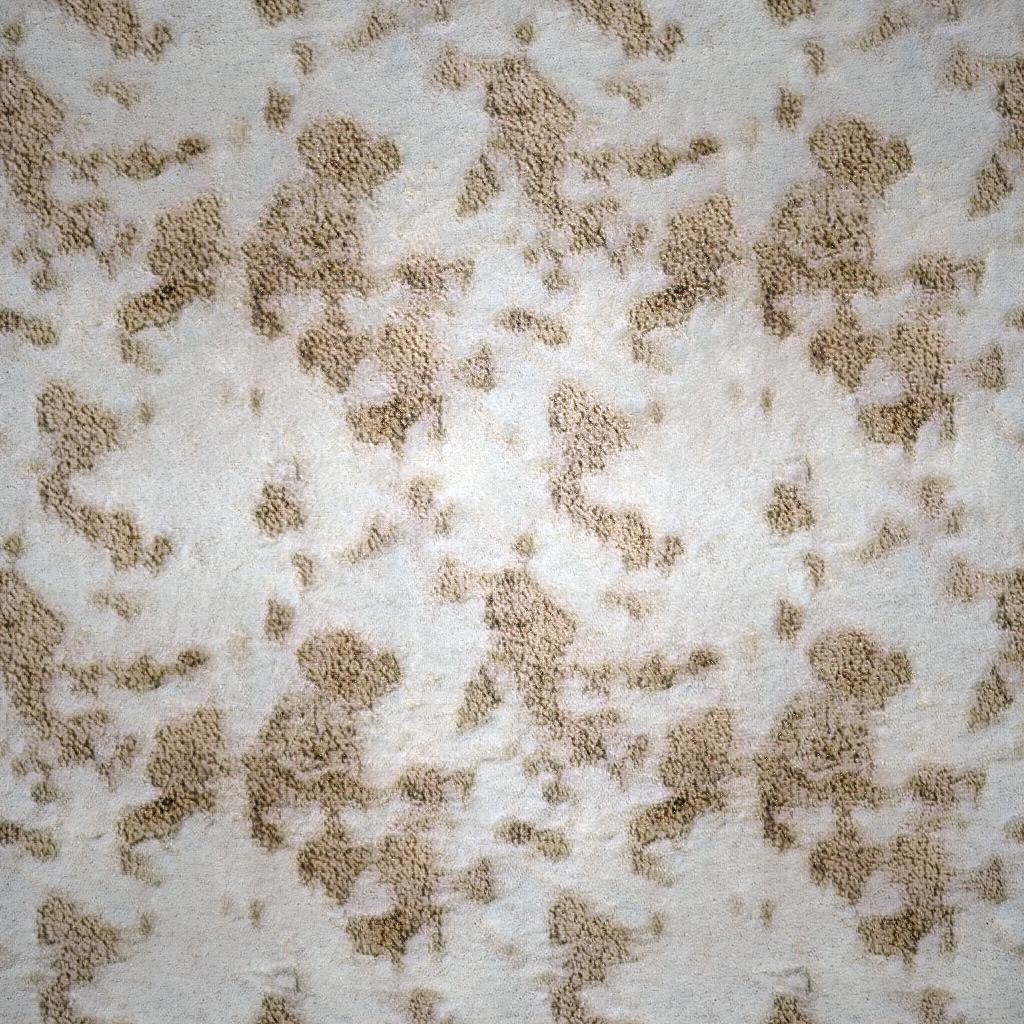}} \\
        
	    \includegraphics[width=0.09\textwidth]{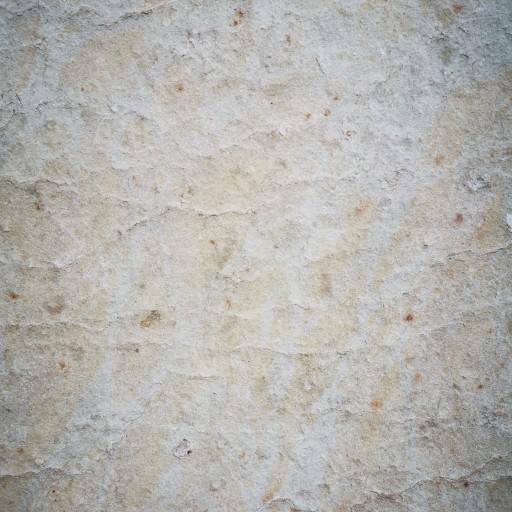}
	    \llap{\frame{\includegraphics[width=0.03\textwidth]{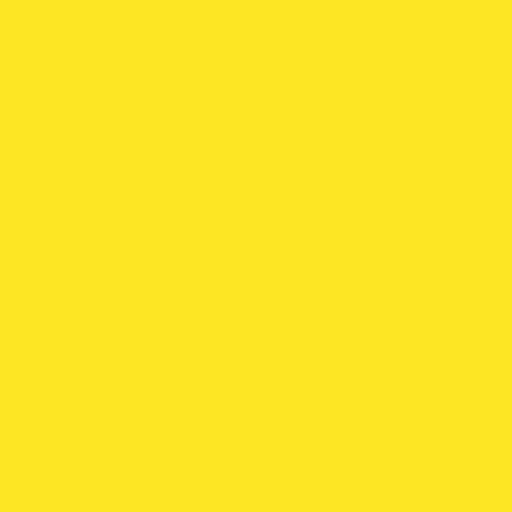}}} &
        \raisebox{20pt}{\scalebox{0.8}{\rotatebox[origin=c]{90}{Transferred}}} &
		\includegraphics[width=0.09\textwidth]{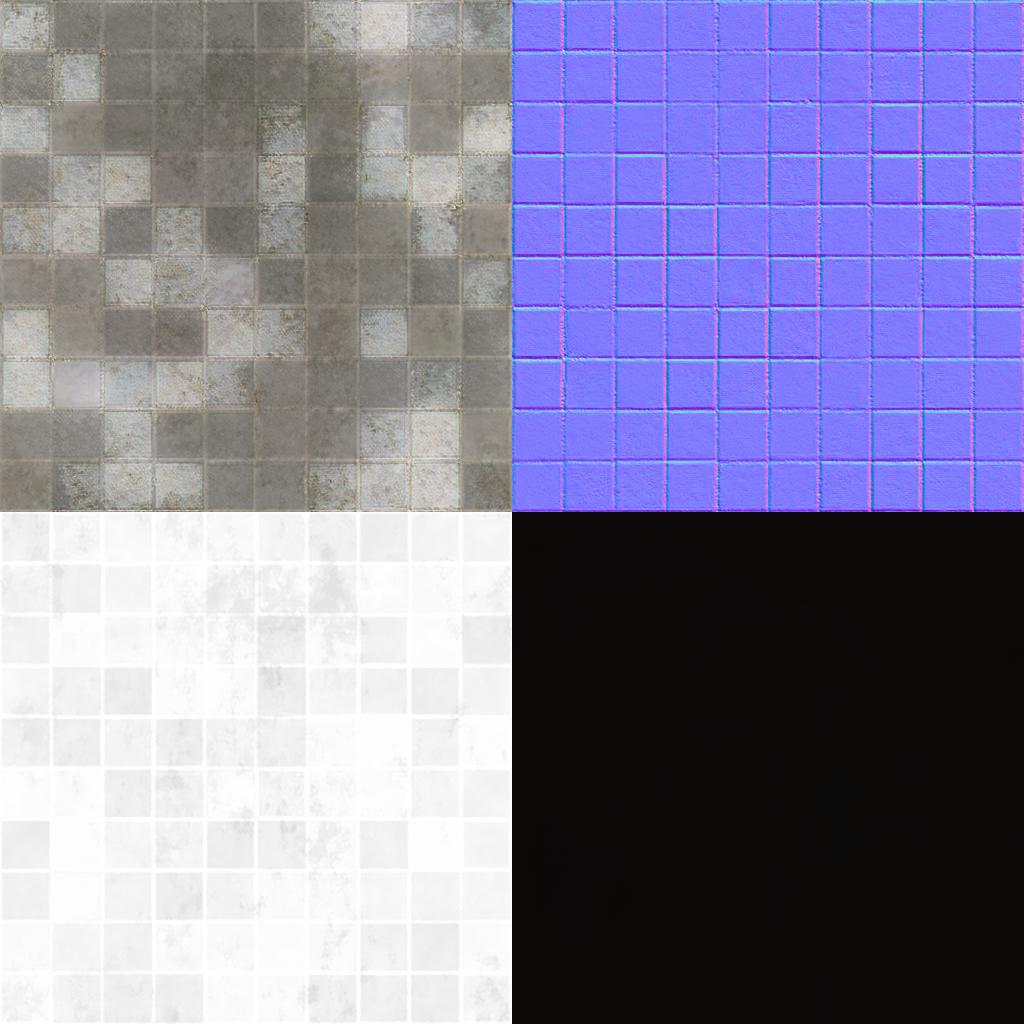} &
        \includegraphics[width=0.09\textwidth]{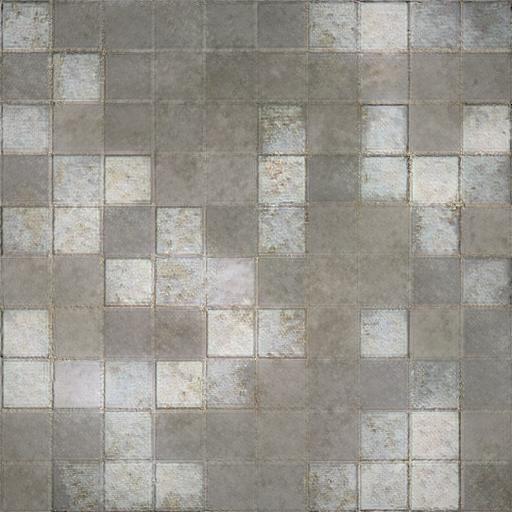} & &
        \includegraphics[width=0.09\textwidth]{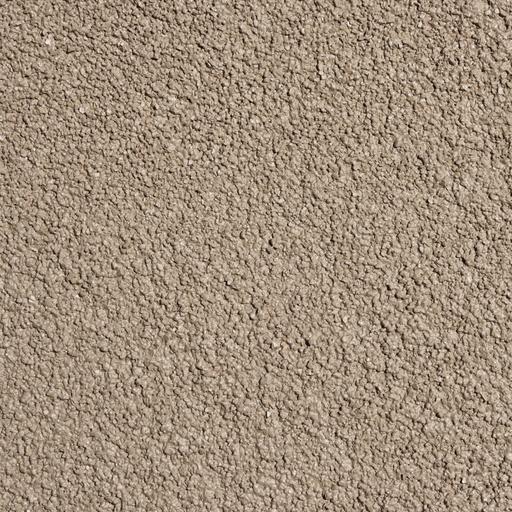}
	    \llap{\frame{\includegraphics[width=0.03\textwidth]{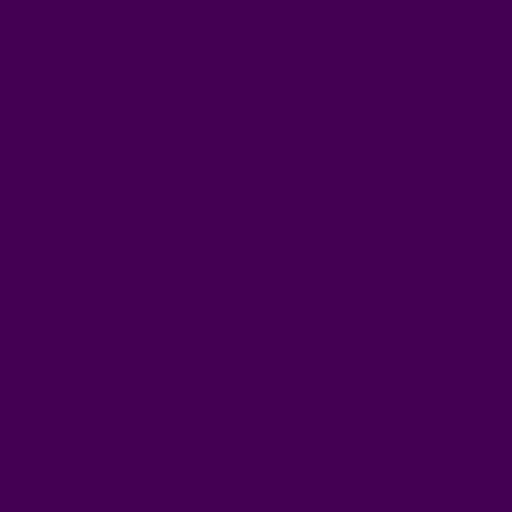}}} &
        \raisebox{20pt}{\scalebox{0.8}{\rotatebox[origin=c]{90}{Transferred}}} &
		\includegraphics[width=0.09\textwidth]{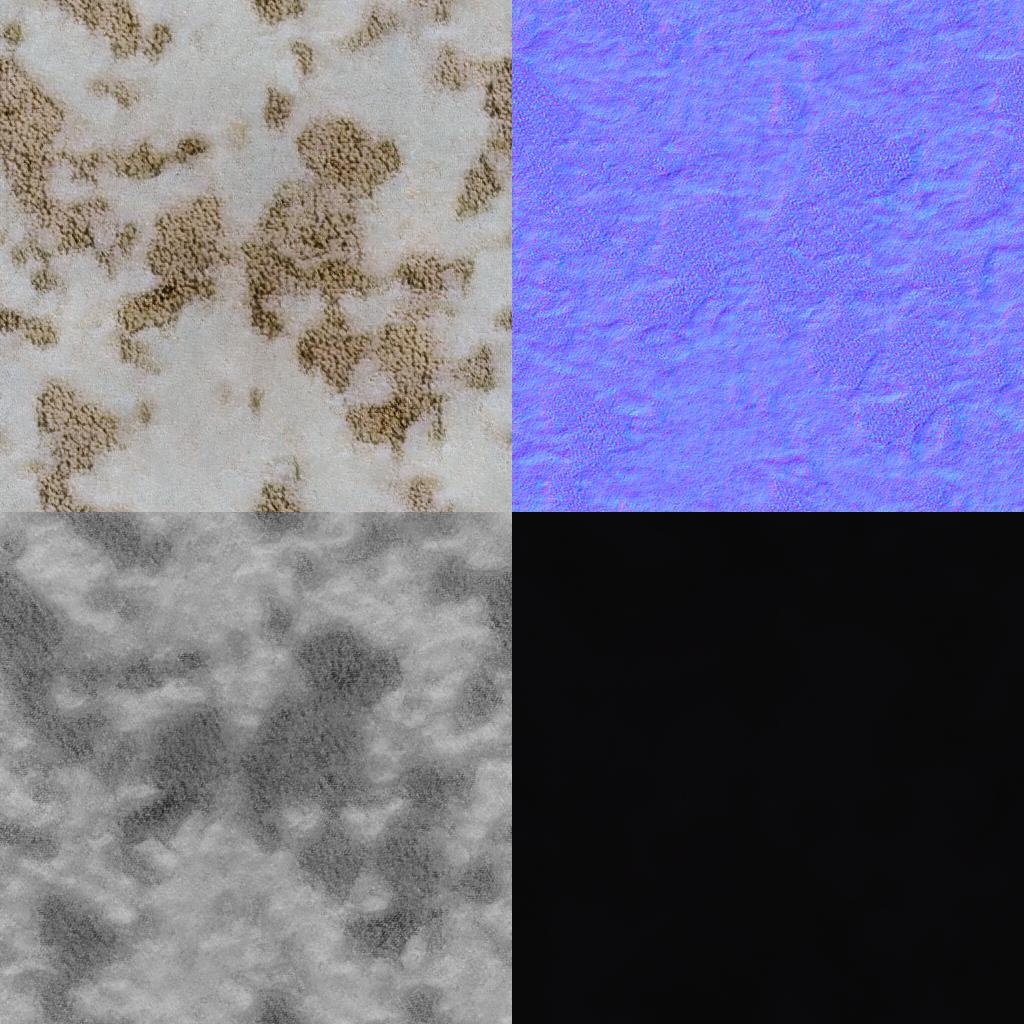} &
        \includegraphics[width=0.09\textwidth]{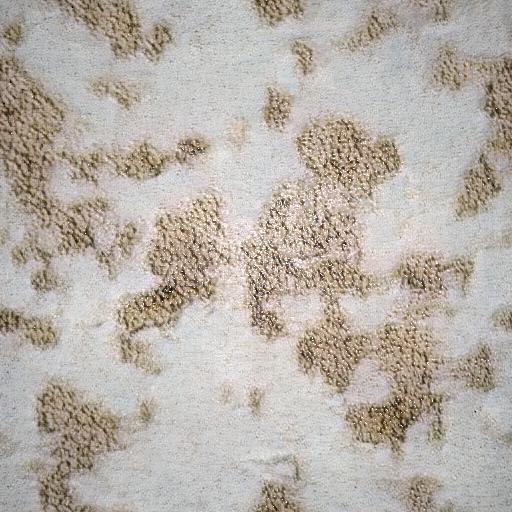} & \\

	\end{tabular}
\vspace{-10pt}
\caption{Multi-target material transfer results of our algorithm from regions in multiple targets to regions in the input material. The newly created material maps remain tileable and can be directly used for content creation. The inset color label maps show the transfer correspondence: we transfer material appearance from and onto regions with same color.}
\label{fig:multi_target}
\vspace{-1mm}
\end{figure*}
\section{Results}
We present single-target material transferred results including material maps and tiled renders in Fig. \ref{fig:results}, showing that our method can faithfully transfer realistic details from real photographs to material maps thanks to material prior provided by our modified MaterialGAN, preserving the tileability of input materials. We also show multi-target transfer results in Fig.~\ref{fig:multi_target} where we are able to transfer material appearance from separate regions from multiple sources. As with a single target, the new material maps preserve tileability allowing to directly use them for texturing. Please see our supplemental material for more results with dynamic lighting.
\begin{figure}
	\centering
	\renewcommand{\arraystretch}{0.6}
	\addtolength{\tabcolsep}{-4pt}
	\begin{tabular}{cccc}
		\includegraphics[width=0.15\textwidth]{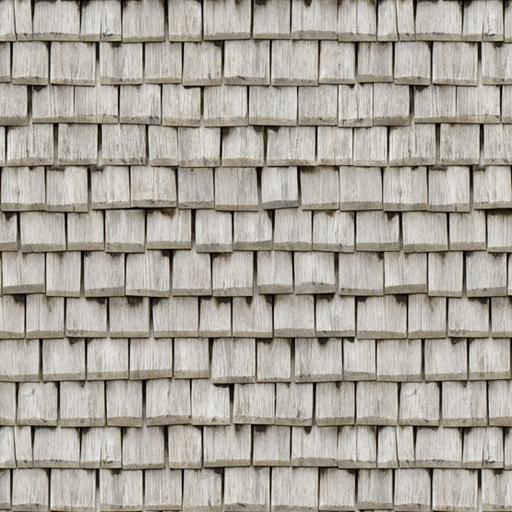} & 
	    \includegraphics[width=0.15\textwidth]{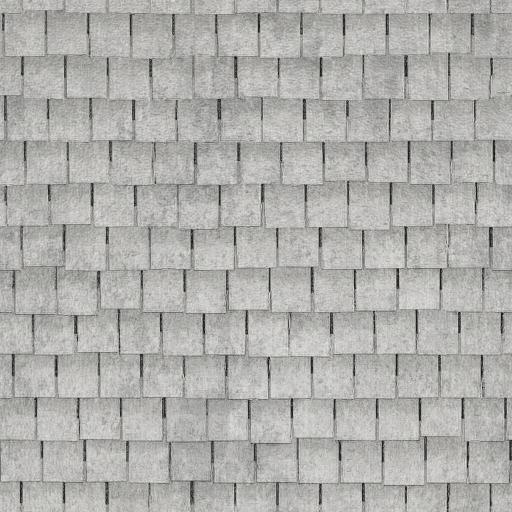}\llap{\frame{\includegraphics[width=0.07\textwidth]{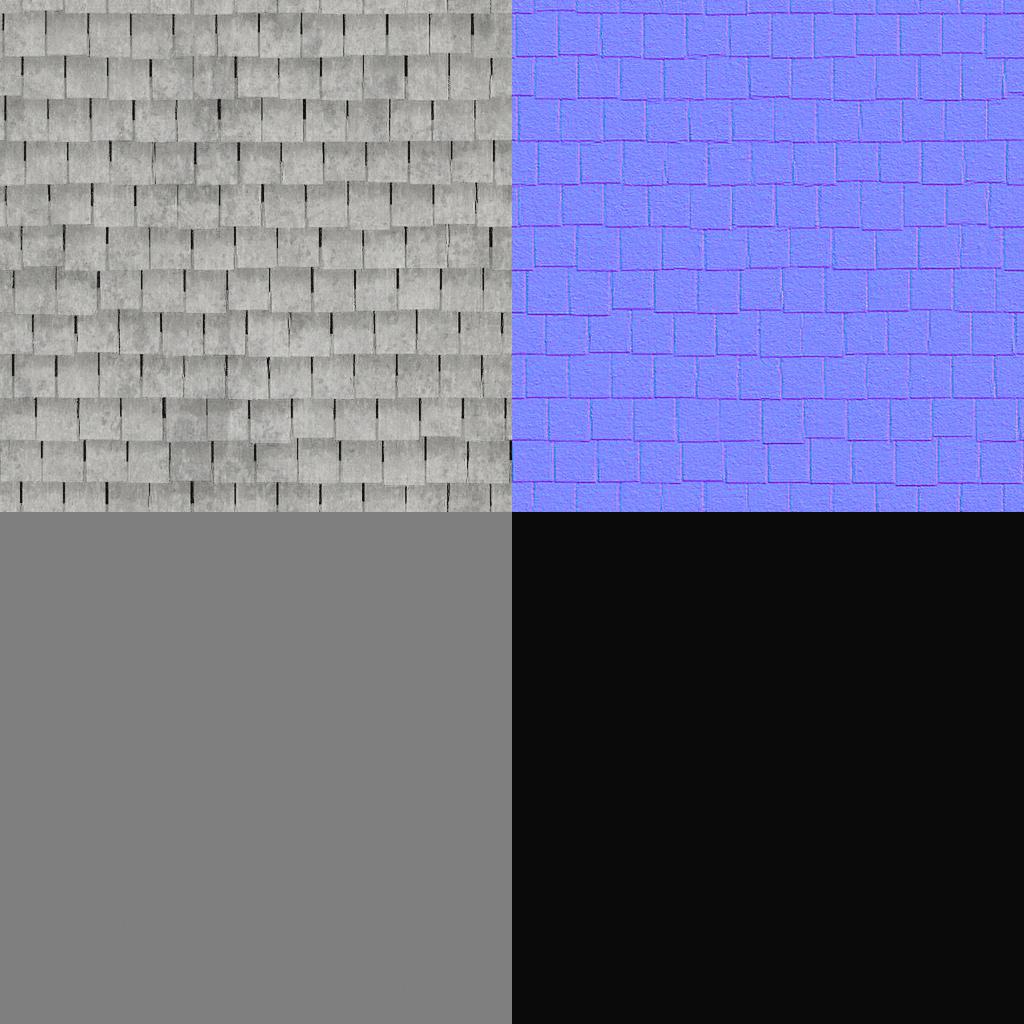}}} &
	    \includegraphics[width=0.15\textwidth]{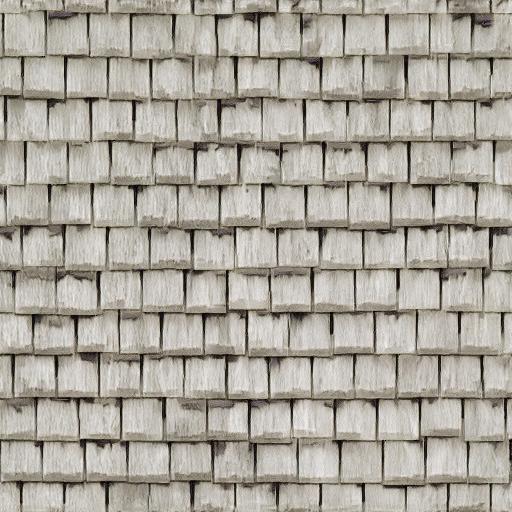}\llap{\frame{\includegraphics[width=0.07\textwidth]{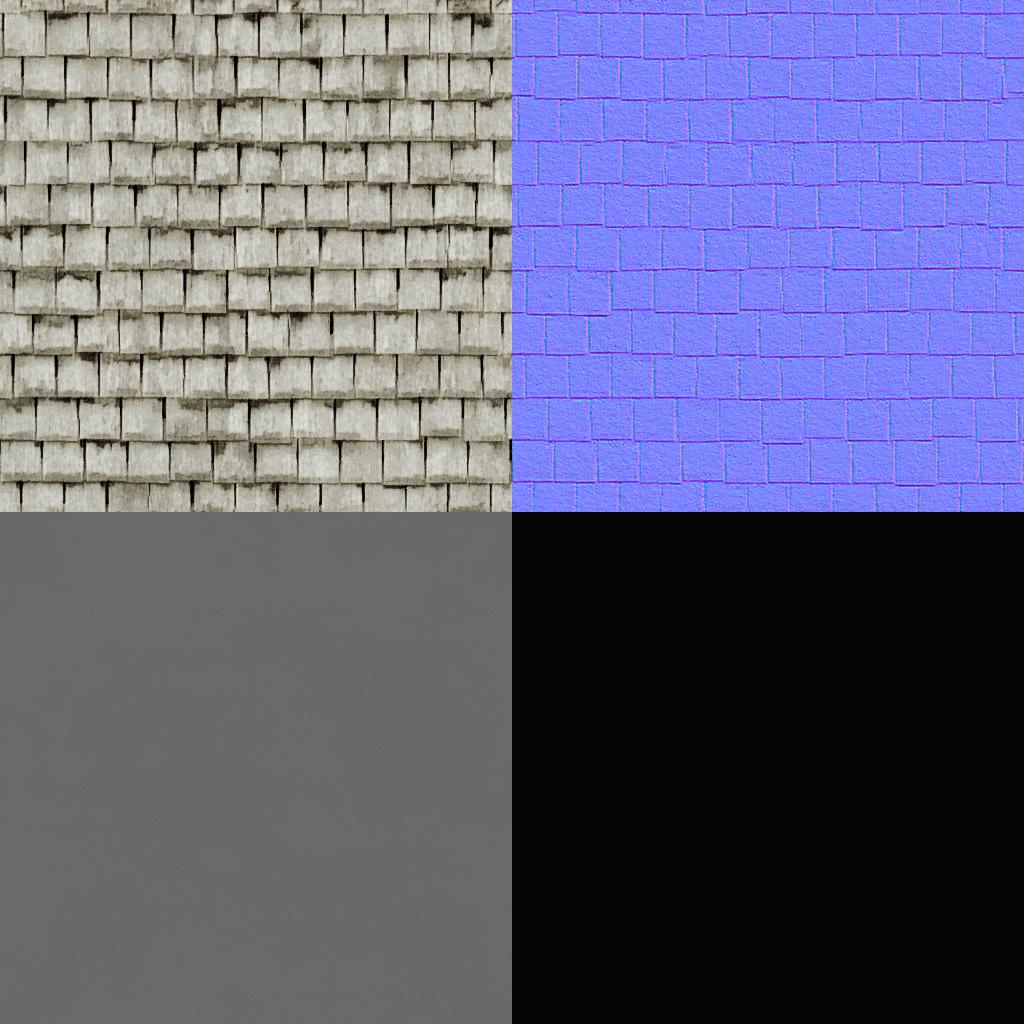}}} \\	    
	    Target & Hu et al. 2019 & Our augmented \\
		\includegraphics[width=0.15\textwidth]{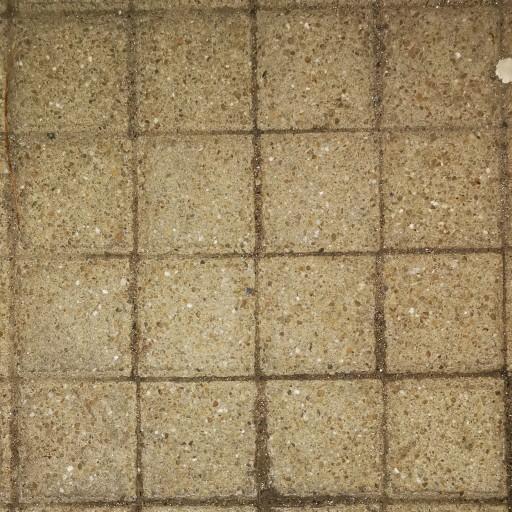} & 
	    \includegraphics[width=0.15\textwidth]{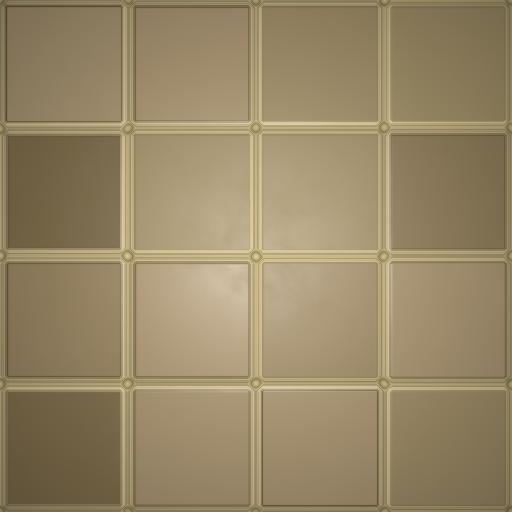}\llap{\frame{\includegraphics[width=0.07\textwidth]{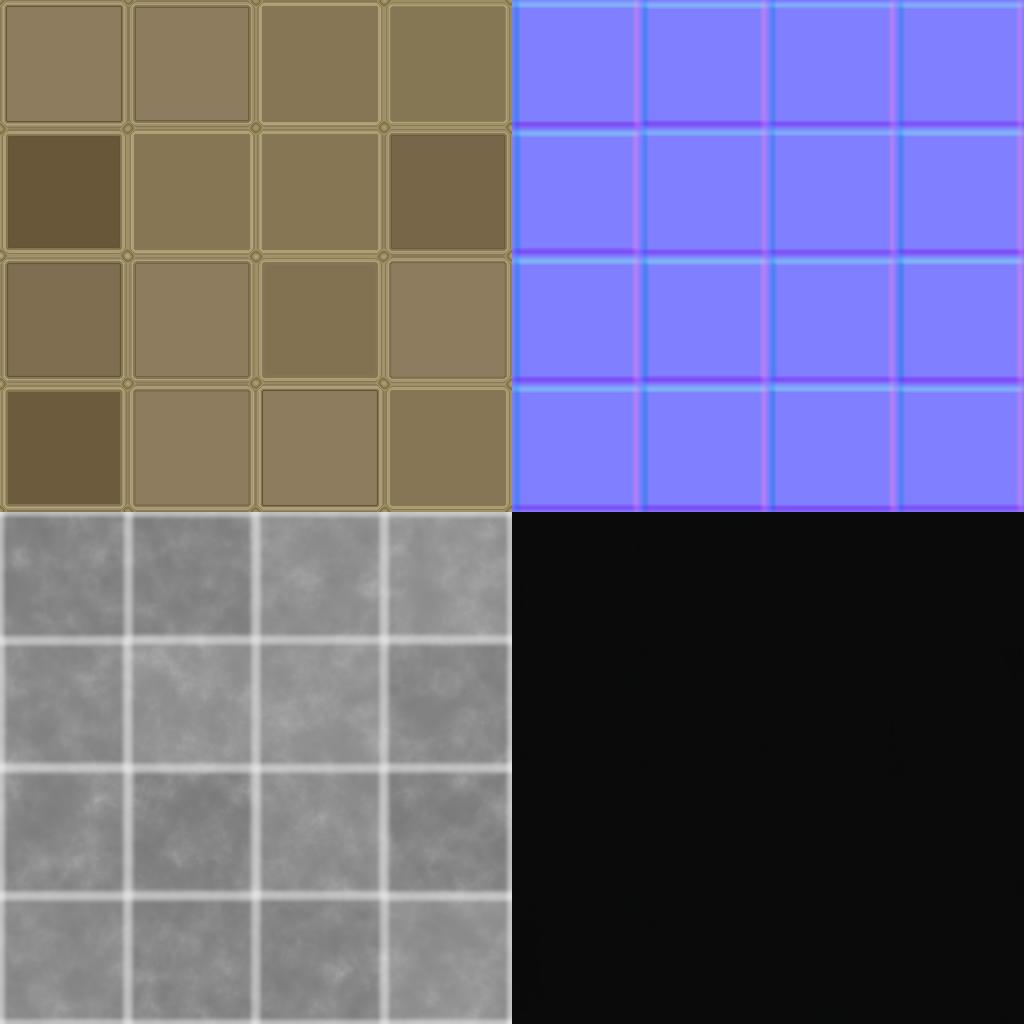}}} &
	    \includegraphics[width=0.15\textwidth]{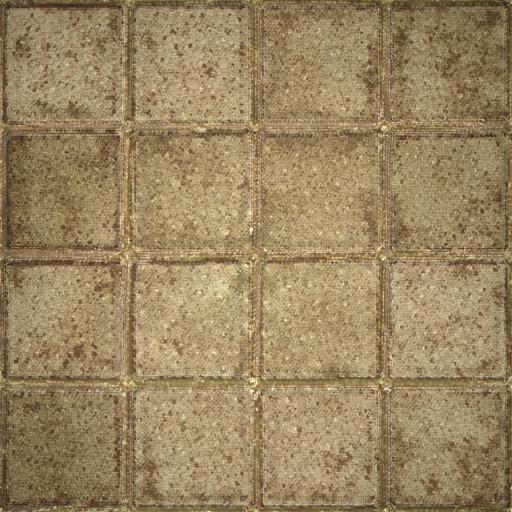}\llap{\frame{\includegraphics[width=0.07\textwidth]{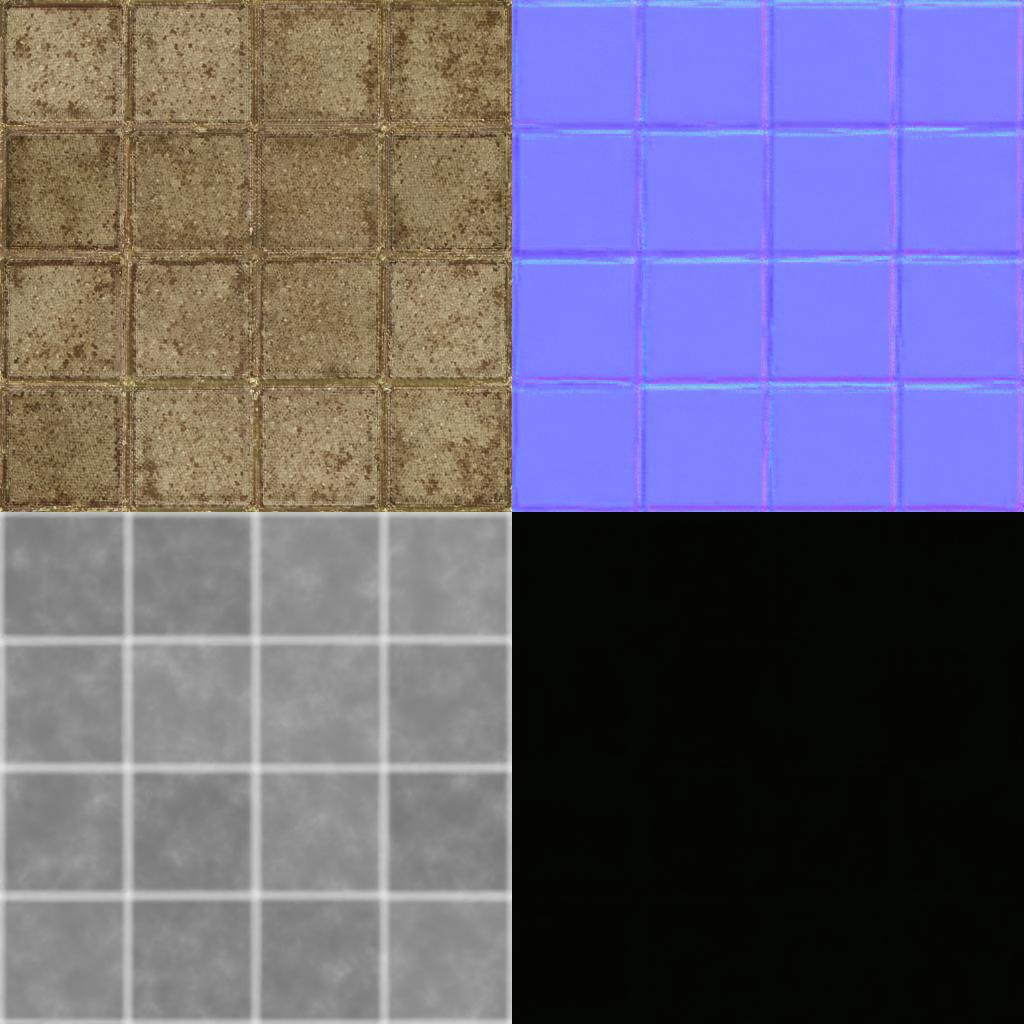}}} \\	    
	    Target & MATch & Our augmented \\
	\end{tabular}
\vspace{-10pt}
\caption{Augmenting inverse procedural results. The procedural materials shown here (middle column) are generated by inverse procedural modeling method \cite{hu2019, Shi20}, but depending on the expressivity of the procedural models, the fitted/optimized material often looks unrealistic. Our method can be applied to improve the realism and augment details of procedural materials generated maps, to better match a target appearance.} 
\label{fig:augment}
\end{figure}
\begin{figure}
	\centering
	\renewcommand{\arraystretch}{0.6}
	\addtolength{\tabcolsep}{-4pt}
	\begin{tabular}{cccc}
	    \includegraphics[width=0.15\textwidth]{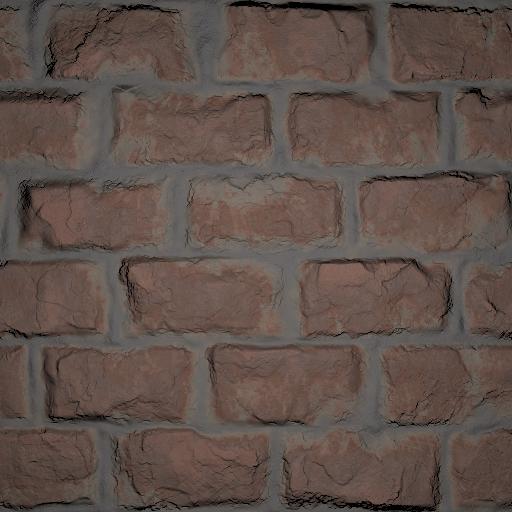}\llap{\frame{\includegraphics[width=0.07\textwidth]{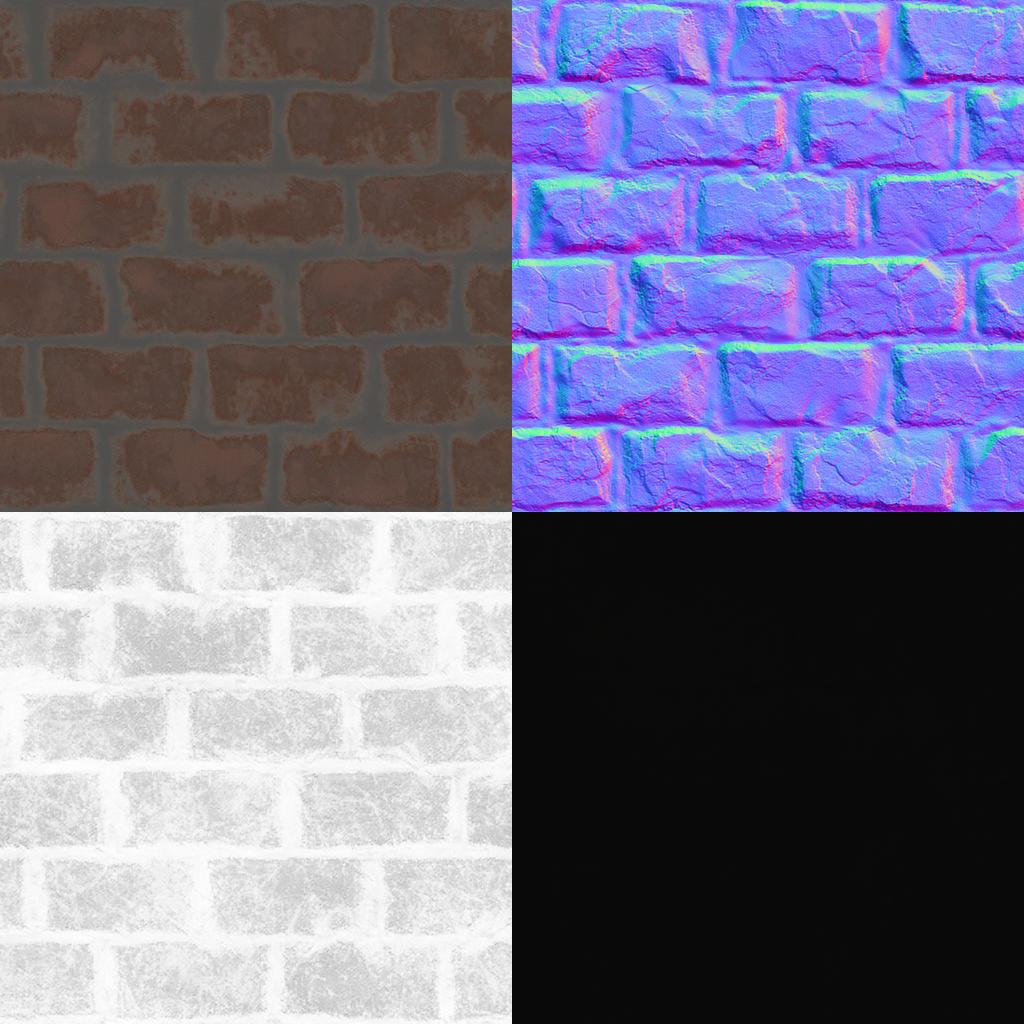}}} &
	    \includegraphics[width=0.15\textwidth]{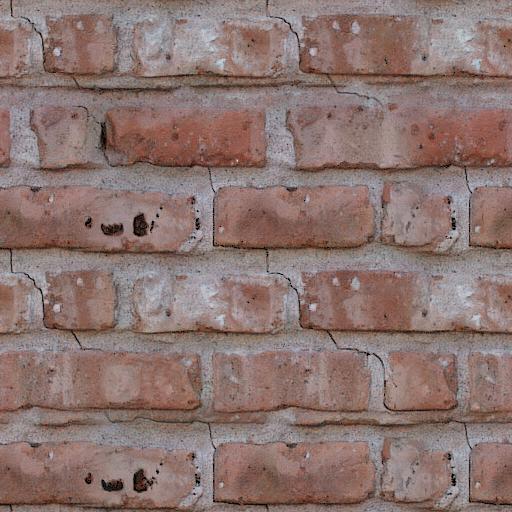}\llap{\frame{\includegraphics[width=0.07\textwidth]{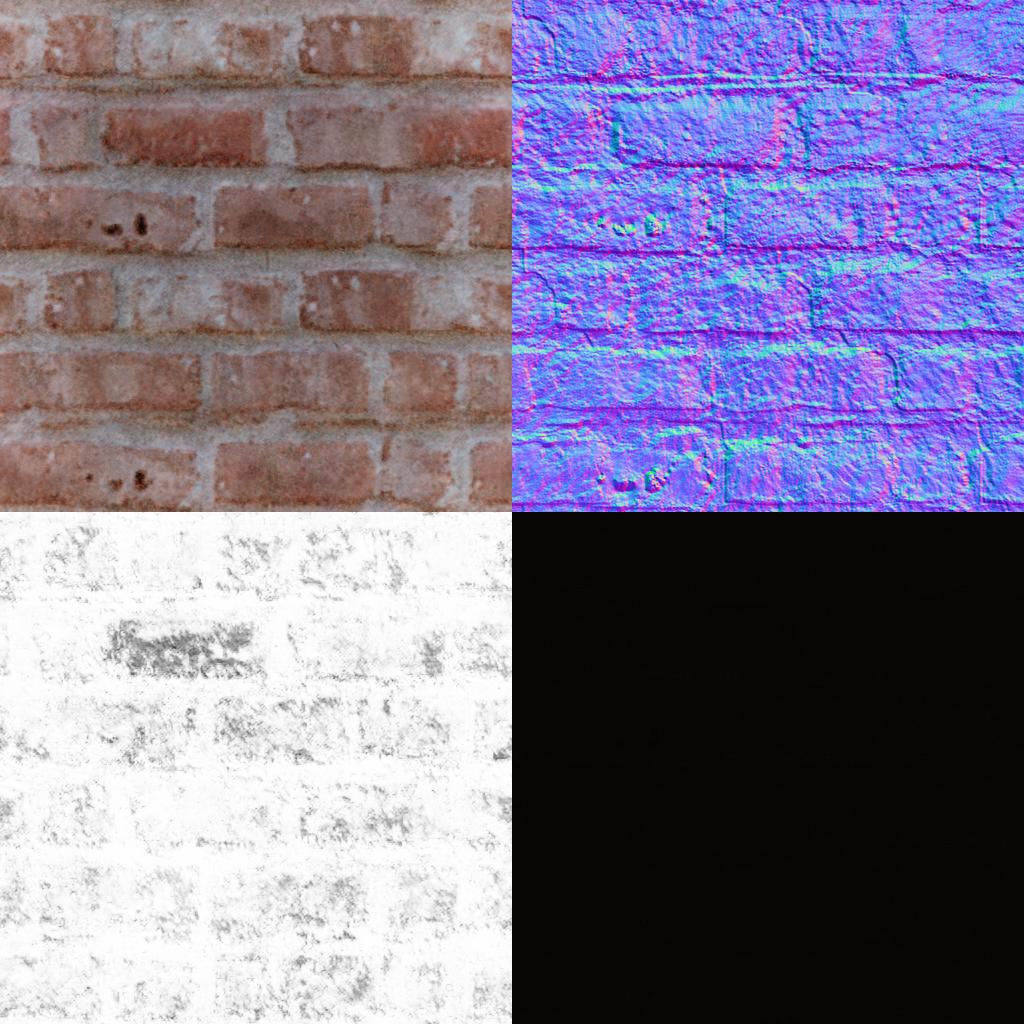}}} &
	    \includegraphics[width=0.15\textwidth]{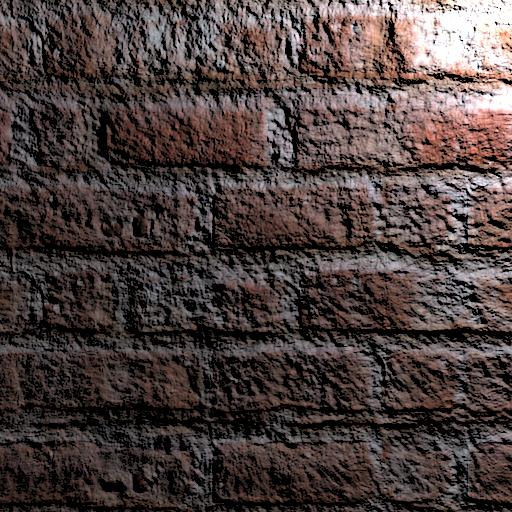}\\
        Input & MaterialGAN & MaterialGAN($20^{\circ}$)\\
	  
		\includegraphics[width=0.15\textwidth]{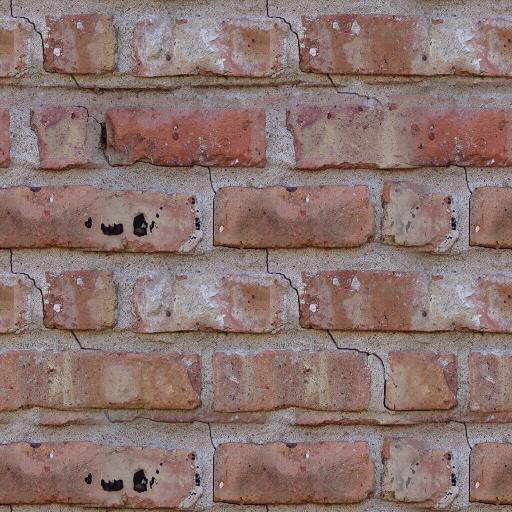} & 
	    \includegraphics[width=0.15\textwidth]{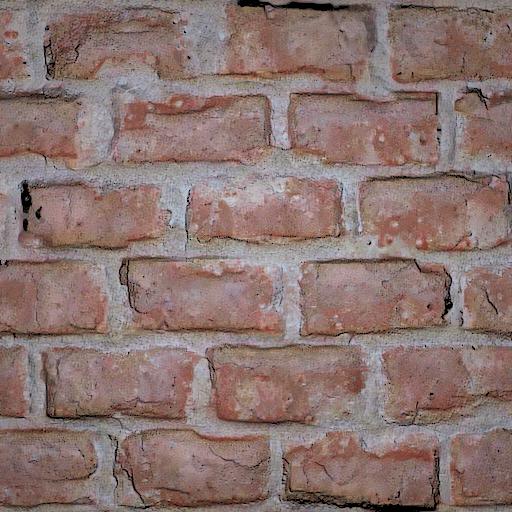}\llap{\frame{\includegraphics[width=0.07\textwidth]{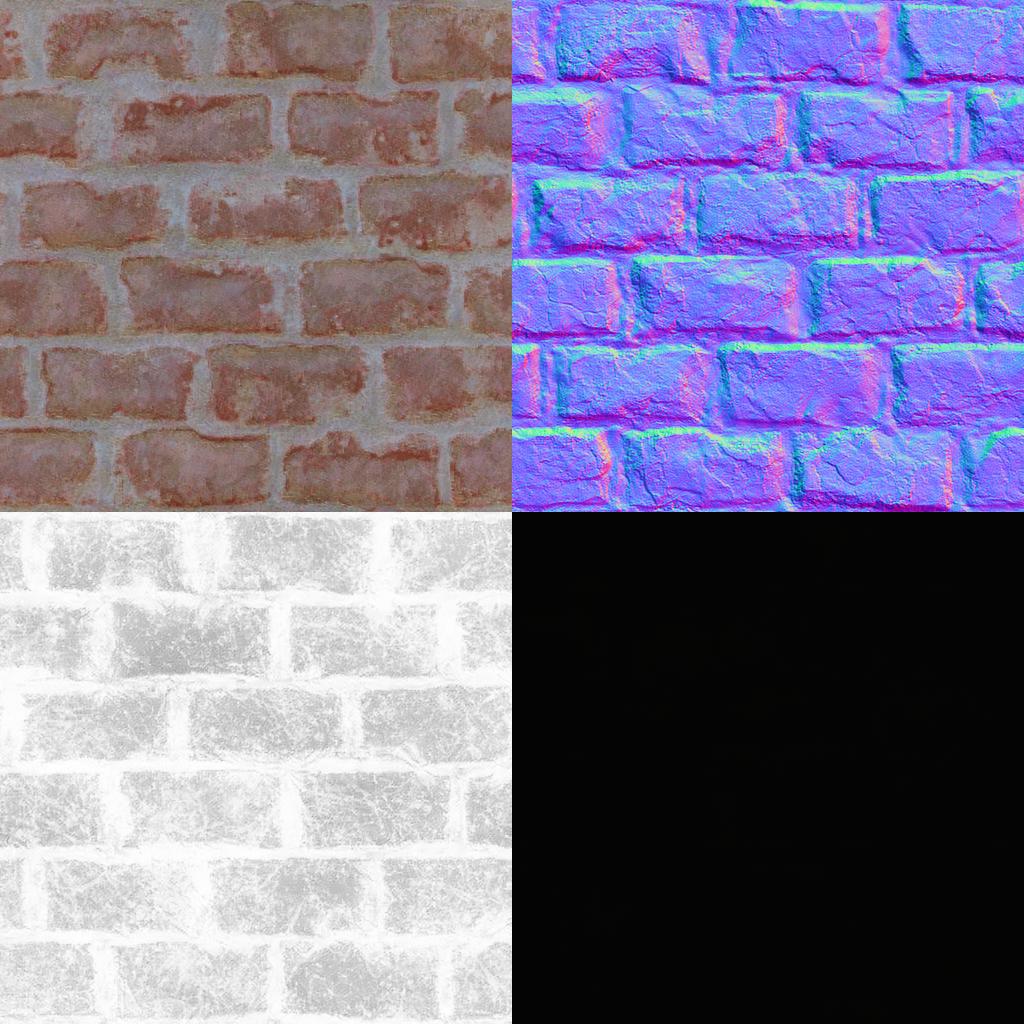}}} &
	    \includegraphics[width=0.15\textwidth]{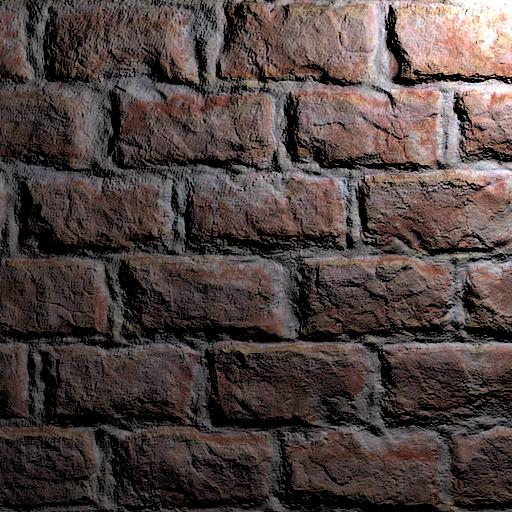}\\
        Target & Ours & Ours($20^{\circ}$)\\\
	\end{tabular}
\vspace{-10pt}
\caption{While we do not aim at reconstructing exact per-pixel material parameters given a photograph, direct single acquisition method could be used when the input material structure is not important. MaterialGAN~\cite{Guo20} achieves high quality per-pixel result on the optimized view but over-fit to it, leading to heavy quality loss under other light-view settings ($20^{\circ}$). Our approach does not suffer from this thanks to the regularization provided by the input material.} 
\label{fig:regularize}
\end{figure}
\subsection{Applications}
Using our material transfer method, we demonstrate different applications:

\textbf{Material Augmentation.} Our method can augment existing material maps based on input photos. This is particularly useful for augmenting procedural materials which are difficult to design realistically. As shown in Fig. \ref{fig:augment}, our method can be applied to minimize the gap between an unrealistic procedural material and a realistic photo, which can be used as an complementary method for existing inverse procedural material modeling systems ~\cite{hu2019, Shi20}.

\textbf{By-example Scenes Design.} As our approach ensures tileability after optimization, the transferred material maps can be directly applied to texture virtual scenes smoothly. Fig. \ref{fig:teaser} shows such an application scenario. Given photographs as exemplars, our method transfers realistic details onto simple-looking materials. With our transferred material maps, we can texture and render the entire scene seamlessly.

\begin{figure}
	\centering
	\renewcommand{\arraystretch}{0.6}
	\addtolength{\tabcolsep}{-4pt}
	\begin{tabular}{cccc}
		\includegraphics[width=0.15\textwidth]{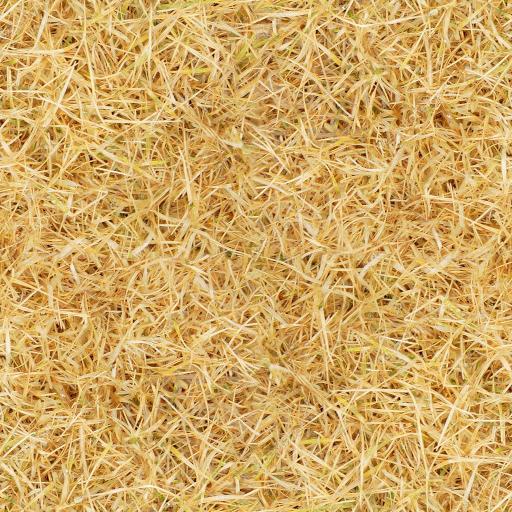} & 
		\includegraphics[width=0.15\textwidth]{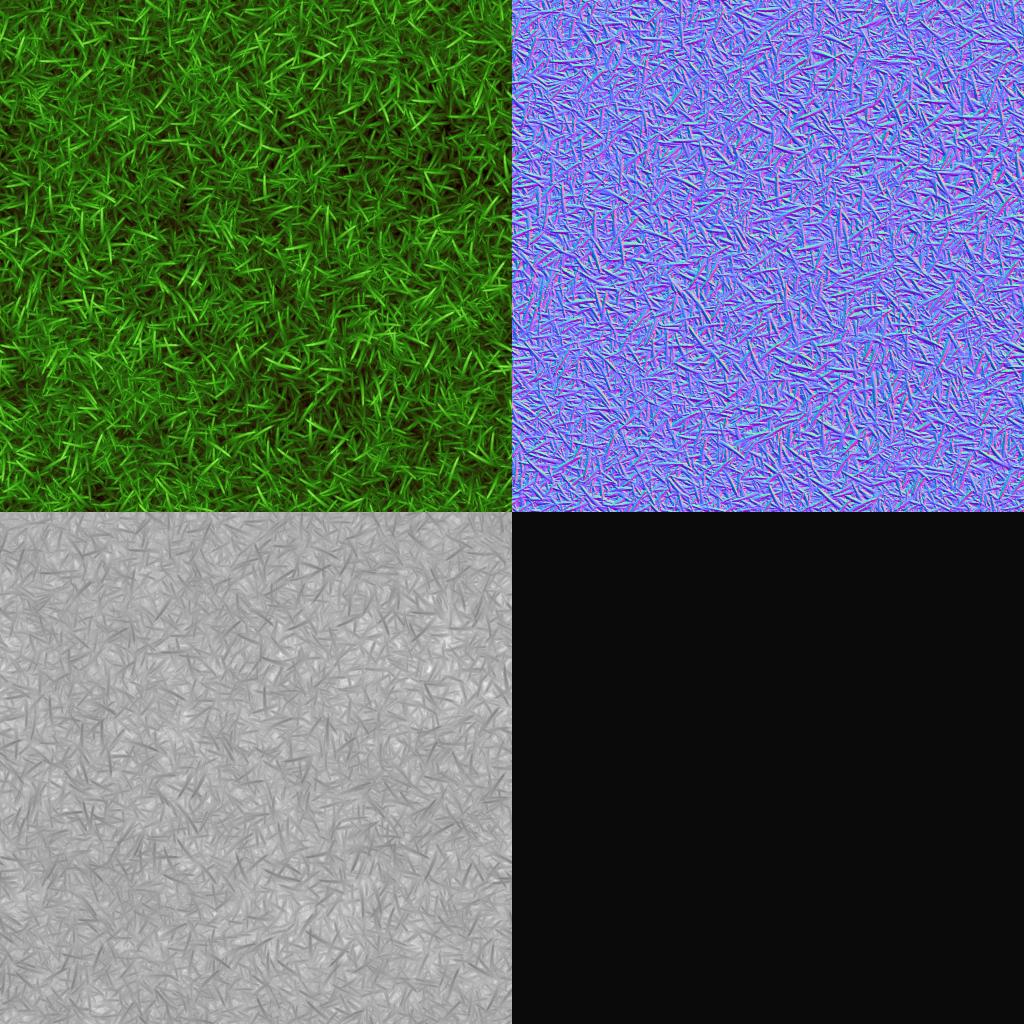} & 
		\includegraphics[width=0.15\textwidth]{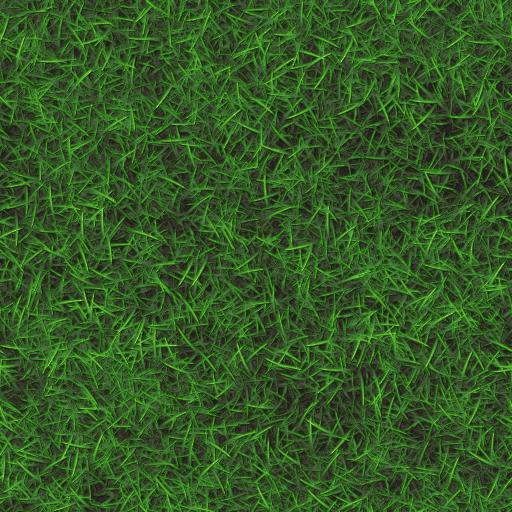} \\
        \scalebox{0.9}{Target}  & \multicolumn{2}{c}{\scalebox{0.9}{Input} }\\
		\includegraphics[width=0.15\textwidth]{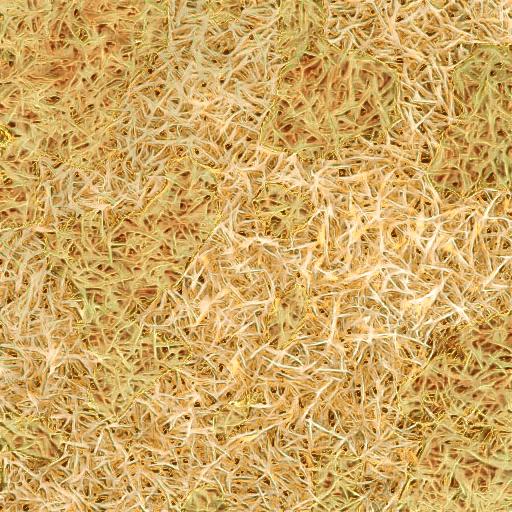} & 
		\includegraphics[width=0.15\textwidth]{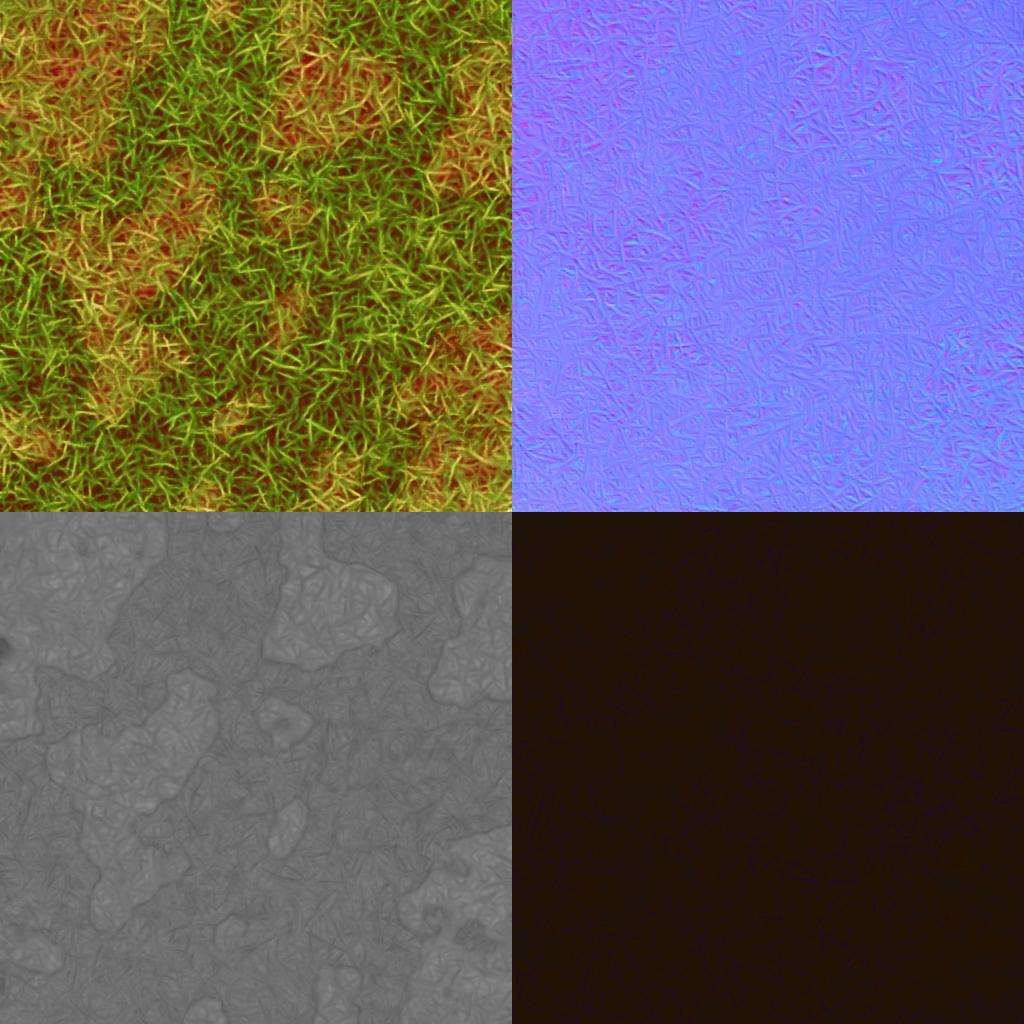} & 
		\includegraphics[width=0.15\textwidth]{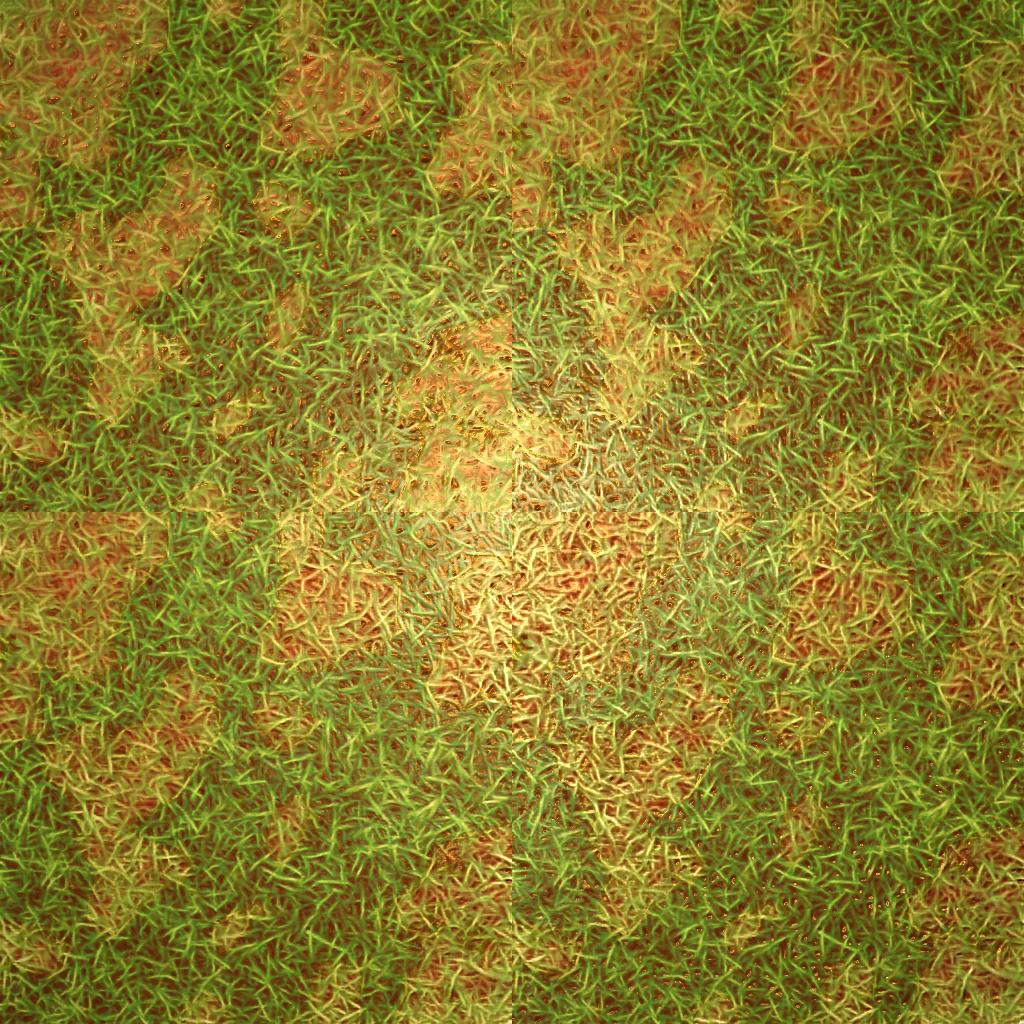} \\
		\multicolumn{2}{c}{\scalebox{0.9}{Deep Image Prior}} & \scalebox{0.9}{Tiled(2x2)} \\
		\includegraphics[width=0.15\textwidth]{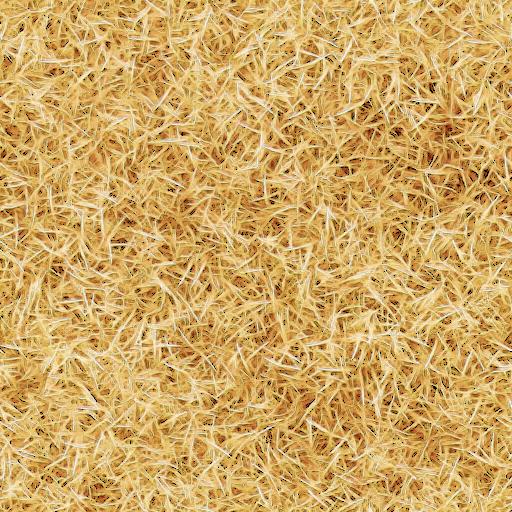} & 
		\includegraphics[width=0.15\textwidth]{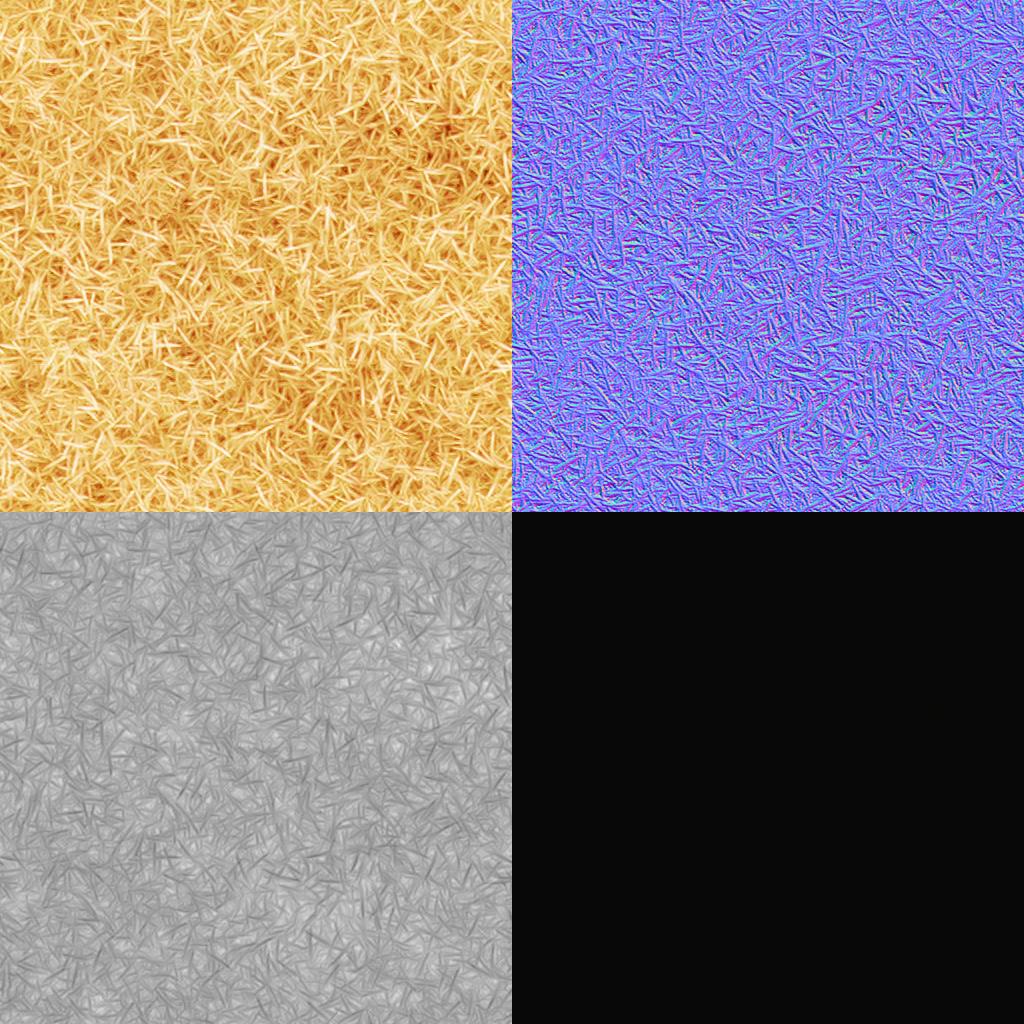} & 
		\includegraphics[width=0.15\textwidth]{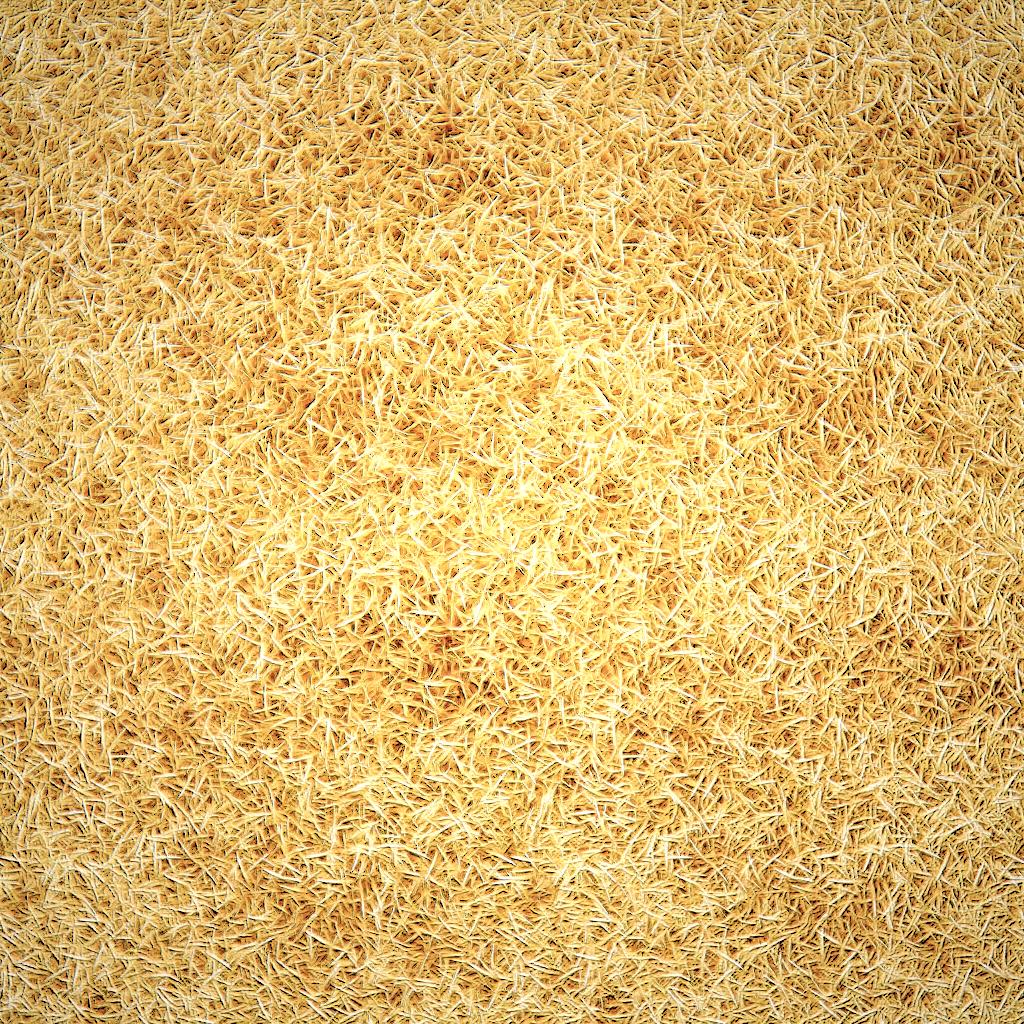} \\
	    \multicolumn{2}{c}{\scalebox{0.9}{Ours} } & \scalebox{0.9}{Tiled(2x2)}  \\
	\end{tabular}
\vspace{-10pt}
\caption{We compare our MaterialGAN prior with a material transfer based on Deep Image Prior\cite{DeepImagePrior}. Compared to a per-pixel optimization, the Deep Image Prior helps avoid local minima to reach the target appearance. However, without material prior and explicit tileable design, Deep Image Prior fails to preserve tileability and to generate view/lighting consistent result. This is illustrated here by the remaining green grass appearing when lighting is not exactly the same as during the optimization. } 
\label{fig:deep_image_prior}
\end{figure}
\subsection{Comparison}
As we cannot compare to traditional style transfer methods since they operate on the image to image domain, we evaluate the benefit of the material prior used in our optimization and compare to two alternative optimization approaches, evaluating a simple per-pixel image style transfer combined with differentiable rendering and using a deep image prior~\cite{DeepImagePrior}. We also evaluate a direct single image material acquisition using MaterialGAN on target images and show that the input material regularizes the result.

\subsubsection{Per-pixel Optimization and Deep Image Prior}
As discussed in Sec. \ref{Sec:tileable-StyleGAN}, directly performing per-pixel optimization on material maps leads to numerous local minima, resulting in artifacts. The optimization is also very sensitive to initialization and learning rate (Fig. \ref{fig:per-pixel}), and loss is prone to diverge if the hyperparameters are not tuned well.

Deep Image prior \cite{DeepImagePrior} is another way to regularize optimization. The image is reparameterized by a neural network, helping to overcome potential local minima. We adapt this idea to our material transfer optimization. As described in the original paper, we use a U-net-like architecture to generate a stacked 9-channel materials map, using a 9-channel set of randomly initialized noise maps as input. During the optimization, the parameters of this network will be optimized while the input noise map is fixed. Different from the original paper, we do not optimize from scratch but first fit the generator network to our input material maps using Eq.~\ref{Eq:project}, otherwise the optimization cannot recover the structure of the input material maps. After fitting the neural network, we perform the material transfer as described in Eq.~\ref{Eq:style-transfer}. Fig.~\ref{fig:deep_image_prior} shows an optimization result generated using this deep image prior version. The neural network prior helps address the local minimum problem encountered in the per-pixel optimization, but its lack of prior on tileable materials results in artifacts and does not preserve tileability. In particular we see in Fig.~\ref{fig:deep_image_prior} that the result looks good with a frontal lighting, but shows significant green coloration when tiled or with varying light.

\subsubsection{Comparison to Material Acquisition}
Our material transfer algorithm is not a material acquisition method because we do not aim at faithfully reconstructing the per-pixel material properties of a single material, but rather to transfer appearance statistics of one or more photographs to the input maps. However, as shown in Fig. \ref{fig:regularize}, compared to a per-pixel material acquisition approach, our optimization is regularized by the initialization with the original material, preventing it from overfitting to a single light/view configuration.

\begin{figure}[t]
    \centering
	\renewcommand{\arraystretch}{0.6}
	\addtolength{\tabcolsep}{-4pt}
    \subfloat[]{
    \label{fig:failure_cases}
    \begin{tabular}{cccc}
		\multirow{2}{*}[30pt]{\includegraphics[width=0.15\textwidth]{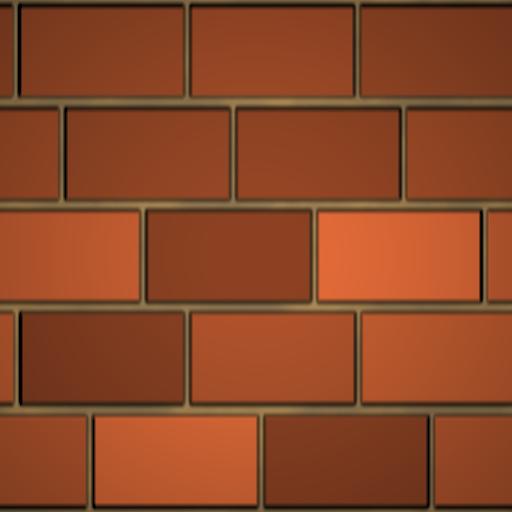}} &
        \includegraphics[width=0.15\textwidth]{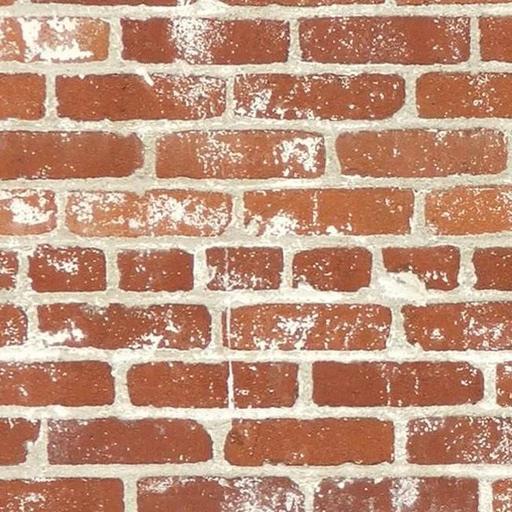} &
        \includegraphics[width=0.15\textwidth]{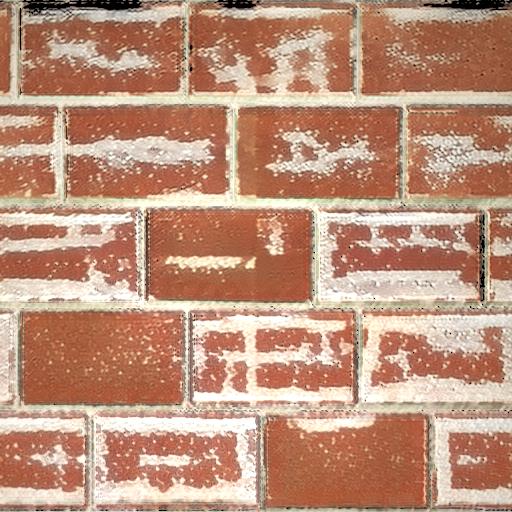} \\
         &
        \includegraphics[width=0.15\textwidth]{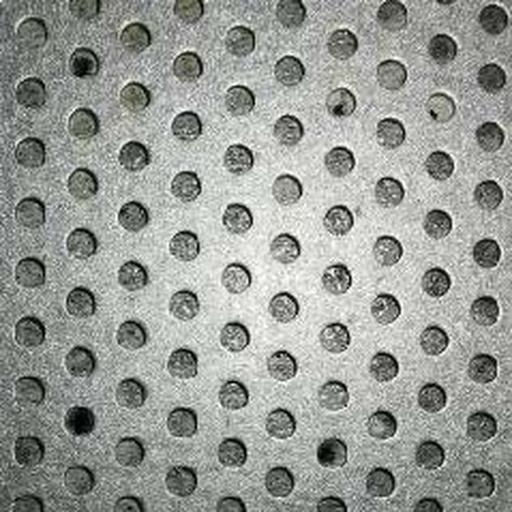} &
        \includegraphics[width=0.15\textwidth]{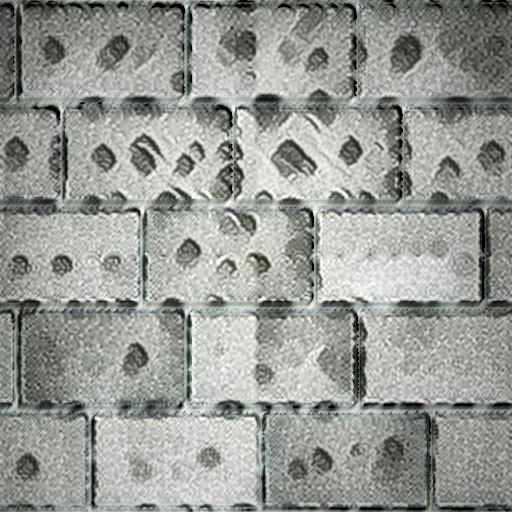} \\
		Input & Target Photos & Transferred 
	\end{tabular}}
	\vspace{-10pt}
    \newline
    
    \subfloat[]{
    \label{fig:conflicting_cues}
	\begin{tabular}{ccc}
		\includegraphics[width=0.15\textwidth]{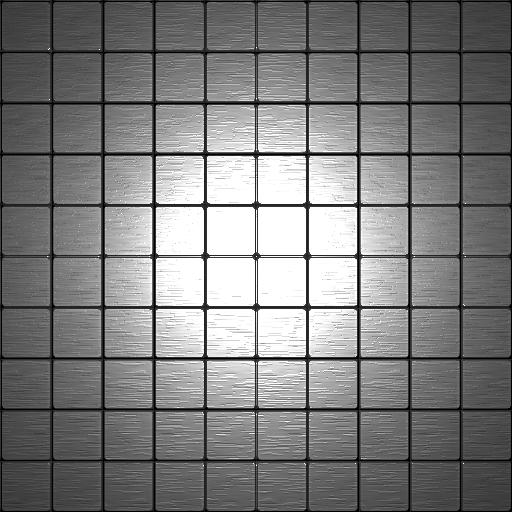}
		\llap{\frame{\includegraphics[width=0.07\textwidth]{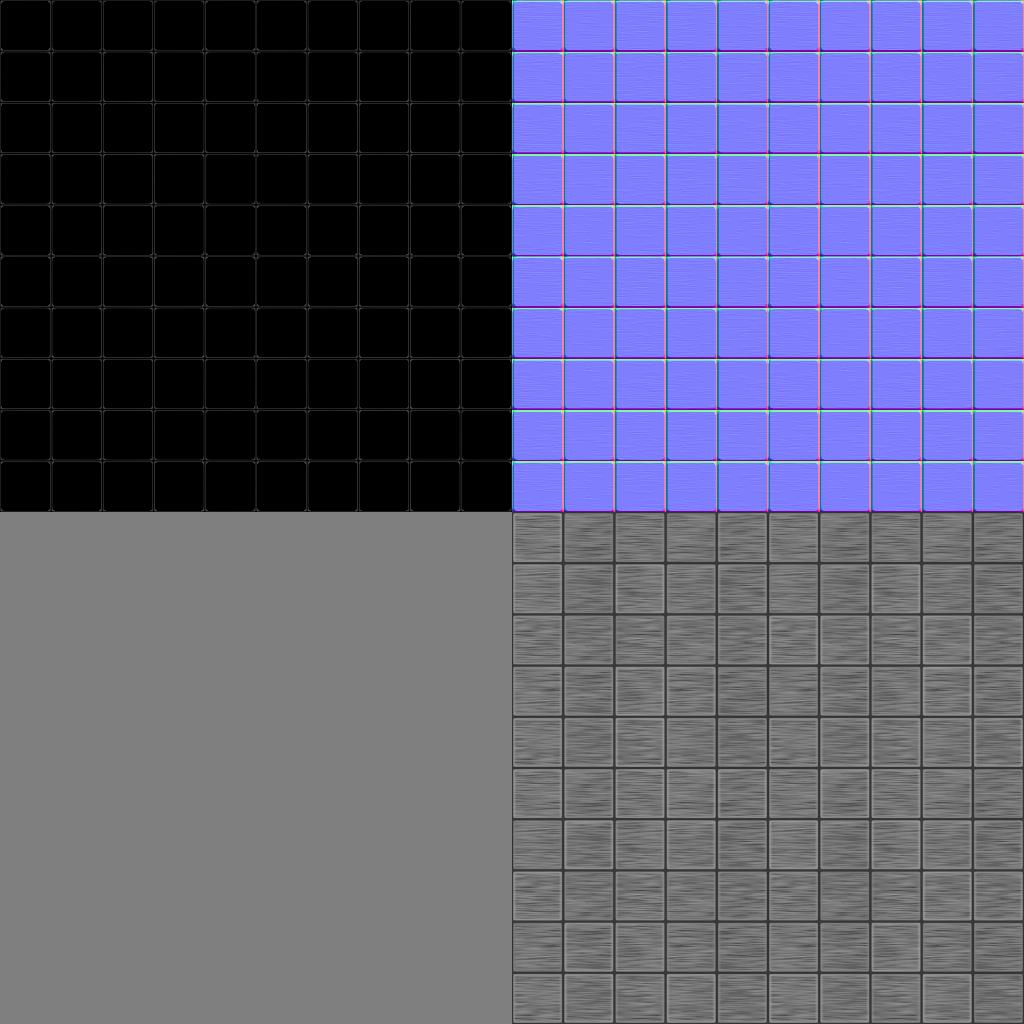}}}  &
        \includegraphics[width=0.15\textwidth]{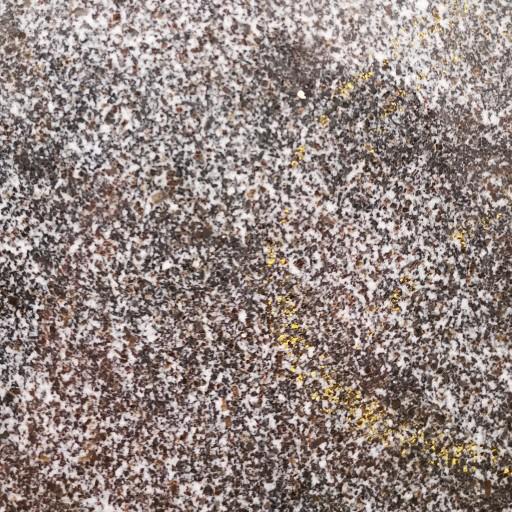} &
        \includegraphics[width=0.15\textwidth]{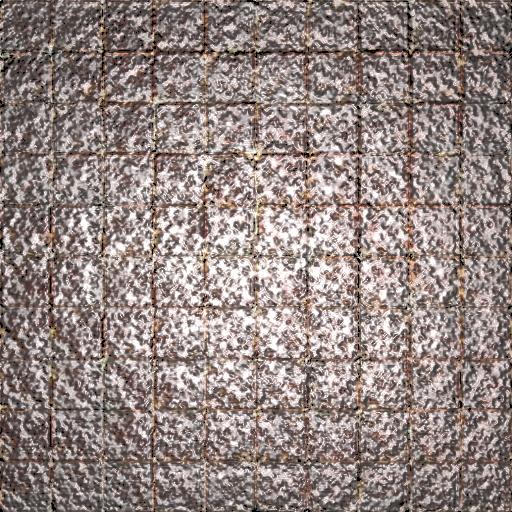}
        \llap{\frame{\includegraphics[width=0.07\textwidth]{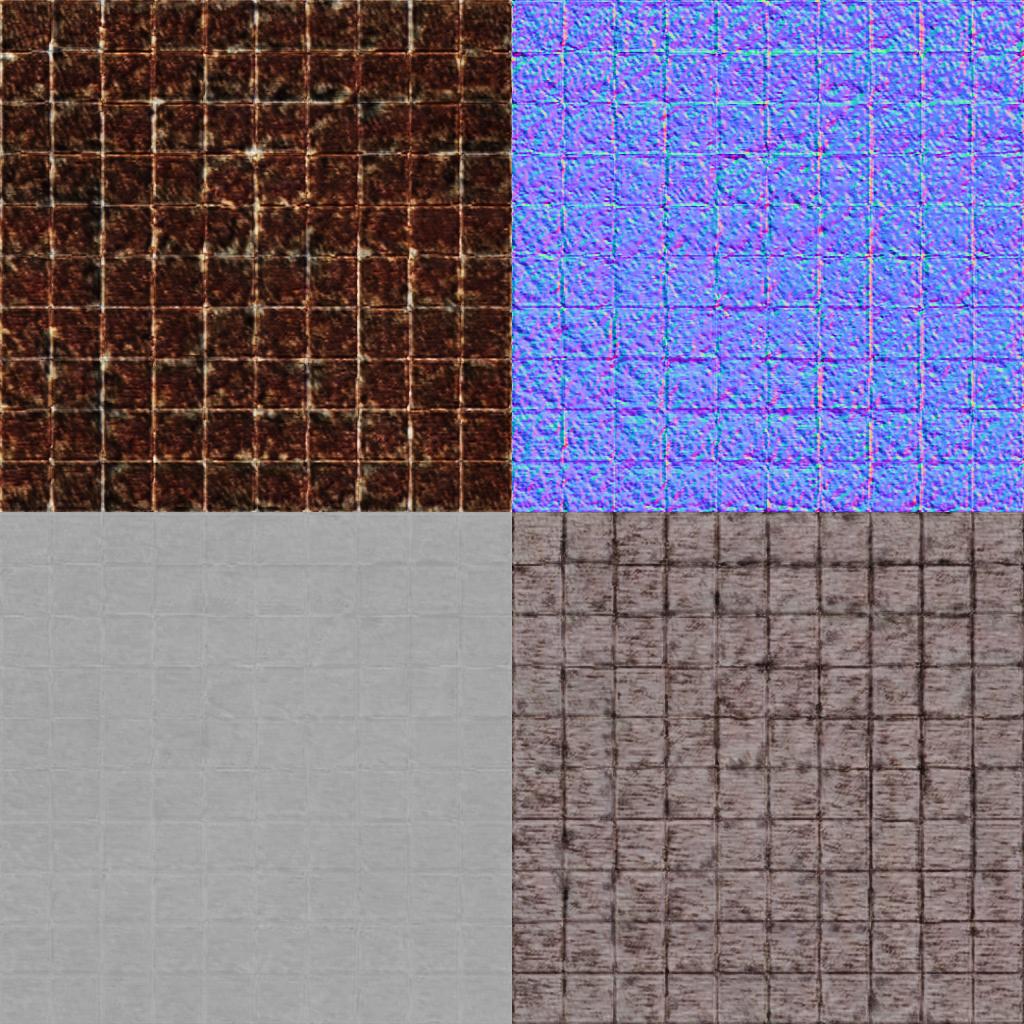}}} \\
        Input & Target & Transferred
	\end{tabular}
    }
    \vspace{-10pt}
    \label{fig:limitations}
    \caption{Limitation: (a) when input material maps and target photos have very different structures or scale, without spatial guidance (label maps), our method cannot find a good way to transfer the high frequency statistics. (b) In the case where the target photo and input material maps have strongly conflicting material properties (here roughness), our method will result in mixed parameters.}
\end{figure}
\begin{figure}[t]
    \centering
	\renewcommand{\arraystretch}{0.6}
	\addtolength{\tabcolsep}{-4pt}
    \subfloat[]{
    \centering
    \label{fig:content_weights}
	\begin{tabular}{cccc}
		\includegraphics[width=0.11\textwidth]{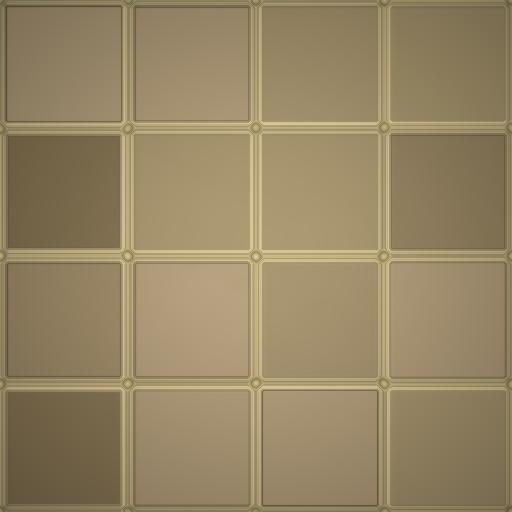} &
        \includegraphics[width=0.11\textwidth]{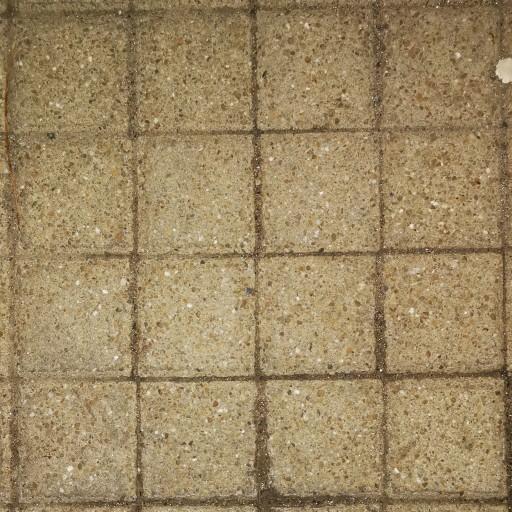} &
        \includegraphics[width=0.11\textwidth]{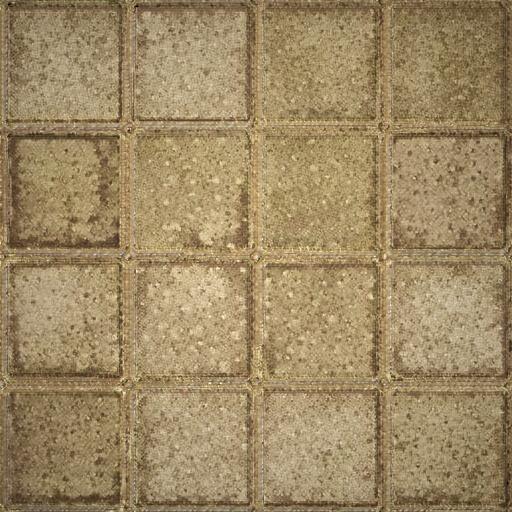} &
        \includegraphics[width=0.11\textwidth]{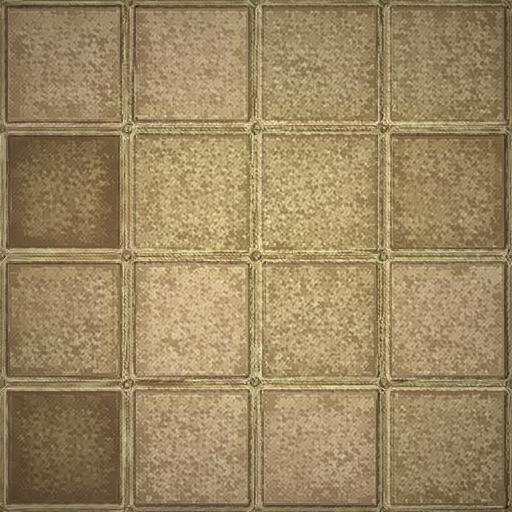} \\
        Input & Target & Our Default & Feature X8 \\
	\end{tabular}
    } 
    \vspace{-10pt}
    \newline
    
    \subfloat[]{
    \label{fig:vgg_layers}
	\begin{tabular}{cccc}
	    \includegraphics[width=0.15\textwidth]{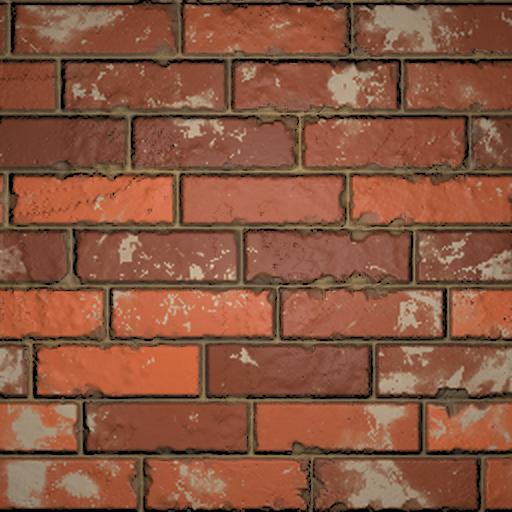} &
	    \includegraphics[width=0.15\textwidth]{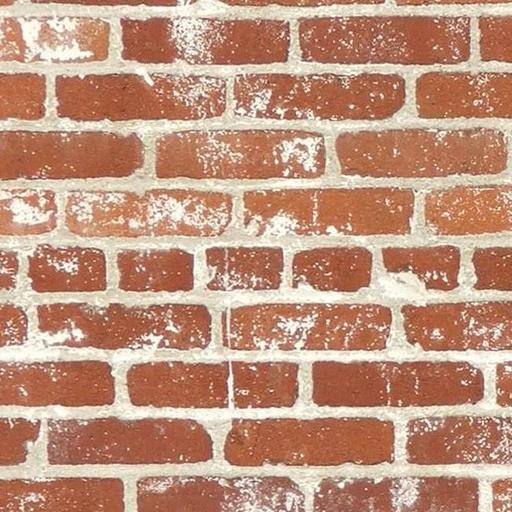} &
	    \includegraphics[width=0.15\textwidth]{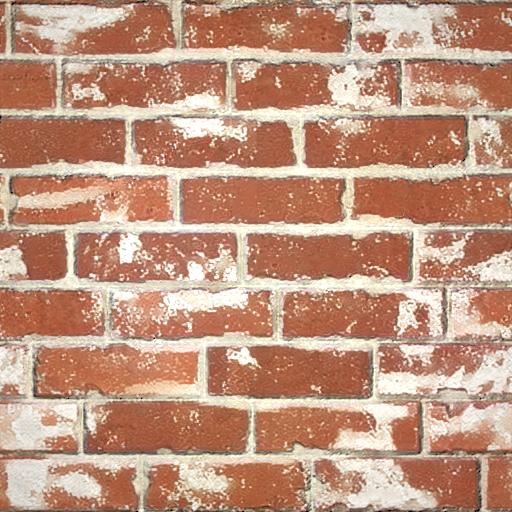}\\
        Input & Target & Our Default\\
	  
		\includegraphics[width=0.15\textwidth]{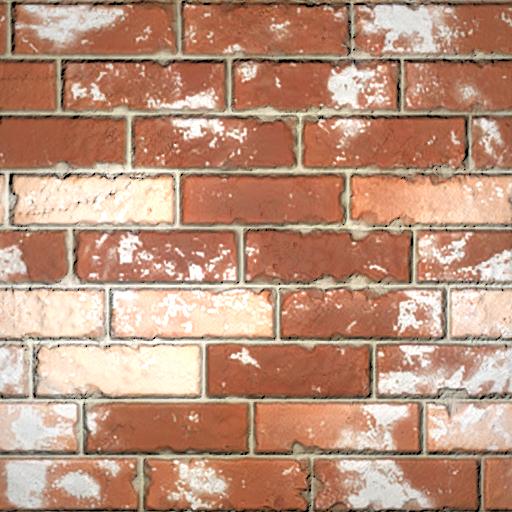} & 
	    \includegraphics[width=0.15\textwidth]{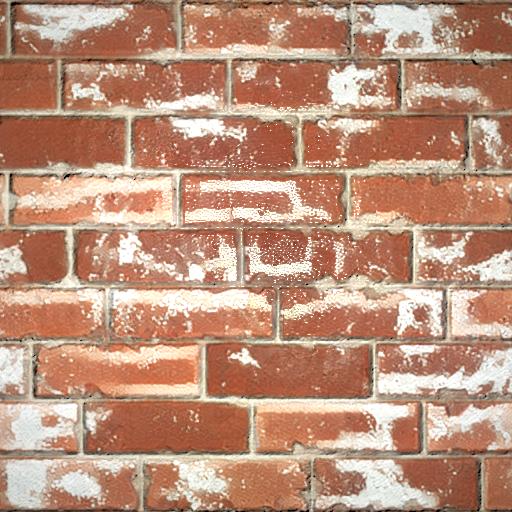} &
	    \includegraphics[width=0.15\textwidth]{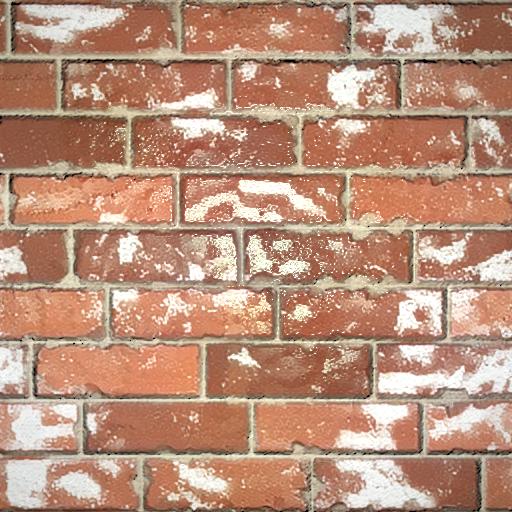}\\
        ReLu1\_1 Only & ReLu2\_1 Only & ReLu3\_1 Only
	\end{tabular}
    }
    \vspace{-10pt}
    \label{fig:weights}
    \caption{(a) Adjusting the relative strength of feature loss and style loss impacts how much statistics is transferred to the source material. As we increase the relative weight of feature loss (x8), the output will look more like the original input. (b) We combine statistics from multiple deep layers of VGG19 \cite{vgg19} for a good micro- and meso-scale transfer (our default). Transferring specific VGG layers (ReLu*\_1 Only) results in different scale of transfer, but doesn't provide precise control.} 
\end{figure}
\section{Discussions, Limitations and Future Work}
Our material transfer provides an efficient approach to control and augment appearance of material maps, showing good transfer results in different scenarios with various targets. Our method however fails to find implicit correspondences between input material maps and target photograph in the case of very different patterns or scales, without explicit spatial guidance (Sec.~\ref{Sec:Spatial-Control}), as shown in Fig.~\ref{fig:failure_cases}. In case of strongly conflicting material properties in the source material and target photo(s), our method tends to mix them. As shown in Fig. \ref{fig:conflicting_cues}, if we transfer a photo of a rough marble texture to a pure metallic material, our method mixes their material parameters.

As shown in Fig.~\ref{fig:content_weights}, to roughly control the impact of the of statistics we want to transfer, we can adjust weights between feature loss $d_0$ and style loss $d_1$. Precise control of the scale at which transfer happens is more challenging. In Fig.~\ref{fig:vgg_layers}, we experiment with transferring statistics from a single VGG layer, showing the different levels of transfer. This however doesn't provide precise transfer scale control. Empirically, we combine statistics from multiple layers (Sec.~\ref{Sec:implementation}) to produce high-quality transfer.

Another limitation of the method comes from the use of a tileable MaterialGAN which is trained at a fixed resolution. Though its base architecture, StyleGAN, has shown great results up to 1K resolution images, the model does not yet trivially support  super-high resolution materials used in large entertainment productions (4K to 8K).

Finally, we use a Cook-Torrance-like material representation, limiting the materials that can be modelled to opaque surfaces, and requiring example photographs under roughly known conditions (e.g flash or sun). An interesting future direction is to explore more complex material effects and allowing material transfer from non-planar in-the-wild objects example.

\section{Conclusion}
We design a novel algorithm to control the appearance of 2D material maps through material appearance transfer. For high-quality material transfer, we train a tileable MaterialGAN, leveraging its learned space as an optimization prior and differentiable rendering to use simple photographs as target appearance. We introduce spatial control with multi-target transfer using a resampled sliced Wasserstein loss and show complex by-example control and augmentation. The newly-synthesized material maps can be used seamlessly in any virtual environment. We believe our approach provides users with a new effective and convenient way to control the appearance of material maps and create new materials, improving the toolbox for virtual content creation. 

\section*{Acknowledgment}
This work was supported in part by NSF Grant No. IIS-2007283.

\bibliographystyle{eg-alpha-doi}  
\bibliography{bibliography} 

\end{document}